\documentclass[11pt]{scrartcl}

\usepackage[a5paper,% 
  left=9mm,%
  right=9mm,%
   top=22mm,%
  bottom=22mm%
]{geometry}

\usepackage{graphicx,amsmath}

\usepackage[shortcuts]{extdash}

\usepackage{ragged2e}

\usepackage{afterpage}

\usepackage{scrlayer-scrpage}
\clearscrheadfoot
\ohead{\pagemark}
\usepackage{lmodern}

\makeatletter
\newcommand{\add@period}[1]{#1.}
\renewcommand{\paragraph}{\@startsection{paragraph}{4}{\z@}
    {3.25ex plus 1ex minus .2ex}{-1em}{\parindent \z@ \raggedright \reset@font\sffamily\bfseries\add@period}}
\makeatother

\usepackage{booktabs}

\deffootnote[2em]{2em}{1em}{\textsuperscript{\thefootnotemark}\,}
\usepackage{authblk}

\usepackage[numbers,sort&compress]{natbib}
\usepackage{graphicx}
\graphicspath{{./figures/}}

\usepackage{lscape}

\usepackage{nicefrac}
\usepackage{amsmath,amssymb}
\usepackage{bm}

\usepackage[separate-uncertainty=true]{siunitx}

\usepackage[font=small,format=plain,%
   labelfont=bf,labelsep=period,%
 ]{caption}

\usepackage{xcolor}
\definecolor{linkColor}{RGB}{0,100,150}
\usepackage[
	colorlinks=true, allcolors = linkColor,
	pdfborder={0 0 0},
	pdfencoding = auto,
]{hyperref}

\usepackage{url}

\newcommand*{\vek}[1]{\mathbf{#1}}
\newcommand*{\dd}{\mathrm{d}}

\renewcommand{\Re}{\mathop{\mathrm{Re}}}

\def\diag{\mathop{\mathrm{diag}}}
\def\Tr{\mathop{\mathrm{tr}}}
\def\Det{\mathop{\mathrm{det}}}

\newcommand*\grad[1]{\nabla_{\kern-0.1em #1}}

\def\snc{s_\mathrm{nc}}
\def\nbar{\mkern 1mu\overline{\mkern-1mu n\mkern-2mu}\mkern 2mu}

\usepackage{framed}
\usepackage{mdframed}

\definecolor{light-gray}{gray}{0.95}
\definecolor{darkgreen}{rgb}{0,0.7,0} 
\definecolor{violet}{rgb}{0.5,0,0.5}
\definecolor{orange}{rgb}{0.8,0.5,0.2}
\definecolor{grey}{rgb}{0.3,0.3,0.3}

\usepackage{newfloat}
\DeclareFloatingEnvironment[%
	fileext=exc,%
	placement={!htb},%
	name=Exercise%
]{exercisefloat}

\newcounter{exercise}

\newenvironment{exercise}[1][!htb]{
	\begin{exercisefloat}[#1]
	\begin{mdframed}[hidealllines=true, backgroundcolor=light-gray]
	\RaggedRight
	\refstepcounter{exercise}
	\color{black}\textbf{Exercise~\theexercise.}
}{
	\end{mdframed}
	\vspace{-0.1cm}
	\end{exercisefloat}
}

\title{Self-organisation of Protein Patterns}

\subtitle{--- Lecture Notes for Les Houches 2018 Summer School on ``Active Matter and Nonequilibrium Statistical Physics'' ---}

\author{Erwin Frey\thanks{frey@lmu.de}}
\author{Fridtjof Brauns}
\affil{Arnold Sommerfeld Center for Theoretical Physics,\\Ludwig--Maximilians--Universit\"at M\"unchen,\\Theresienstra\ss e 37, D-80333 M\"unchen, Germany}
\date{}

\begin{document}

\maketitle

\clearpage
\tableofcontents

%% ================================
%% INTRODUCTION
%% ================================
\section{Introduction}
\label{sec:introduction}

One of the most striking manifestations of biological self-organisation is the spatial organisation of cells. 
Key cellular and intracellular processes, like cell division and migration, positioning of organelles, as well as differentiation and proliferation, depend on a cell's ability to dynamically organise its intracellular space. 
A prerequisite for all of these processes is a non-uniform spatial distribution of proteins, which acts as an organising template for downstream processes. 
Remarkably, even the interior of bacterial cells --- lacking a nucleus and other organelles found in eukaryotic cells --- is highly organised~\cite{Laloux:2013a,Shapiro:2009a}.
For example, so-called Min proteins in \emph{E. coli} cells oscillate back and forth between the two cell poles and thereby establish a time-averaged concentration minimum of MinC at midcell~\cite{deBoer:1989a,Raskin:1999a}.
This gradient (among possible other factors) guides the localisation of the FtsZ ring to midplane~\cite{Walker_etal:2020}, which initiates cell wall synthesis by recruiting the cell division machinery there. This positioning mechanism ensures division into equally sized daughter cells~\cite{Erickson:2010a}. 
Apart from the placement of the cell division site, protein pattern formation in bacterial cells also plays a crucial role in correct chromosome and plasmid segregation and the positioning of chemotactic protein clusters and flagella~\cite{Shapiro:2009a}.

\begin{figure}[!t]
\centering
\includegraphics[width=0.55\linewidth]{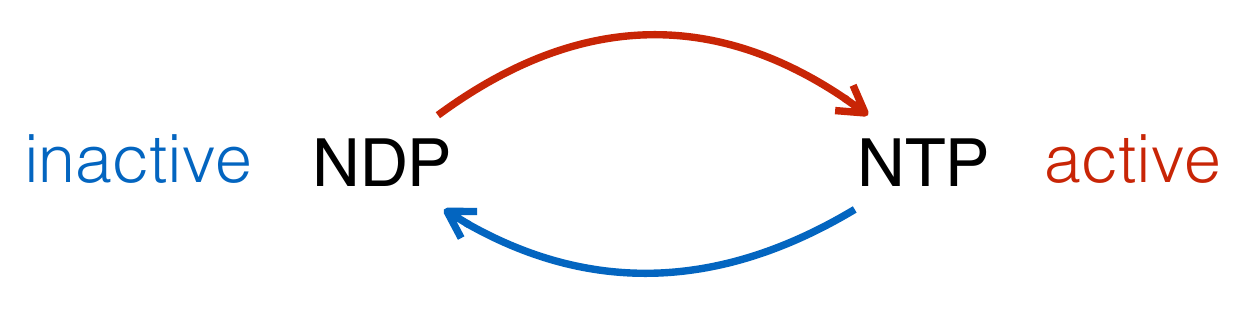} 
\caption{
\textbf{Illustration of an NTPase switch/cycle.} NTPases consume free energy to switch between active (NTP-bound) and inactive (NDP-bound) states. Proteins that bind and hydrolyse nucleoside tri-phosphates (NTP) to nucleoside di-phosphates (NDP) are crucial for almost all aspects of life~\protect\cite{Leipe:2002a}. 
}
\label{fig:NTPase_cycle}
\end{figure}

Generic design features shared by the diverse biochemical interaction networks underlying protein pattern formation in cells include: 
(i)~The dynamics (approximately) \textit{conserves the mass of each individual protein species}: on the time scale of pattern formation neither protein production nor protein degradation are significant processes. 
(ii)~The biochemical networks contain one or several NTPases that can switch between active (NTP-bound) and inactive (NDP-bound) states driven by the chemical energy provided by NTP hydrolysis~\cite{Alberts:Book,Phillips:Book}; see Fig.~\ref{fig:NTPase_cycle}. These chemical processes continually drive the system away from thermal equilibrium (i.e.\ they \emph{break detailed balance}). Therefore, (stationary) protein patterns are non-equilibrium steady states.
(iii)~The biochemical reactions are characterised by (positive and negative) \emph{feedback mechanisms} such that the chemical rate equations describing the dynamics of these reactions are generically \emph{nonlinear}.
(iv)~The proteins are either transported by \emph{diffusive fluxes} or by molecular motors along cytoskeletal filaments. In these lecture notes we will (mostly) confine ourselves to diffusive dynamics. Then, the spatiotemporal dynamics of protein patterns is described by \emph{mass-conserving reaction-diffusion} (MCRD) equations. This will be the central topic of these lecture notes.

What are the principles underlying self-organisation in such reaction-diffusion processes that result in diverse protein patterns?
Though the term `self-organisation' is frequently employed in the context of complex systems, just as we do here, we would like to emphasise that there is no universally accepted theory of self-organisation that explains how in general order and structure emerge from the interaction between a system's components.
The field which has arguably contributed most to a deeper understanding of emergent phenomena is Nonlinear Dynamics, especially with concepts such as `catastrophes'~\cite{Thom:Book}, `Turing instabilities'~\cite{Turing:1952a}, and `nonlinear attractors'~\cite{Guckenheimer:Book}.
However, although pattern formation and its underlying concepts have found their way into textbooks~\cite{CrossGreenside:Book,Murray:BookII}, we are far from answering the above question in a comprehensive and convincing way. 
Mass conservation, which is generic for intracellular protein dynamics, is typically not considered in the classical literature on pattern formation.
These lecture notes give a fresh perspective on pattern formation in mass-conserving systems by formulating their spatiotemporal dynamics in terms of geometric concepts in phase space.

We will highlight some of the recent progress in the field, but also address some of the fascinating questions that remain open. 
In the following section we will start with giving some biological background on protein-based pattern formation at a rather conceptual level. 
This is followed by introducing the general mathematical formulation of reaction-diffusion systems in cellular geometry.
Section~3 serves a twofold purpose. 
First, we discuss protein reaction kinetics of well-mixed systems. 
Second, we introduce the basic mathematical methods for analysing nonlinear systems (ordinary differential equations) in phase space.
Even though this is largely textbook material, we include a concise presentation and put a special emphasis on mass-conserving systems. 

In Section~4, we analyse spatially extended two-component reaction--dif\-fu\-sion systems. 
We start with a concise recapitulation of the pioneering work by Alan Turing~\cite{Turing:1952a}.
This serves to introduce the method of linear stability analysis for spatially extended systems, but also to highlight the differences between systems that do and do not conserve total protein mass.
The remainder of the section gives a comprehensive analysis of two-component MCRD systems, which borrows from a recent detailed analysis of such systems~\cite{Brauns.etal2020c}. 
There, we will show that the dynamics of these systems can be understood on the basis of a single underlying principle: moving local (reactive) equilibria --- controlled by local protein masses --- lead to the formation of concentration gradients which in turn drive the diffusive redistribution of globally conserved masses. 
Moreover, we will discuss how this dynamic interplay can be embedded in the phase plane of the reaction kinetics. 
In this phase plane, reaction and diffusion are represented by simple geometric objects: the reactive nullcline and the diffusive flux-balance subspace.
On this phase-space level, physical insight can be gained from geometric criteria and graphical constructions.
In particular, we will show that the pattern-forming `Turing instability' in MCRD systems is a mass-redistribution instability and that the features and bifurcations of stationary patterns can be obtained by a graphical construction in the phase-space of the reaction kinetics.

An important aspect not captured in the discussion presented in Section~4 is the role of the spatially extended cytosol. Generically, the attachment--detachment kinetics at the membrane surface lead to cytosolic gradients normal to the membrane.
Section~5 discusses the basic aspects of this bulk-boundary coupling, and introduces the linear stability analysis of laterally homogeneous steady states in systems with extended bulk. Moreover, we show that cytosolic gradients normal to the membrane can lead to geometry-induced pattern formation.
In Section~6, we will discuss the dynamics in control space, i.e.\ the space of conserved protein species, exemplified by the dynamics of Min proteins in reconstituted systems. 
We will show how the ideas of geometrically characterising the dynamics of MCRD systems can be generalised to complex biochemical systems with more than one conserved protein species.
The final section will give a concise overview over the main theoretical concepts introduced in these lecture notes and provide an outlook on how to generalise the theoretical framework of MCRD systems and apply these ideas to other non-equilibrium pattern forming systems. 

%% ================================
%% PROTEIN PATTERNS
%% ================================

\section{Protein patterns}
\label{sec:protein_patterns}

In this section, we first discuss some biological background for a set of paradigmatic, intracellular protein patterns, focusing on the underlying biochemical networks. 
This is followed by an overview of what we believe are the general design principles common to all these systems.
On that basis we introduce a general mathematical formulation of mass-conserving reaction-diffusion equations (MCRD) in cellular geometry. 

\subsection{Intracellular protein patterns}
\label{sec:intracellular_protein_patterns}

\begin{figure}
	\includegraphics{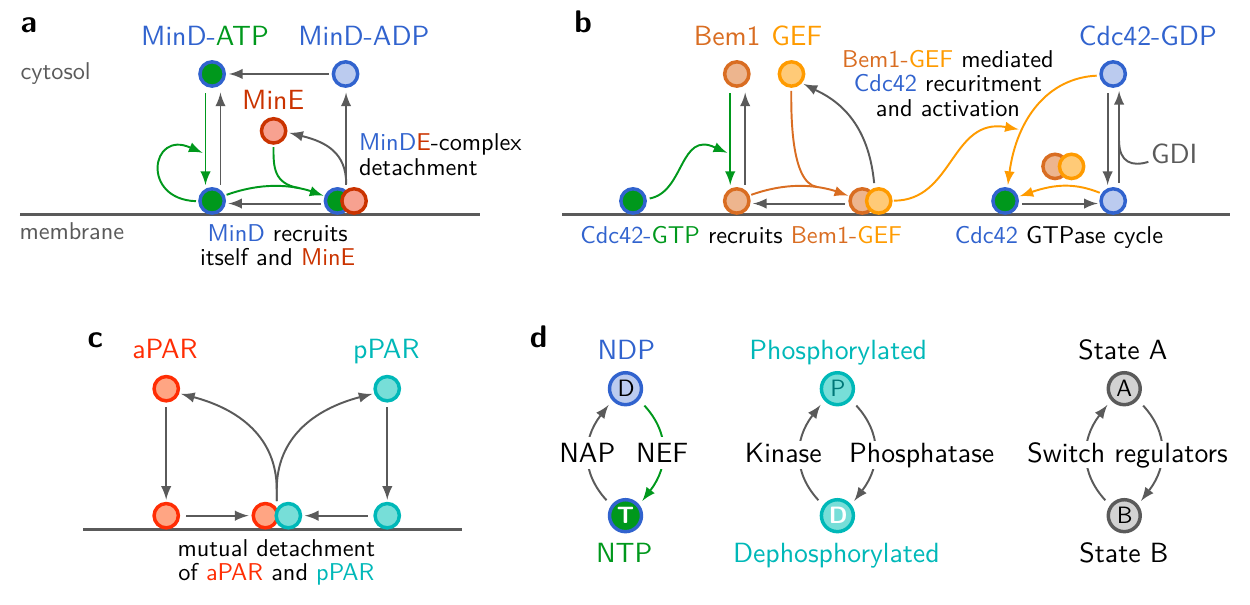} 
	\caption{Biochemical interaction networks of three model systems for self-organised intracellular pattern formation. (a) The Min system of \textit{E.~coli} \protect\cite{Huang:2003a,Halatek:2012a}. (b) Cdc42 system of \textit{S.~Cerevisiae} \protect\cite{Bi:2012a,Freisinger:2013a,Klunder:2013a}. (c) PAR system of \textit{C.~elegans} \protect\cite{Goehring:2011a}. (d) Switching between two conformal states of the proteins involved is a recurring theme in the biochemical networks (a--c). Cycling between membrane-bound and cytosolic states is driven by the ATPase/GTPase cycle of MinD and Cdc42 respectively, while the PAR-proteins each cycle between different phosphorylation states. In general we expect switching between distinct conformal states --- catalysed by ``switch regulators'' such as NTPase-activating proteins (NAPs), nucleotide exchange factors (NEFs), phophatases, and kinases --- to be a core element of biochemical networks that mediate intracellular pattern formation. Adapted from \protect\cite{Halatek:2018b}.}
\label{fig:biochemical-networks}
\end{figure}

\paragraph{The Min system in {\itshape E.~coli}}
Cell division in \textit{E.~coli} requires a mechanism that reliably directs the assembly of the Z-ring division machinery (FtsZ) to midcell \cite{Lutkenhaus:2007a}. 
How cells solve this task is one of the most striking examples for intracellular pattern formation: the pole-to-pole Min protein oscillation \cite{Raskin:1999a}. 
The Min protein system consists of three proteins, MinD, MinE, and MinC.
In its ATP-bound form, the ATPase MinD associates cooperatively with the cytoplasmic membrane (see Fig.~\ref{fig:biochemical-networks}a). Membrane-bound MinD forms a complex with MinC, which inhibits Z-ring assembly. Thus, to form a Z-ring at midcell, MinCD complexes must accumulate in the polar zones of the cell but not at midcell. 
The dissociation of MinD from the membrane is mediated by its ATPase Activating Protein (AAP) MinE, which is also recruited to the membrane by MinD, forming MinDE complexes. In this complex, MinE triggers the ATPase activity of MinD, thus initiating the detachment of both MinD-ADP and MinE. Subsequently, MinD-ADP undergoes nucleotide exchange in the cytosol such that its ability to bind to the membrane is restored (see Fig.~\ref{fig:biochemical-networks}a). The joint action of MinD and MinE gives rise to oscillatory dynamics: MinD accumulates at one cell pole, detaches due to the action of MinE, diffuses, and accumulates at the opposite pole. The oscillation period is about one minute, and during that time almost the entire mass of MinD and MinE is redistributed through the cytosol from one end of the cell to the other and back.

\paragraph{The Cdc42 system in {\itshape S.~cerevisiae}}
Budding yeast (\textit{S.~cerevisiae}) cells are spherical and divide asymmetrically by growing a daughter cell from a localised bud. The GTPase Cdc42 spatially coordinates bud formation and growth via its downstream effectors. To that end, Cdc42 must accumulate within a restricted region of the plasma membrane (a single Cdc42 cluster) \cite{Johnson:1999a}. Formation of a Cdc42 cluster, i.e.\ cell polarisation, is achieved in a self-organised fashion from a uniform initial distribution even in the absence of spatial cues (symmetry breaking) \cite{Chant:1991a}.
Like all other GTPases, Cdc42 switches between an active GTP-bound state, and an inactive GDP-bound state. Both active and inactive Cdc42 forms associate with the plasma membrane, with Cdc42-GTP having the higher membrane affinity. Furthermore, Cdc42-GDP is preferentially extracted from the membrane by its Guanine Nucleotide Dissociation Inhibitor (GDI) Rdi1, which enables it to diffuse in the cytoplasm (see Fig.~\ref{fig:biochemical-networks}b) \cite{Koch:1997a,Johnson:2009a}. Switching between GDP- and GTP-bound states is catalysed by two classes of proteins: Guanine nucleotide Exchange Factors (GEFs) catalyse the replacement of GDP by GTP, switching Cdc42 to its active state; GTPase Activating Proteins (GAPs) enhance the slow intrinsic GTPase activity of Cdc42, i.e.\ hydrolysis of GTP to GDP \cite{Bi:2012a}. Cdc42 in budding yeast has only one known GEF, Cdc24, and four GAPs: Bem2, Bem3, Rga1, and Rga2. A key player of the Cdc42 polarisation machinery is the scaffold protein Bem1 which is recruited to the membrane by Cdc42-GTP, and itself recruits Cdc42's GEF (Cdc24) to form a Bem1--GEF complex (Fig.~\ref{fig:biochemical-networks}b) \cite{Bose:2001a,Butty:2002a,Irazoqui:2003a}. In turn, Bem1--GEF complexes recruit Cdc42 to the membrane and activate it there, thus closing a positive feedback look (mutual recruitment) that drives Cdc42 polarisation.

\paragraph{The PAR system in {\slshape C.~elegans}}
So far we have discussed examples for intracellular pattern formation in unicellular prokaryotes (Min oscillations in \emph{E.\ coli}) and in eukaryotes (Cdc42 polarisation in \emph{S.\ cerevisiae}). 
A well studied instance of intracellular pattern formation in multicellular organisms is the establishment of the anterior-posterior axis in the \textit{C.~elegans} zygote \cite{Goehring:2011a,Hoege:2013a,Munro:2004a}. 
The key players here are two groups of PAR proteins: The aPARs (PAR-3, PAR-6, and aPKC) localise in the anterior half of the cell. The pPARs (PAR-1, PAR-2, and LGL) localise in the posterior half. 
In the wild type, polarity is established upon fertilisation by cortical actomyosin flow oriented towards the posterior centrosomes, in other words by active transport of pPAR proteins \cite{Munro:2004a,Goehring:2011a}. 
After polarity establishment, this flow ceases, but polarity is maintained. In addition, it has been shown that polarity can be established without flow \cite{Goehring:2011a}. These results suggest that PAR protein polarity in \textit{C.~elegans} is based on a reaction-diffusion mechanism.
The protein dynamics are based on the antagonism between membrane-bound aPAR and pPAR proteins, mediated by mutual phosphorylation which initiates membrane detachment at the interface between aPAR and pPAR domains near midcell (see Fig.~\ref{fig:biochemical-networks}c).
Thus, PAR-based pattern formation is driven by (mutual) detachment where opposing zones come into contact, and is therefore quite different than the attachment (recruitment) based systems discussed above. 

\subsection{General biophysical principles of intracellular pattern formation}
\label{sec:biochemical_principles}

In all examples discussed above the biological function associated with the respective pattern is mediated by membrane-bound proteins alone, in other words: an important class are \emph{intracellular patterns are membrane-bound patterns}. 
Furthermore, the diffusion coefficients of membrane-bound proteins are generically at least two orders of magnitude lower than those of their cytosolic counterparts, e.g.\ a typical value for diffusion along a membrane would be between \SI{0.01}{\micro m^2.s^{-1}} and \SI{0.1}{\micro m^2.s^{-1}}, while a typical cytosolic protein has a diffusion coefficient of about \SI{10}{\micro m^2.s^{-1}}, see e.g.\ \cite{Bendezu:2015a,Meacci:2006a}. 

The key unifying feature of all protein interaction systems is switching between different protein states or conformations. The conformation (state) of a protein can change as a consequence of interactions with other biomolecules (lipids, nucleotides, or other proteins). Likewise, the interactions available to a protein are determined by  its conformation. This can be summarised as the switching paradigm of proteins (Fig.~\ref{fig:biochemical-networks}d), which is best exemplified for NTPases such as MinD or Cdc42 whose dynamics are in essence driven by deactivation and reactivation through nucleotide exchange. The phosphorylation and dephosphorylation of PAR proteins by kinases and phosphatases, respectively, exemplifies the same principle. In all these cases, switching is tied to membrane affinity, and thus to the flux of proteins into and out of the cytosol.

\begin{figure}[tb]
\centering
\includegraphics{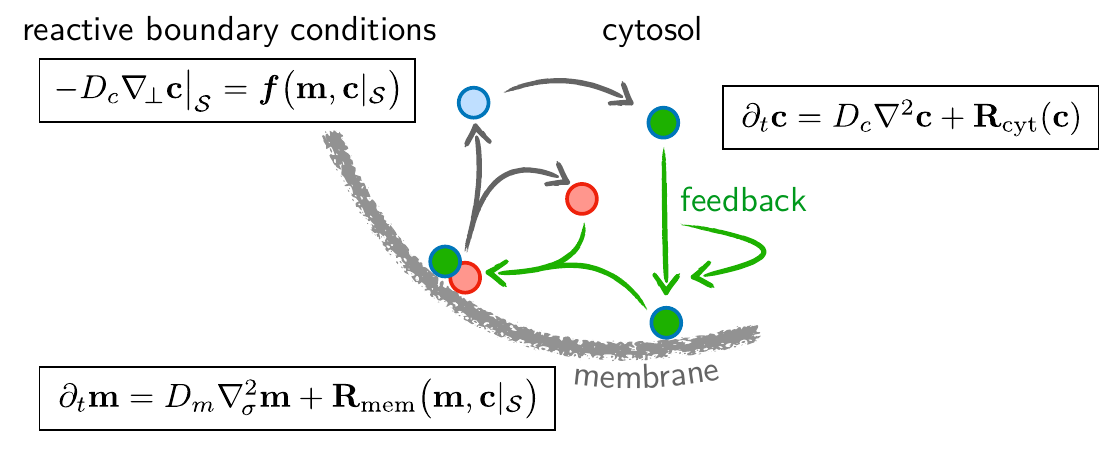}
\caption{
\textbf{Illustration of general biophysical principles of intracellular pattern formation}.
At the heart of intracellular pattern-forming systems are NTPases that can switch between active and inactive states; here indicated by green-filled and blue-filled spheres.
Proteins cycle between membrane-bound and cytosolic states conserving the total mass of each protein species. 
Local conservation of protein mass requires reactive boundary conditions such that the net reactive fluxes of a given protein state between membrane and cytosol equal the diffusive fluxes of the respective proteins in the cytosol.  
In each of the compartments, the protein dynamics is described by a set of coupled reaction--diffusion equations, as indicated in the graph.
The reaction kinetics is mass-conserving: on the time scale of pattern formation both protein production and protein degradation are negligible.
}
\label{fig:switch_paradigm}
\end{figure}

Dynamics based on conformational switching conserve the copy number of the protein. Therefore, intracellular protein dynamics are generically represented by mass-conserving reaction--diffusion systems --- and pattern formation in a mass-conserving system can only be based on transport (redistribution), it cannot depend on production or degradation of proteins. In the absence of active transport mechanisms (such as vesicle trafficking) the only available transport process is molecular diffusion. 
Given that membrane-bound proteins barely diffuse, we can assert that the biophysical role of the cytosol in these systems is that of a (three-dimensional) `transport layer'. 
Hence, the (functionally relevant) membrane-bound protein pattern must originate from redistribution via the cytosol, i.e.\ the coupling of membrane detachment in one spatial region of the cell to membrane attachment in another region, through the maintenance of a diffusive flux in the cytosol (Fig.~\ref{fig:switch_paradigm}).  
However, transport by diffusion eliminates concentration gradients. Hence, if a diffusive flux is to be maintained, a gradient needs to be sustained. Note that due to fast cytosolic diffusion, this  gradient can be rather shallow and still induce the flux necessary to establish the pattern (the flux is simply given by the diffusion coefficient times the gradient).

\subsection{Reaction-diffusion equations in cellular geometry} 
\label{sec:MCRD_cell_geometry}

\def\nD{n_\mathrm{D}^{}}
\def\nE{n_\mathrm{E}^{}}

\def\cD{c_\mathrm{D}^{}}
\def\cDD{c_\mathrm{DD}^{}}
\def\cDT{c_\mathrm{DT}^{}}
\def\cE{c_\mathrm{E}^{}}

\def\md{m_\mathrm{d}^{}}
\def\mde{m_\mathrm{de}^{}}

\def\kD{k_\mathrm{D}^{}}
\def\kdD{k_\mathrm{dD}^{}}
\def\kdE{k_\mathrm{dE}^{}}
\def\kde{k_\mathrm{de}^{}}

Based on the above general biophysical principles we now formulate a general set of mass-conserving reaction-diffusion equations in cellular geometry. 

\paragraph{Cellular geometry: membrane and cytosol}

\begin{figure}[!h]
\centering
\includegraphics[scale=1.25]{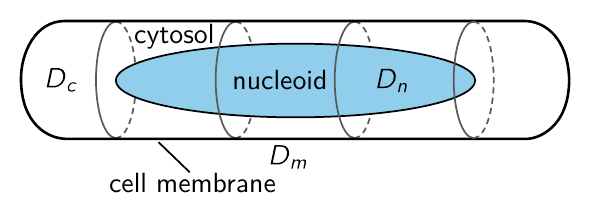}
\caption{\textbf{Schematic representation of the geometry of a rod-shaped bacterial cell.} There are three main cellular compartments: cell membrane, cytosol, and nucleoid. The diffusion constants in these compartments are in general different. Adapted from \protect\cite{Frey:2018a}.
}
\label{fig:cell_geometry}      
\end{figure}

Figure \ref{fig:cell_geometry} illustrates the geometry of a rod-shaped prokaryotic cell. 
It is comprised of three main compartments: the cell membrane, the cytosol, and the nucleoid. 
There are two major facts that are relevant for intracellular pattern formation. 
First, the diffusion constants in the cytosol and on the cell membrane are vastly different. Typical values are of the order of $D_c \approx \SI{10}{\micro m.s^{-1}}$, in the cytosol and $D_m \approx  \SI{0.01}{\micro m.s^{-1}}$ on the membrane. 
Second, due to the rod-like shape, the ratio of cytosolic volume to membrane area differs markedly between polar and midcell regions. 
Beyond this local variation of volume to surface ratio, the overall ratio of cytosol volume to membrane area depends on the shape of the cell.

Consider now a set of proteins that can take different conformational states.
As an example, think of the Min system that has two proteins, MinD and MinE, each of which can be in chemically different conformations, for instance, MinD-ATP, MinD-ADP, and the MinDE heterodimer.
As proteins maybe either membrane-bound or cytosolic, we explicitly distinguish cytosolic states with concentrations $\vek{c}(\bm{x},t) = \{c_\alpha^{}(\bm{x},t)\}_{\alpha}^{}$ and states bound to the membrane surface $\mathcal{S}$ with the concentrations $\vek{m}(\bm{\sigma},t) = \{m_\beta^{} (\bm{\sigma},t)\}_\beta^{}$ with $\bm{\sigma} \in \mathcal{S}$.
The indices $\alpha$ and $\beta$ indicate a protein in a certain conformational state in the cytosol and on the membrane respectively. 
We will collectively refer to the protein concentrations as $\vek{u} = (\vek{m},\vek{c})$.

\paragraph{Dynamics in the cytosolic volume (3d)} 
In a general form, the reaction-diffusion equations for proteins diffusing in the cytosol read
\begin{equation}
	\partial_t 	
	\vek{c} (\vek{x},t)
	= 
	D_{c} \nabla^{2} 
	\vek{c} (\vek{x},t) 
	+ 
	\vek{R}_\mathrm{cyt}
	\left( \vek{c} (\vek{x},t) \right)
	\, ,
\end{equation}
where for simplicity we have assumed that the diffusion constants of all proteins in the cytosol have the same value $D_c$.
The terms collected in the vector $\vek{R}_\mathrm{cyt}$ characterise the chemical reactions taking place in the cytosol.
Typically, these reactions are only between different cytosolic proteins, and therefore the functions $\vek{R}_\mathrm{cyt}$ depend only on the cytosolic concentrations $\vek{c}$.
As an example we take the biochemical reaction scheme for Min proteins as shown in Fig.~\ref{fig:biochemical-networks}a.
In this scheme there is only one MinE conformation with volume concentration $\cE (\vek{x},t)$, and two MinD conformations corresponding to active MinD-ATP and inactive MinD-ADP with volume concentrations $\cDT (\vek{x},t)$, and $\cDD (\vek{x},t)$, respectively.
There is only one cytosolic reaction, namely the reactivation of cytosolic MinD-ADP by nucleotide exchange (with rate $\lambda$) to MinD-ATP.
Hence, the set of reaction-diffusion equations read:
\begin{subequations}
\label{eq:Min-cytosl-dyn}
\begin{align}
	\partial_{t} \cDD (\vek{x},t)
	&= D_{c}\nabla^{2} \cDD -
	   \lambda \, \cDD  \, ,
	\label{eq:de1} \\
		\partial_{t}\cDT (\vek{x},t)
	&= D_{c}\nabla^{2}\cDT + 
	   \lambda \, \cDD  \, ,
	\label{eq:de2}\\
	\partial_{t}\cE (\vek{x},t) 
	&= D_{c}\nabla^{2}\cE  \, .
	\label{eq:de3}
\end{align}	
\end{subequations}
A more complex and biochemically more realistic reaction scheme could include two different cytosolic states of MinE, a reactive state and a latent state~\cite{Denk:2018a}.
Then, Eq.~\eqref{eq:de3}, generalises to 
%\begin{subequations}
\begin{align*}
	\partial_{t}
	c_\text{Er}^{} (\vek{x},t)  
	& =  
	D_{c}\nabla^{2} 
	c_\text{Er}^{}
	- \mu \, c_\text{Er}^{} \, , \\
	\partial_{t}
	c_\text{El}^{} (\vek{x},t) 
	& =  
	D_{c} \nabla^{2}
	c_\text{El}^{}  
	+ \mu \, c_\text{Er}^{}  \, .
\end{align*}
%\end{subequations}
This extension of the reaction scheme actually has important implications on the robustness of the patterns formed~\cite{Denk:2018a}.

\paragraph{Dynamics on the membrane surface (2d)} 
The reaction-diffusion equations for membrane-bound proteins are more complex for two reasons. 
First, diffusion is constrained to the membrane surface $\bm{\sigma} \in \mathcal{S}$ with $\grad{\mathcal{S}}^{2}$ the Laplace--Beltrami operator on that surface.
Second, the reactions in general depend on both the concentration of proteins on the membrane $\vek{m} (\bm{\sigma},t)$ and the cytosolic concentrations in the immediate vicinity of the membrane surface, $\vek{c}|_\mathcal{S} := \vek{c}(\vek(x),t)|_{\vek{x} \in\mathcal{S}}$: 
\begin{equation}
	\partial_t 	
	\vek{m} (\bm{\sigma},t)
	= 
	D_m \grad{\mathcal{S}}^{2} 
	\vek{m} (\bm{\sigma},t) 
	+ 
	\vek{R}_\mathrm{mem}
	\big(
	  	\vek{m} (\bm{\sigma},t),
	  	\vek{c}|_\mathcal{S} (\bm{\sigma},t)
	\big)
	\, ,
\end{equation}
where we have for simplicity assumed that the diffusion constants $D_m$ of all membrane-bound conformations of all proteins are equal.
As an example, we again take the Min reaction network illustrated in Fig.~\ref{fig:biochemical-networks}a.
There are two different protein conformations on the membrane: active MinD-ATP with areal density $\md (\bm{\sigma},t)$, and heterodimers comprised of active MinD and MinE with areal density $\mde (\bm{\sigma},t)$.
As illustrated in Fig.~\ref{fig:biochemical-networks} the chemical reactions are~\cite{Huang:2003a,Halatek:2012a}:

\begin{itemize} 
\item \textit{Spontaneous attachment} of active MinD to the membrane (with attachment rate $k_\mathrm{D}$) and \textit{recruitment of cytosolic MinD-ATP} by already membrane-bound active MinD (with rate $\kdD \md$): 
\begin{equation*}
	R_\mathrm{d}^+ 
	=  
	\left(
	\kD + \kdD \, \md
	\right) \cDT |_\mathcal{S}^{}
	\, .
\end{equation*}
The superscript $(+)$ indicates that this reaction leads to an increase of proteins on the membrane, i.e.\ we have a protein flux from the cytosol to the membrane.
In the following we call $\kdD$ the recruitment rate.
\item \textit{Recruitment of cytosolic MinE} by membrane-bound MinD-ATP (with rate $\kdE \md$):
\begin{equation*}
	R_\mathrm{de}^+ 
	= 
	\kdE \, 
	\md \, \cE|_\mathcal{S}^{}
	\, .
\end{equation*}
Similar as above this is a flux towards the membrane. We assume that after recruitment  MinD-ATP and MinE form a  membrane-bound MinDE complex.
\item As MinE is an ATPase activating protein (AAP) it stimulates \textit{ATP hydrolysis} (with hydrolysis rate $k_\mathrm{de}$) turning active MinD-ATP into inactive MinD-ADP which leads to detachment and decay of the membrane-bound MinDE complexes into cytosolic MinD-ADP and MinE:
\begin{equation*}
	R_\mathrm{de}^- 
	= 
	\kde \, \mde
	\, .
\end{equation*}
As this amounts to a loss of proteins from the membrane into the cytosol we indicate this with a superscript $(-)$.
\end{itemize}
Taken together, the reaction-diffusion equations on the membrane read
\begin{subequations} \label{eq:Min-membrane-dyn}
\begin{align}
	\partial_{t} 
	\md (\bm{\sigma},t)
	&= D_{m} \grad{\mathcal{S}}^{2} m_\mathrm{d\phantom{e}}
	 + R_\mathrm{d\phantom{e}}^+(\vek{m},\vek{c}|_\mathcal{S})
	 - R_\mathrm{de}^+(\vek{m},\vek{c}|_\mathcal{S}), \, 
	\\
	\partial_{t}\mde (\bm{\sigma},t) 
	&= D_{m}\grad{\mathcal{S}}^{2}\mde + 
	   R_\mathrm{de}^+(\vek{m},\vek{c}|_\mathcal{S}) 
	   - R_\mathrm{de}^-(\vek{m},\vek{c}|_\mathcal{S})
	  \, . 
\end{align}
\end{subequations}
Extending this so called \emph{skeleton model} of the Min system to include the switching of MinE between two cytosolic states gives~\cite{Denk:2018a}
\begin{equation*}
	R_\mathrm{de}^+ 
	\to 
	R_\mathrm{recruit, El}^+ + R_\mathrm{recruit, Er}^+
	= 
	\big( 
	k_\mathrm{dE}^\mathrm{l} \, c_\mathrm{El}^{}|_\mathcal{S}^{}
	+
	k_\mathrm{dE}^\mathrm{r} \, c_\mathrm{Er}^{}|_\mathcal{S}^{}
	\big) \, \md 
	\, ,
\end{equation*}
where $k_\mathrm{dE}^\mathrm{r}$ and $k_\mathrm{dE}^\mathrm{l}$ are the recruitment rates for reactive and latent MinE respectively.

\begin{figure}[tb]
\centering
\includegraphics{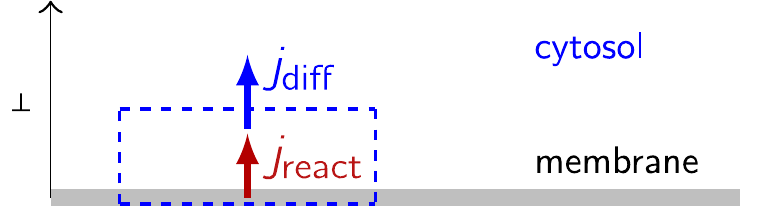}
\caption{
\textbf{Illustration of reactive boundary conditions.} 
For a given protein state, the net reactive flux from the membrane into the cytosol, $j_\text{react}$, has to equal the cytosolic diffusive flux of that protein state right at the membrane, $j_\text{diff}$. This guarantees local mass conservation: an equal number of proteins flow into and out of the infinitesimal volume indicated by the blue dashed box. The arrow signified with $\perp$ indicates the inward normal on the membrane.
}
\label{fig:reactive_bc}
\end{figure}

\paragraph{Reactive boundary conditions}
The dynamic equations given above for the cytosol and the membrane are not complete. 
One needs to specify the boundary conditions that capture the coupling between cytosol and membrane by chemical reactions that involve the attachment and detachment of proteins from the cytosol onto the membrane and vice versa.
These boundary conditions are defined by the condition that the net reactive fluxes $j_\mathrm{react}$ equal the diffusive flux due to cytosolic gradients normal to the membrane, $j_\mathrm{diff}$ in order to guarantee local particle number conservation; see Fig.~\ref{fig:reactive_bc} for an illustration.
Mathematically, this condition can be formulated using Gauss' divergence theorem by integrating the bulk dynamics over an infinitesimal volume (dashed box in the figure) at the membrane, where the reactive flow is represented by an additional source/sink located at the boundary. Thus one obtains
\begin{equation*}
	-j_\mathrm{react} = 
	\lim_{\varepsilon \rightarrow 0} \int_0^\varepsilon \dd x_\perp \big[D_c \nabla_{\!\perp}^2 c -\partial_t c \big] =
	- \lim_{\varepsilon \rightarrow 0} D_{c}\nabla_{\!\perp} c|^{}_\varepsilon = j_\mathrm{diff}
\end{equation*}
where we have introduced $\nabla_{\!\perp} = \vek{n} \cdot \nabla$, the gradient operator acting along the membrane's inward normal vector $\vek{n}$.
Evaluating the integral, we used that the system is closed, i.e.\ $\nabla_{\!\perp} c |^{}_0 = 0$, and $\lim_{\varepsilon \rightarrow 0} \int_0^\varepsilon \dd x_\perp \, \partial_t c = 0$.x
The above equation states that any exchange of proteins between the membrane and the cytosol leads to diffusive fluxes and thereby to protein gradients in the cytosol since the membrane effectively acts as a sink or source of proteins.
In general, one has 
\begin{equation} \label{eq:bulk-surface-general}
	\vek{j}_\mathrm{diff}^{} 
	= - D_c \nabla_{\!\perp} \vek{c} \big|_{\mathcal{S}} 
	= \bm{f}(\vek{m},\vek{c}|_\mathcal{S})
\end{equation}
where $\bm{f}$ denotes the corresponding net reactive flux from the membrane into the cytosol.
For the Min system, we have explicitly
\begin{subequations}
\label{eq:Min-boundary-coupling}
\begin{align}
	j_\text{diff,DD} 
	= - D_{c}\nabla_{\!\perp} \cDD \big|_{\mathcal{S}} 
	& = \hphantom{-} R_\mathrm{de}^-(\vek{m},\vek{c}|_\mathcal{S})
	  \, , \label{eq:bc1} \\
	j_\text{diff,DT} 
	= - D_{c}\nabla_{\!\perp} \cDT \big|_{\mathcal{S}} 
	& =  - R_\mathrm{d}^+(\vek{m},\vek{c}|_\mathcal{S})
	\, , \\
	j_\text{diff,E\phantom{E}} 
	= \kern0.62em - D_{c}\nabla_{\!\perp} 
	c_\text{E}^{} \big|_{\mathcal{S}} 
	& =  - R_\mathrm{de}^+(\vek{m},\vek{c}|_\mathcal{S}) + R_\mathrm{de}^-(\vek{m},\vek{c}|_\mathcal{S})
	\, .
\end{align}
\end{subequations}
While hydrolysis, described by the term $R_\text{de}^-$, leads to a protein flux off the membrane, recruitment and attachment of MinD as well as recruitment of MinE, described by $R_\text{d}^+$ and $R_\text{de}^+$, respectively, induce protein fluxes from the cytosol onto the membrane.
These reactive fluxes have to be equal to the diffusive fluxes of the corresponding protein species. 
Equation~\eqref{eq:bc1} states that detachment of MinD-ADP following hydrolysis on the membrane, $j_\mathrm{react,DD} = R_\mathrm{de}^- = \kde \, \mde > 0$, is balanced by gradients of MinD-ADP in the cytosol, $j_\text{diff,DD} = - D_{c}\nabla_{\!\perp} \cDD |_{\mathcal{S}}$. This means that there is a negative gradient of $\cDD$ from the membrane into the cytosol, i.e.\ a surplus of inactive MinD close to the membrane.
For active MinD there is a depletion zone close to the membrane as attachment and recruitment imply a flux of proteins from the cytosol onto the membrane. 
Finally, the diffusive fluxes of MinE in the cytosol equals the difference in the reactive flux due to hydrolysis and the reactive flux corresponding to MinE recruitment by MinD.
As the reactive fluxes for the respective protein states in the cytosol play a similar role as the reaction terms $\vek{R}_\mathrm{mem}$ for the protein states on the membrane, we have introduced the notation: $\vek{j}_\text{diff}$.
Accounting for the reactive and latent MinE states individually the last of the above boundary conditions generalises to
%\begin{subequations}
\begin{align*}
  	- D_{c}\nabla_{\!\perp} 
  	c_\text{Er}^{} \big|_{\mathcal{S}} 
	& = - R_\mathrm{recruit,Er}^+(\vek{m},\vek{c}|_\mathcal{S})
	+ R_\mathrm{de}^-(\vek{m},\vek{c}|_\mathcal{S})  
	\, , \\
	- D_{c}\nabla_{\!\perp} 
  	c_\text{El}^{} \big|_{\mathcal{S}} 
	& = - R_\mathrm{recruit,El}^+(\vek{m},\vek{c}|_\mathcal{S})
	\, .
\end{align*}
%\end{subequations}
In most of the following, we do not account for the extended bulk but study dynamics where gradients normal to the membrane can be neglected. 
Still, it is important to keep in mind that this is not always possible.
We will later, in Sec.~\ref{sec:bulk-boundary-coupling}, come back to the discuss situations where these gradients, induced by the bulk-boundary coupling, play an important role.

\paragraph{Mass-conservation}
The coupled reaction-diffusion dynamics on the membrane and in the cytosol conserve the protein numbers
\begin{subequations}
\begin{align}
	N_\mathrm{D} &= \int_\mathcal{V} \dd^3 x \, \cDD + \cDT + 
	\int_\mathcal{S} \dd^2 x \, \md + \mde \, , \\
	N_\mathrm{E} &= \int_\mathcal{V} \dd^3 x \, \cE + 
	\int_\mathcal{S} \dd^2 x \, \mde \, .
\end{align}
\end{subequations}
It will turn out later, that mass conservation will be essential in arriving at a systematic understanding of the mechanisms leading to pattern formation.

\begin{figure}[!t]
\centering 
\includegraphics{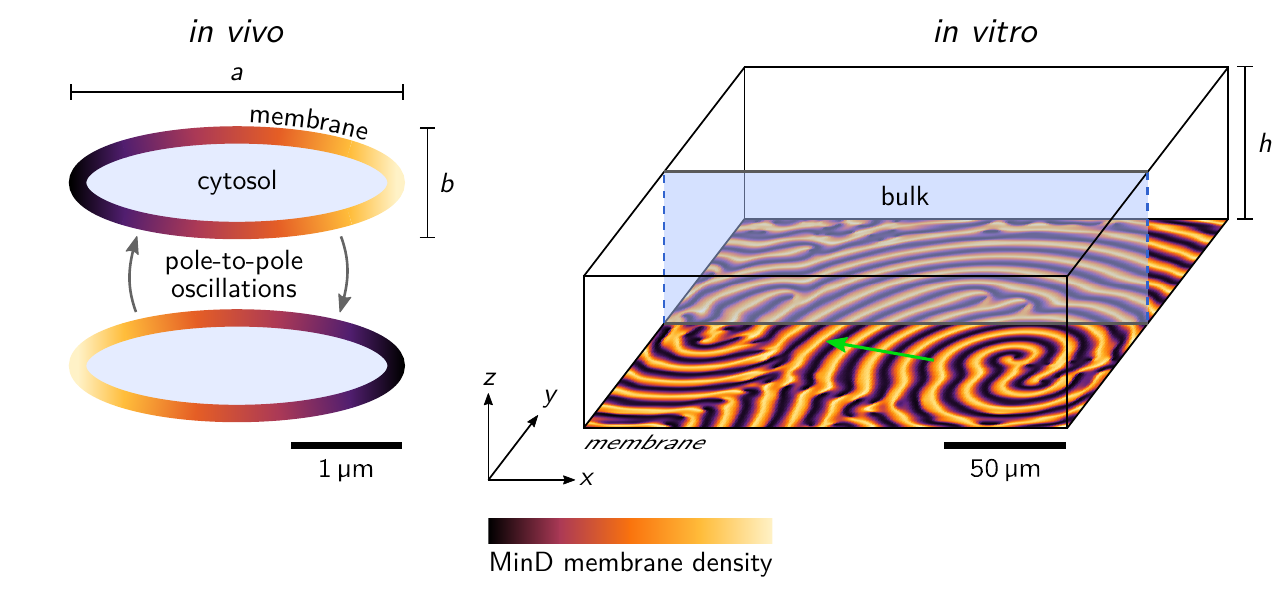}
\caption{
Simulation snapshots of the Min system (performed in COMSOL Multiphysics using the parameters given in Table~\ref{tab:Min-parameters}).
The rod shape of \textit{E.~coli} is approximated by an ellipse in a reduced two-dimensional geometry, where the third spatial dimension is integrated out.
The green arrow indicates the propagation direction of traveling waves \textit{in vitro}.
Note the vast difference in spatial scale between the two scenarios. 
For details and movies of the dynamics see \protect\cite{Halatek:2012a} and~\protect\cite{Halatek:2018a} respectively.
}
\label{fig:Min-simulation-examples}
\end{figure}

\paragraph{Finite element simulations} 
Reaction--diffusion models with bulk-surface coupling can be simulated numerically using finite element methods. 
For illustration, Fig.~\ref{fig:Min-simulation-examples} shows snapshots from simulation results of the Min system [Eqs.~\eqref{eq:Min-cytosl-dyn}, \eqref{eq:Min-membrane-dyn}, \eqref{eq:Min-boundary-coupling}] for a reduced two-dimensional \emph{in vivo} geometry~\cite{Halatek:2012a} and a three-dimensional \emph{in vitro} box geometry~\cite{Halatek:2018a}. 
The model parameter used for the \emph{in vitro} setup are listed in Tab.~\ref{tab:Min-parameters}.  

The main insights obtained from the numerical analysis of the effective two\-/dimensional \emph{in vivo} model were the following~\cite{Halatek:2012a}:
Four molecular processes --- membrane recruitment of MinD, formation of MinDE complexes by recruitment of MinE, detachment of MinDE complexes, and nucleotide exchange of MinD in the cytosol --- suffice to reproduce all oscillatory patterns as well as their observed temperature dependence. The essential nonlinearities in the system come from cooperative recruitment of cytosolic MinD and MinE to the membrane by membrane-bound MinD. Two conditions turn out to be crucial for robust pattern formation. First, MinD recruitment cannot be too weak in comparison to MinE recruitment. This gives rise to a mechanism we termed ``canalised transfer'' of MinD. The interplay between strong MinD recruitment and nucleotide exchange enables early growth of new polar zones and thereby drives the transition from pole-to-pole oscillation to striped oscillations in filamentous cells. The second condition is explicit inclusion of cell geometry via bulk-boundary coupling. Stable stripes were only obtained if the full bulk geometry was taken into account.
 
 On the right in figure~\ref{fig:Min-simulation-examples} illustrates the three-dimensional geometries that represent the experimental setup for \emph{in vitro} Min protein pattern formation~\cite{Loose:2008a}: A lipid bilayer fixed to the bottom of a large three-dimensional box~\cite{Halatek:2018a}. These simulations allow to study how pattern formation is affected by the height of the cytosolic volume above the reactive membrane. It is found that there is a Turing-type instability at a minimal bulk height. However, the emerging standing wave pattern (Turing pattern) lose stability after a long transient to a spatiotemporal chaotic attractor. Most remarkably, these numerical studies show that driving the system further from the onset of the Turing instability leads to a reorganisation of the chaotic attractor, characterised by continuously increasing spatial correlation. This reorganisation culminates in a transition to long-range correlated traveling wave patterns as observed in the experiments~\cite{Loose:2008a}. These waves are strikingly robust and maintained at arbitrarily large bulk heights. 

\begin{table}
\centering
\renewcommand\arraystretch{1.1}
\begin{tabular}{@{}clll@{}}
		\toprule
		Symbol & Unit & Value & Description \\
		\midrule
		$h$ & \si{\micro m} & 30 & Bulk height\\
		$\nbar_\mathrm{D}$ & \si{\micro m^{-3}} & 638 & Total MinD density \\
		$\nbar_\mathrm{E}$ & \si{\micro m^{-3}} & 410 & Total MinE density \\
		$D_m$ & \si{\micro m^2.s^{-1}} & 0.013 & Membrane diffusion \\
		$D_c$ & \si{\micro m^2.s^{-1}} & 60 & Cytosol diffusion\\
		$\lambda$ & \si{s^{-1}} & 6 & Nucleotide exchange\\
		$\kD$ & \si{\micro m.s^{-1}} & 0.065 & Spontaneous MinD attachment \\
		$\kdD$ & \si{\micro m^3.s^{-1}} & 0.098 & MinD self-recruitment \\
		$\kdE$ & \si{\micro m^3.s^{-1}} & 0.126 & Recruitment of MinE by MinD \\
		$\kde$ & \si{s^{-1}} & 0.34 & MinDE complex dissociation\\
		\addlinespace[0.1em]	
		\bottomrule
		\addlinespace[0.2em]	
	\end{tabular}
\caption{Standard parameters for the Min skeleton model in the \textit{in vitro} setting.}
\label{tab:Min-parameters}
\end{table}

Numerical simulations play an important role in the study of nonlinear systems, where analytic approaches are mostly restricted to special cases, like the vicinity of fixed points and homogeneous steady states. 
They are key to gain intuition into the phenomenology of a given system, which is often the first step of a deeper analysis.
However, the results from numerical simulations remain inherently limited to the specific model and the set of parameters simulated.
Sampling large parameter sets is often prohibitively time consuming (computationally costly).
Moreover, without an understanding of the underlying principles, the results cannot be generalised, remain specific to the model and parameter studied. 
A theoretical framework is required to gain such an understanding and find general principles.
In the following sections, we will present the central elements of such a framework for nonlinear systems.

%% ================================
%% REACTION KINETICS (well mixed systems)
%% ================================

\section{Protein reaction kinetics}

\def\cD{c_\mathrm{D}^{}}
\def\mT{m_\mathrm{T}^{}}
\def\mTsq{m_\mathrm{T}^{2}}
\def\mD{m_\mathrm{D}^{}}
\def\koff{k_\mathrm{off}}
\def\kon{k_\mathrm{on}}
\def\kfb{k_\mathrm{fb}}
\def\kGEF{k_\mathrm{GEF}^{}}
\def\kGAP{k_\mathrm{GAP}^{}}
\def\kbasal{k_\mathrm{basal}}
\def\kauto{k_\mathrm{auto}}

This chapter serves as an introduction to the most important concepts of dynamic system theory. 
It builds on material discussed in standard textbooks \cite{Strogatz:Book,Izhikevich:Book}, but tries to put an emphasis on those concepts that are needed for the analysis of mass-conserving systems. 

\subsection{Rate equations for well-mixed biochemical systems}

A network of biomolecular reactions typically consists of a set of elementary reactions including processes like degradation ($\mathrm{A} \xrightarrow{} \emptyset$), production ($\emptyset \xrightarrow{} \mathrm{A}$), birth/autocatalysis ($\mathrm{A} \xrightarrow{} 2 \mathrm{A}$), dimer formation ($\mathrm{A} + \mathrm{B} \xrightarrow{} \mathrm{AB}$) and conformational changes ($\mathrm{A} \xrightarrow{} \mathrm{A}^*$).
Here we are interested in protein reaction networks of the type illustrated in Fig.~\ref{fig:biochemical-networks}.
In this section we will discuss how to analyse the dynamics of such networks in well-mixed systems, i.e.\ we assume that the size of the reaction compartment is much smaller than all diffusive length scales. 
Then, given a set $S$ of chemical species with concentrations $\vek{u} = \{ u_i (t)\}_{i \in S}$, the dynamics is given by a system of coupled ordinary differential equations (ODEs)
\begin{equation}
	\partial_t \vek{u} (t) 
	=
	\vek{f} 
	\big(\vek{u}(t); \vek{\mu}\big)
	\, ,
	\label{eq:general_ode}
\end{equation}
where the parameters $\vek{\mu} \in \mathbb{R}^{N_\mathrm{p}}$ denote the kinetic rate constants for the various chemical processes; $N_\mathrm{p}$ is the number of kinetic parameters.
For a given biochemical reaction scheme, one can readily find these equations, called \textit{chemical rate equations}, using the \emph{law of mass action}.\footnote{The law of mass action by Guldberg-Waage assumes that the rate of a chemical reactions is directly proportional to the product of the densities of the reacting species. In general, it is only valid if correlations can be neglected.
} 
We have seen examples already when we discussed the reaction scheme for the Min system in Section~\ref{sec:MCRD_cell_geometry}.

There is an elaborate mathematical theory, called \textit{dynamic system theory}, that allows to analyse systems of coupled nonlinear ordinary differential equations.
The basic idea of this theory, going back to the pioneering work of Poincar\'e \cite{Poincare:Book}, is to characterise the system's dynamics in terms of geometric structures in the phase space spanned by the set of dynamic variables $\{ u_i (t) \}$.
In the following we will give a concise overview of dynamic system theory, restricting ourselves to simple systems with only one or two dynamic variables. 
The interested reader will find further information in introductory textbooks~\cite{Izhikevich:Book,Strogatz:Book} or more advanced monographs~\cite{Wiggins:Book,Guckenheimer:Book}.

\begin{figure}[tb]
\centering
\includegraphics{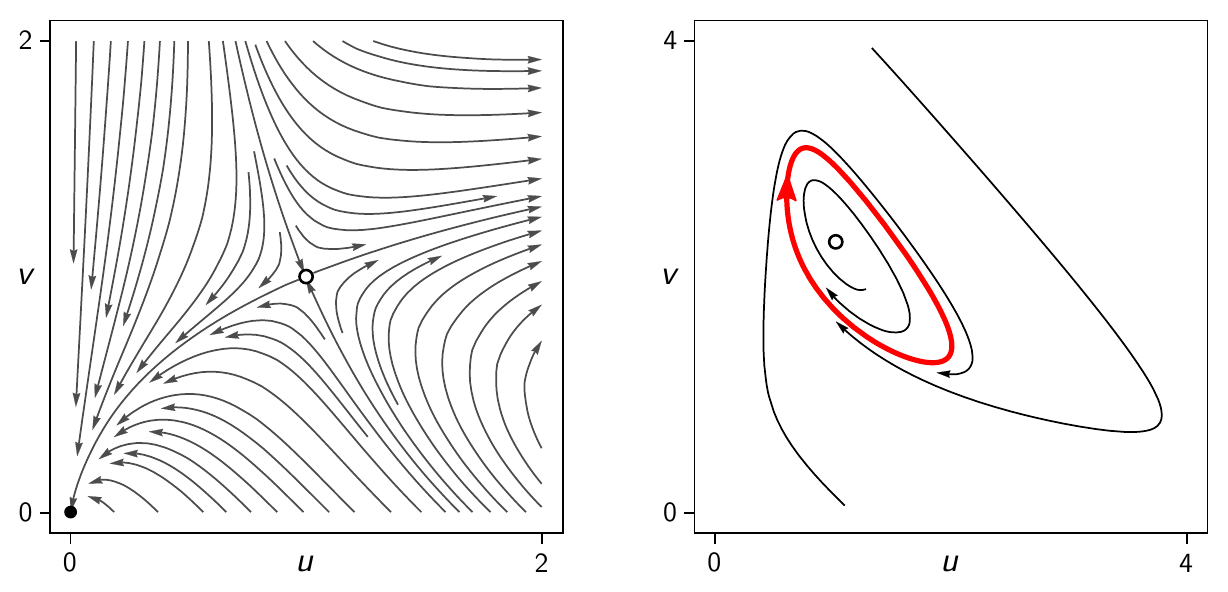}
\caption{
Visualisation of the phase space flow for two-variable dynamic systems produced by Mathematica's \texttt{StreamPlot} function.
(\textit{Left}) The system defined by $\partial_t u = -u + u^2 v$, and $\partial_t v = u-v$ has a saddle point at $(u,v)=(1,1)$ and a stable fixed point at $(u,v)=(0,0)$.
(\textit{Right}) The `Brusselator' defined by $\partial_t u = \mu + u^2 v - (\lambda + 1) u$ and $\partial_t v = \lambda u - u^2 v$ with $\mu = 1$ and $\lambda = 2.3$ exhibits limit cycle oscillations (red line) and has an unstable fixed point at $(u,v) = (1, \lambda/\mu)$.
}
\label{fig:2d_flow}
\end{figure}

A solution $\vek{u} (t)$ of the ordinary differential equation Eq.~\eqref{eq:general_ode} is a curve in $\mathbb{R}^{|S|}$ parametrised by time $t$ (also called \emph{orbit} in \emph{phase space}). 
One may specify an initial condition $\vek{u} (t_0) = \vek{u}_0$ at some time $t_0$, which is often (for autonomous systems) conveniently chosen as $t_0=0$. A set of curves $\vek{u} (t)$ corresponding to a set of different initial conditions is called a \emph{flow} in phase space, cf.\ examples shown in Fig.~\ref{fig:2d_flow}. 
The goal of dynamic systems theory is to find a \emph{geometrical characterisation} of the flow in phase space, which is sometimes also called the \emph{phase portrait}. 
In other words, one would like to answer questions of the type: 
How does an orbit $\vek{u}(t)$ depend on the initial condition $\vek{u}_0$?
How does the phase portrait change \emph{qualitatively} under variation of control parameters like the kinetic rates $\vek{\mu}$?
What `types' of flow profiles are possible, i.e.\ can we geometrically classify the phase portraits? 
What is the asymptotic behaviour of the orbits as $t \to \infty$, i.e.\ what are the `attractors' of the dynamics?
Can one characterise and classify transitions between attractors?

\subsection{One-component systems}
\label{sec:1d_well-mixed}

\begin{figure}[!b]
\centering 
\includegraphics{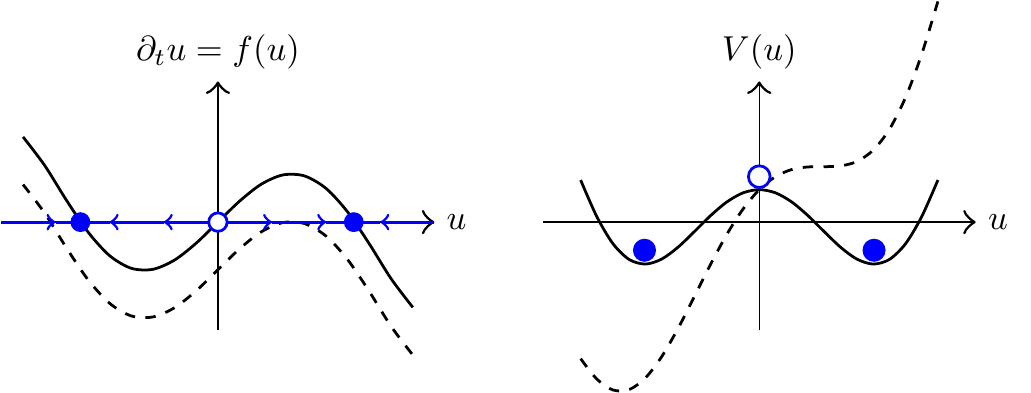}
\caption{Illustration of flow and fixed points for a one-component dynamical system. (\textit{Left}) \emph{Flow diagram}: The blue arrows indicate the direction of the velocity $\partial_t u$ of the dynamic variable $u$. Filled and open symbols correspond to stable and unstable fixed points (equilibria), respectively. (\textit{Right}) \emph{Potential landscape}: The dynamics can be interpreted as particles rolling into the minima of a potential $V(u)$. Changes in the control parameter $\mu$ lead to abrupt and qualitative changes in the flow of the dynamic variable $u$ (the particle's dynamics) at some threshold values, called bifurcation points (see dashed curves).
}
\label{fig:find-fp}
\end{figure}

To familiarise ourselves with some basic concepts of nonlinear dynamics we study one-dimensional systems, i.e.\ ordinary differential equations with a single dynamical variable $u$ and a single control parameter $\mu$:
\begin{equation}
	\partial_t u(t) = f\big(u(t),\mu\big) \, ,
\label{eq:1d_ode_general}
\end{equation}
where $f(u,\mu)$ is some nonlinear function; an example is shown in Fig.~\ref{fig:find-fp}.
Graphically it is trivial to find the \emph{fixed points (equilibria)} of Eq.~\eqref{eq:1d_ode_general}, and characterise their stability: 
The fixed points $u^*$ are given by the intersections of $f(u)$ with the $u$-axis, $f(u^*)=0$, and it is simply the sign of $f(u)$ which determines whether the dynamic variable $u$ decreases or increases; for an illustration see Fig.~\ref{fig:find-fp}.
Generically, at the fixed points, the function $f$ has a finite slope $f'(u^*) \neq 0$. Only at specific values of the control parameter $\mu$, the first or higher order derivatives of $f$ at fixed points may vanish. 
These special parameter values mark \emph{bifurcations} where the flow changes qualitatively.
This is illustrated by the dashed curve in Fig.~\ref{fig:find-fp}. At this special point the function $f$ is tangential to the $u$-axis such that $f'(u^*) = 0$. Upon further shifting the curve $f$ down the two rightmost fixed points are lost. This is called a saddle-node bifurcation and will be discussed in more detail below.

Before we continue with a discussion of the possible bifurcation scenarios, let us briefly remark that one-dimensional systems are special as the dynamics can always be written in the form 
\begin{equation*}
	\partial_t u = -\partial_u V(u) \, .
\end{equation*} 
where $V(u) = - \! \int \! \dd u f(u)$. This equation can be interpreted as the dynamics of an overdamped particle (with friction coefficient $\zeta =1$) moving in a potential landscape given by $V(u)$, c.f.\ Fig.~\ref{fig:find-fp}. 
Locally stable and unstable fixed points then correspond to local minima and maxima of the potential $V(u)$. In general, this analogy only holds for one-dimensional systems. For higher dimensional systems certain conditions have to be met for the dynamics to be formulated in terms of a dynamics in a potential landscape. If such a formulation is possible, then the dynamics is called \emph{relaxational dynamics}. 

\begin{figure}[t!]
\centering 
\includegraphics{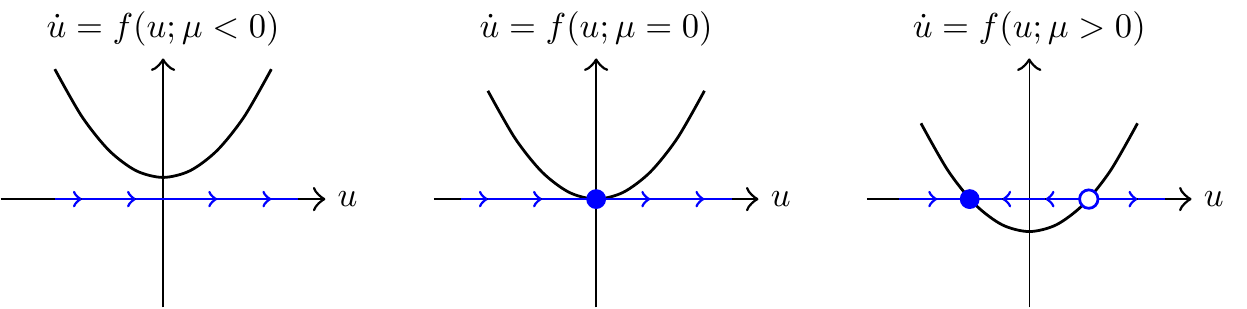}
\caption{\textbf{Normal form of a \emph{saddle-node} bifurcation}, $f(u,\mu)=-\mu+u^2$, for control parameters $\mu<0$ (\textit{left}), $0$ (\textit{center}), and $\mu>0$ (\textit{right}). The flow changes qualitatively at the bifurcation point $\mu_\mathrm{c}=0$.}
\label{fig:saddle_node}
\end{figure}

\paragraph{Saddle-node bifurcation}
Consider the ordinary differential equation
\begin{equation}
{\partial_t u} = -\mu + u^2\, ,
\label{eq:normal_form_saddle_node}
\end{equation}
with $f(u,\mu)=-\mu+u^2$ shown in Fig.~\ref{fig:saddle_node} for different values of the control parameter $\mu$. 
While for $\mu<0$ there are no intersections of $f(u)$ with the $u$-axis and hence no fixed points, there are two fixed points at $u^* = \pm \sqrt{\mu}$ for $\mu>0$. 
At $\mu = 0$, these fixed points coalesce into a half-stable fixed point at $u^* = 0$. 
We say that a bifurcation occurs at the threshold (or critical) value $\mu_c=0$ since the vector field in phase space is qualitatively different for $\mu>0$ and $\mu<0$.   

\begin{figure}[b!]
\centering
\includegraphics{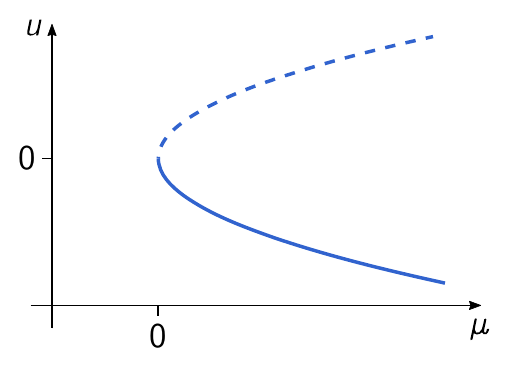}
\caption{\textbf{Bifurcation diagram of a saddle-node bifurcation}. While for $\mu<0$ there are no fixed points, two branches of fixed points emerge at the bifurcation point $\mu_\mathrm{c} = 0$. Branches of stable and unstable fixed points are indicated by a solid and a dashed line respectively.}
\label{fig:bifurcation_saddle-node}
\end{figure}

The stability of the fixed points can either be directly read off from the sign of $f(u)$ in Fig.~\ref{fig:saddle_node}, or by performing a \emph{linear stability analysis}. 
To this end we consider small deviations $\delta u := u - u^*$ from a given fixed point $u^*$ and ask whether the dynamics drives the system back to this fixed point or away from it. Upon Taylor expanding $f(u)$ close to the fixed point one finds 
\begin{equation*}
	\partial_t \, \delta u 
	= f'(u^*) \, \delta u + {\mathcal O} (\delta u^2)
	\, ,
\end{equation*}
where $f'(u^*) = \partial_u f \big|_{u^*} 
	          = 2u^* 
	          = \pm 2 \sqrt{\mu}$.
Hence $u^*=-\sqrt{\mu}$ is a linearly stable, and $u^*=\sqrt{\mu}$ a linearly unstable fixed point as the corresponding values of $f'(u^*)$ are negative and positive, respectively. 
This type of bifurcation at $\mu_\mathrm{c}=0$ is called a \emph{saddle-node bifurcation}, since at the bifurcation point a saddle-node emerges. 
The corresponding \emph{bifurcation diagram} showing the fixed points and their stability as a function of the control parameter is shown in Fig.~\ref{fig:bifurcation_saddle-node}.

Saddle-node bifurcations are actually a rather generic type of bifurcation. 
They occur if at some threshold value $\mu_\mathrm{c}$ of a control parameter the derivate of $f(u,\mu)$ at a fixed point $f(u_\mathrm{c},\mu_\mathrm{c}) = 0$ vanishes, $ \partial_u f |_{u_\mathrm{c},\mu_c} = 0$, but higher order derivatives are finite.
Simply imagine that you are shifting the curve shown in Fig.~\ref{fig:find-fp} vertically up or down until the maximum or the minimum touches the $x$-axis. Then, close to such a point the Taylor expansion reads 
\begin{align*}
f(u,\mu) 
&\approx (\mu-\mu_c) \partial_\mu f \big|_{u_\mathrm{c},\mu_\mathrm{c}} 
+ \frac{1}{2}(u-u_c)^2 \partial_u^* f \big|_{u_\mathrm{c},\mu_\mathrm{c}}
\nonumber \\ 
&:= a \, \delta \mu + b \, \delta u^2 \, ,
\end{align*}
where $\delta \mu = \mu-\mu_\mathrm{c}$ and $\delta u = u-u_\mathrm{c}$, and we have neglected terms of order $\delta u^3$ and $\delta \mu^2$. 
Hence, locally the nonlinear dynamics is of the same functional form as the normal form of a saddle-node bifurcation, Eq.~\eqref{eq:normal_form_saddle_node}, with $a,b\neq 0$. 

Because a saddle-node bifurcation requires tuning of one parameter, it is a so-called codimension-one bifurcation.
In fact, it is the only generic codimension-one bifurcation in one-dimensional systems.
There are two other bifurcations, the pitchfork bifurcation and the transcritical bifurcation, which require special circumstances (like symmetries) or tuning of parameters. These will be discussed further below.

\paragraph{The cusp bifurcation, bistability, and catastrophes}

Next, we discuss an example of a codimension-two bifurcation, the so called cusp, whose normal form is given by
\begin{equation} \label{eq:cusp-normal-form}
\partial_t u = ru - u^3 + h\, ,
\end{equation}
where $r$ and $h$ are control parameters.
As we will see, it has a close relationship with the saddle-node bifurcation.
One may also rewrite the dynamics Eq.~\eqref{eq:cusp-normal-form} in terms of a potential $V(u)$ which has the form of a Landau free energy for an Ising model in an external magnetic field $h$: $V(u) = -\frac{1}{2}ru^2 + \frac{1}{4}u^4 - hu$.

\begin{figure}[tb]
\centering
\includegraphics{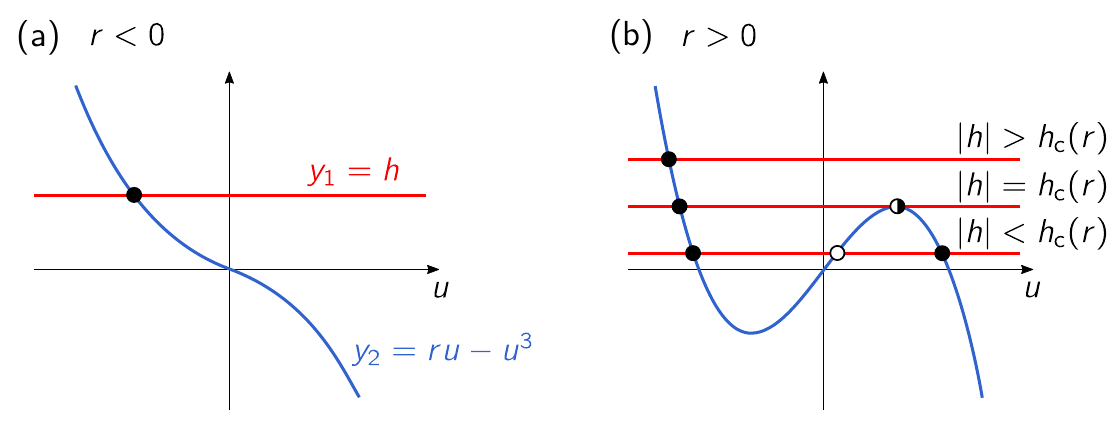}
\caption{
\textbf{Graphical analysis of fixed points for the cusp normal form.} The intersection of the graphs for $y_1 (u)=-h$ (red) and $y_2(u) =ru-u^3$ (blue) determine the location of the fixed points, and stability can be inferred from the slope of $y_2(u)$.
(a) For $r < 0$, there is only one stable fixed point (filled circle), independent of the magnitude of $h$.
(b) For $r \geq 0$, there are up to three fixed points depending on the magnitude of $h$. While for $|h|>h_\mathrm{c}(r)$, there is only one stable fixed point, there are three fixed points in the regime $|h|<h_\mathrm{c}(r)$: stable fixed points on the left and right branch $u_\pm^* (h)$ indicated by the solid circles, and an intermediate unstable fixed point $u_0^* (h)$, indicated by the open circle. At $|h|=h_\mathrm{c}(r)$ there are saddle-node bifurcations.
}
\label{fig:cusp-graphical-construction}
\end{figure}

The fixed points of Eq.~\eqref{eq:cusp-normal-form} can be determined graphically as the intersections between the functions $y_1 (u)=-h$ and $y_2 (u) = ru-u^3$, as illustrated by the red and blue curves in Fig.~\ref{fig:cusp-graphical-construction}, respectively. 
For $r<0$ (`paramagnetic phase'), the function $y_2 (u) = ru - u^3$ is monotonically decreasing in $u$, and hence there is only one stable fixed point $u^*$, given by $u^*\approx -h/r$ for small $h$. 
In contrast, for $r>0$, there may be up to three fixed points depending on the magnitude of the control parameter $h$. 
For large values of $|h|$, there is only one fixed point and it is stable. 
Upon lowering $h$, there are saddle-node bifurcations when $y_1 (u)=-h$ becomes tangential to $y_2(u)=ru-u^3$. This condition determines two lines of saddle-node bifurcations, $h = \pm h_\mathrm{c}(r)$, in the parameter plane, with $h_\mathrm{c} (r) = 2\left(r/3\right)^{\nicefrac{3}{2}}$.

In the parameter regime $|h| < h_\mathrm{c}(r)$, the dynamics is bistable with two stable fixed points $u_\pm^* (h)$ separated by an unstable fixed point $u_0^* (h)$. The bistable regime ends in a \emph{cusp} at $(r,h)=(0,0)$, where the two lines $h = \pm h_\mathrm{c}(r)$ meet. 
\begin{figure}
\centering
\includegraphics{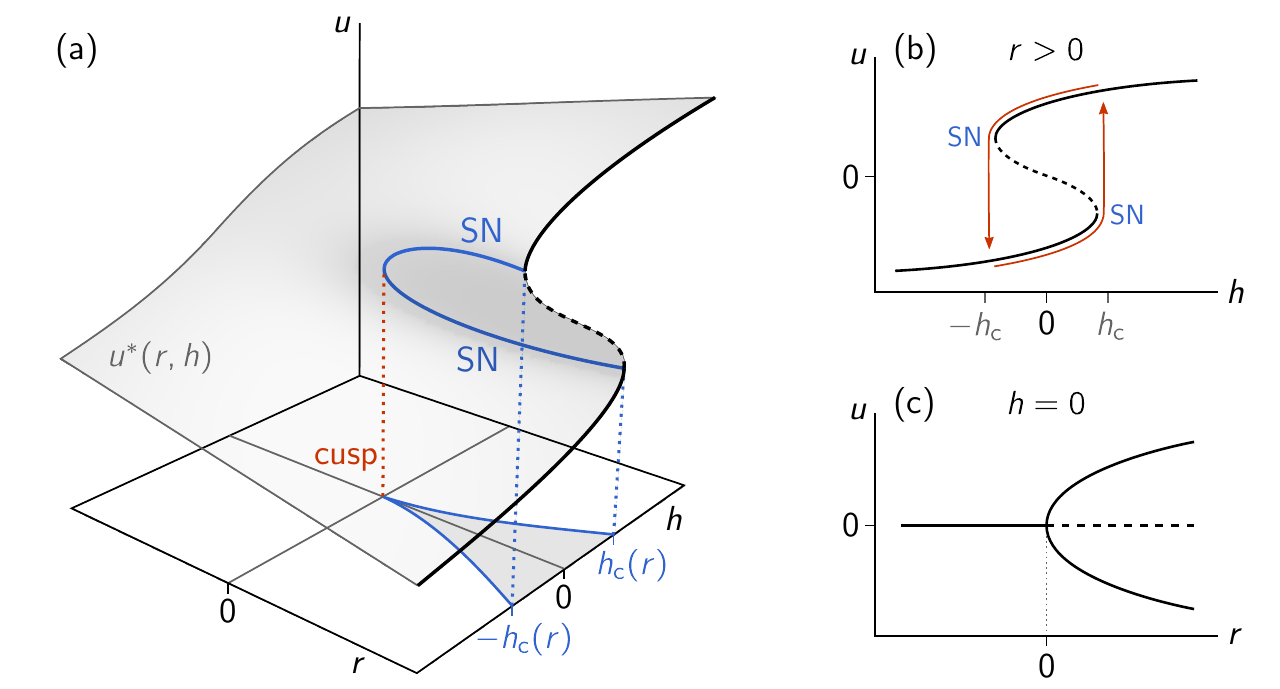}
\caption{\textbf{Cusp bifurcation scenario.} (a) Surface of fixed point $u^*(r,h)$. Two lines of saddle-node bifurcations (SN, blue lines) emanate from the cusp point at $(r,h) = (0,0)$. In the regime enclosed by the saddle-node bifurcations (shaded in gray in the parameter plane), the system is bistable, while it is monostable everywhere else.
(b) Bifurcation diagram $u^* (h)$ for $r > 0$: For $|h|>h_\mathrm{c}(r)$ the system is monostable, and becomes bistable in the domain $|h|<h_\mathrm{c}(r)$ with two stable fixed points separated by an unstable fixed point. Upon first increasing $h$ through one of the SN bifurcations and then back again, there is hysteresis as indicated by the red arrows. 
(c) Pitchfork bifurcation in $r$ for the special case $h = 0$, where the system is symmetric under $u \rightarrow -u$. In general, a system undergoes a pitchfork bifurcation if one passes through the cusp tangentially to the SN lines meeting there.
}
\label{fig:cusp-scenario}
\end{figure}
This bifurcation scenario is visualized in Fig.~\ref{fig:cusp-scenario}a, showing the surface of fixed points $u^*(r,h)$ over the $(r,h)$ parameter plane. 
The line of saddle-node bifurcations (blue line) is where the slope of the surface becomes vertical. The cusp point is where the line of saddle-node bifurcations itself becomes vertical.

While for $r<0$ (`paramagnetic phase') one may continuously change the fixed point value from positive to negative values upon lowering $h$, this is not possible for $r>0$ (`ferromagnetic phase').
Here, starting with a fixed point on the upper branch and lowering h there is a threshold value $-h_\mathrm{c} (r)$ where the upper branch disappears (saddle-node bifurcation), and as a consequence the dynamic variable changes abruptly to the lower branch (see Fig.~\ref{fig:cusp-scenario}b). This is sometimes also called a \emph{`catastrophe'}~\cite{Thom:Book}.
Increasing $h$ again after the catastrophe, the system will remain in the lower branch up to the saddle-node bifurcation at $h_\mathrm{c} (r)$. This behaviour is called \emph{hysteresis}.

A new type of bifurcation, called pitchfork bifurcation, takes place if one passes exactly through the cusp point, tangentially to the saddle-node lines meeting there in the parameter plane. 
Here, this corresponds to keeping $h=0$ constant and varying $r$ (see Fig.~\ref{fig:cusp-scenario}c). 
In this case there is inversion symmetry in $u$. Any $h \neq 0$ breaks this symmetry such that the system undergoes a saddle-node bifurcation instead of a pitchfork bifurcation (this is sometimes referred to as `imperfect pitchfork bifurcation'~\cite{Strogatz:Book}).
In general, a pitchfork bifurcation requires fine tuning or the presence of a symmetry.
In the cusp normal form Eq.~\eqref{eq:cusp-normal-form}, tuning $h=0$ encompasses both, as it corresponds to a situation where a symmetry ($u \to - u$) is present.

\paragraph{Transcritical bifurcation}

Consider the following reaction scheme
%\begin{subequations} %\label{eq:contact_process_scheme}
\begin{align*}
	&\text{M} + \text{C} \xrightarrow{\,\lambda\,} \text{M} + \text{M} \, ,\\
	&\text{M}     \xrightarrow{\,\delta\,} \text{C} \, .
\end{align*}
%\end{subequations}
This may be viewed as the dynamics of a one-protein system, where the protein can be in two distinct states M and C. 
These states may be considered as active and inactive states or as membrane-bound (M) and cytosolic (C) states. 
In the latter case, one may read the first reaction as a recruitment process where membrane-bound proteins recruit cytosolic proteins to the membrane with a rate $\lambda$, and the second reaction as detachment of these membrane-bound proteins back into the cytosol with rate $\delta$.
The corresponding rate equations read (assuming a well-mixed compartment)
%\begin{subequations}
\begin{align*}
	\partial_t m(t) &= \hphantom{-} \lambda \, m \, c - \delta \, m 
	\, , \\
	\partial_t c(t) &= - \lambda \, m \, c + \delta \, m 
	\, ,
\end{align*}
%\end{subequations}
where $m$ and $c$ denote the number of proteins on the membrane and in the cytosol, respectively. 
The process defined by the reaction scheme above can also be interpreted as a  \emph{contact process}, a model for the dynamics of an infection.\footnote{Instead of a disease spreading in a population you may also consider the spreading of opinions. Of course, spreading of diseases and opinions in a population is affected by a plethora of factors, e.g.\ the spatial distribution of people, how they are connected through social networks, and physical constitution resp. personality.} 
Then C is a healthy person infected by a sick person M, $\lambda$ is the infection rate, and $\delta$ is the recovery rate.

\begin{figure}[tb]
\centering
\includegraphics{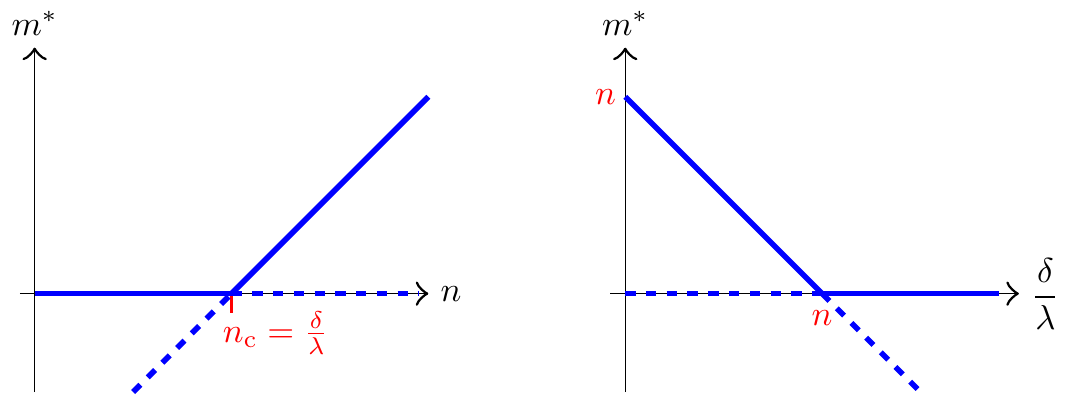}
 \caption{\textbf{Bifurcation diagram of the contact process.} (\textit{Left}) As a function of population size the infection spreads only if the size exceeds a threshold size given by $n_\mathrm{c} = \delta/\lambda$. (\textit{Right}) For a given population size $n$, the infection spreads if the recovery rate is smaller than the infection rate, $\delta/\lambda < n$.
 }
 \label{fig:contact_process_bfc}
\end{figure}

The dynamics conserves the total number of individuals, as the total number $n = m + c$ remains invariant: $\partial_t n = 0$. 
Thus the dynamics reduces to a single equation
\begin{equation*}
	\partial_t m 
	= 
	\lambda \, m (n-m) -\delta \, m =: f(m)
	\, .
\end{equation*}
The fixed points ($\partial_t m = 0$) are given by
\begin{equation*}
	m^*_1 = 0 \,, \quad 
 	m^*_2 = n - \frac{\delta}{\lambda}\, ,
\end{equation*}
corresponding to a state where there are no proteins on the membrane and a state where a finite fraction of proteins is on the membrane.
We may easily check their stability upon calculating the first derivative $f^\prime (m) = \lambda \,(n-2m)-\delta$ at the respective fixed points. One finds 
%\begin{subequations}
\begin{align*}
	f^\prime(m^*_1 ) 
	&= \lambda \, n -\delta \, , \\
	f^\prime(m^*_2)  
	&= \delta-\lambda \, n \, ,
\end{align*}
%\end{subequations}
such that $m^*_2$ is stable while $m^*_1$ is unstable for $n > \delta/\lambda$ and vice versa for $n < \delta/\lambda$, i.e.\ the fixed points interchange their stability at $n_\mathrm{c} = \delta/\lambda$.
This is called a \textit{transcritical bifurcation}; see Fig.~\ref{fig:contact_process_bfc}.

This result may be interpreted in various ways. 
First, let's say that the recruitment (infection) rate and the detachment (recovery) rate are both fixed. 
Then $n_\mathrm{c} = \delta / \lambda$ denotes a threshold value for the protein number  (population size), above which proteins start to attach to the membrane (an infection spreads in the population).
The number of membrane-bound proteins (sick people) is then given by  $m^* = n - {\delta}/{\lambda}$. 
Below the threshold $n_\mathrm{c}$, the whole population will eventually become healthy. 
Second, for a given protein number (population size) $n$, it depends on the ratio of detachment (recovery) to recruitment (infection) rate whether proteins attach to the membrane (the infection may spread in the population). 
While for low detachment (recovery) rate $\mu /\lambda < n$ protein bind to the membrane (the infection spreads), the membrane remains devoid of proteins (the infection will be eliminated) for high detachment (recovery) rate, $\delta /\lambda > n$. 

\begin{exercise}[h] \label{ex:conservative-brusselator}
Perform a bifurcation analysis of the set of equations
\begin{align*}
	\partial_t m(t) &= \hphantom{-} \, \lambda \, m^2 \, c - \delta \, m \\
	\partial_t c(t) &= - \, \lambda \, m^2 \, c + \delta \, m 
\end{align*}
where a stronger nonlinearity than above has been assumed for the recruitment term. 
What kind of bifurcation does the system exhibit?		
\end{exercise}

\begin{exercise}[h] \label{ex:limited-resouces_simple}
Consider a basic model of a growing microtubule with a limited amount $L$ of tubulin dimers in the confined volume of a cell. To study the dynamics of the microtubule length $l$ in an elementary scenario, assume that the depolymerisation rate $\delta$ is constant while the polymerisation rate is resource limited, $\gamma (l) \,{=}\, \gamma_0 ( L - l )$. Then the growth dynamics of the microtubule is given by
\begin{align*}
	\partial_t \, l(t) = 
	\gamma (l) - \delta
	\, .
\end{align*}
Discuss the dynamics as a function of the reaction rates $\gamma_0$ and $\delta$ as well as the amount of limited resources (tubulin dimers) $L$. 
What kind of bifurcation does the system exhibit? What does this imply for the possibility of controlling the length of a microtubule?
\end{exercise}

\paragraph{General two-component systems with mass conservation}

We consider a general two-component system with mass conservation
\begin{subequations}
\label{eq:MCRD_general_well-mixed_twocomp}
\label{eq:2compR}
\begin{align} 
	\partial_t m (t) &=  + f(m,c) \, ,\\
	\partial_t c (t) &=  - f(m,c) \, .
\end{align}
\end{subequations}
In the biological context of cell polarisation, the nonlinear kinetics term $f(m,c)$ is typically of the form 
\begin{equation*} 
	f (m,c) = a(m) \, c - d(m) \, m \, ,
\end{equation*}
where the non-negative terms $a(m)$ and $d(m)$ denote the rates of attachment of proteins from the cytosol to the membrane and detachment back into the cytosol, respectively.
For example, one may assume a reaction kinetics with autocatalytic recruitment (Michaelis--Menten kinetics with Hill coefficient $2$) and linear detachment~\cite{Mori:2008a}
\begin{equation}
	f(m,c) 
	= \left(\kon + \kfb \, \frac{m^2}{K_\mathrm{d}^2+m^2}\right) c - \koff \, m
	\, .
\end{equation}
Measuring time in units of the inverse off-rate $\koff^{-1}$, and densities in units of $K_\mathrm{d}$, this expression can be made non-dimensional
\begin{equation} 
	\label{eq:wave-pinning-nondim} 
	f(m,c) 
	= 
	\left(
	\kappa
	+ 
	\kappa_\mathrm{fb}^{}\, 
	\frac{m^2}{1+m^2}
	\right) c 
	- m \,,
\end{equation}
with $\kappa_\mathrm{fb} := \kfb/\koff$ and $\kappa := \kon/\koff$. 
In the following, we will for illustration purposes often use the case $\kappa_\mathrm{fb} = 1$, leaving $\kappa$ as the only free kinetic parameter.

\begin{figure}
	\centering
	\includegraphics[scale=1.25]{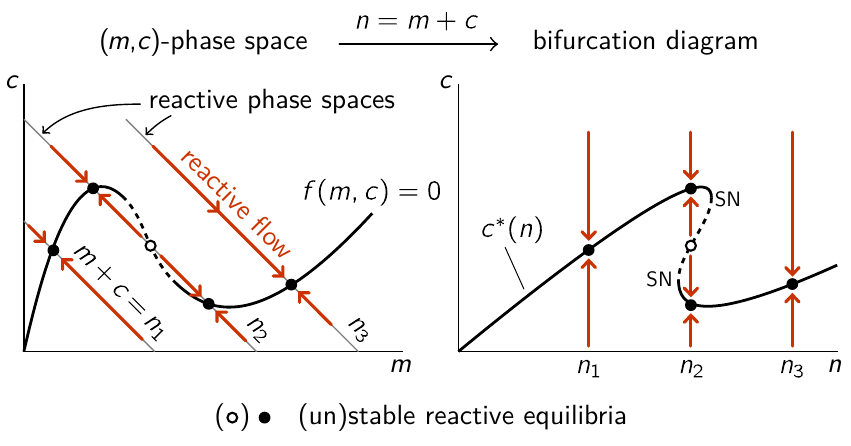}
\caption{
\textbf{Phase space and bifurcation structure of a well-mixed, mass-conserving two-component system.} The total density $n$ is a control parameter of mass-conserving reactions. The properties (number, position, and stability) of the reactive equilibria depend on the total density (mass) in a well-mixed compartment. The conservation law $m + c = n$ is geometrically represented by 1-simplexes in phase space, where each value of the total density corresponds to a unique 1-simplex. We refer to these subspaces as \emph{reactive phase spaces} (local phase spaces in context of spatially extended systems). Local reactions interconvert the conformational states of the proteins and hence change their densities, giving rise to a flow in phase space (red arrows) which, due to mass conservation, is confined to the reactive phase spaces. The flow vanishes along the \emph{reactive nullcline} ${f(m,c)=0}$ (black line) which is a line of \emph{reactive equilibria}. Each intersection of a reactive phase space with the reactive nullcline is a reactive equilibrium $(m^*(n),c^*(n))$ for a given total density $n$ (shown as disks, $\bullet/\circ$, for three different values $n_1$, $n_2$, and $n_3$). The $(m,c)$-phase portrait can be transformed into a bifurcation diagram $c^*(n)$ by the (skew) transformation $n = m +c$. Because of the conservation law, the well-mixed system has only one degree of freedom, so the only possible bifurcations are saddle-node bifurcations (SN) where the reactive nullcline is tangential to a reactive phase space. Adapted from \protect\cite{Brauns.etal2020c}.} 
	\label{fig:total-density-bifurcation-structure}
\end{figure}

The reaction kinetics, Eq.~\eqref{eq:MCRD_general_well-mixed_twocomp}, can be analysed in $(m,c)$-phase space using geometric reasoning; see Fig.~\ref{fig:total-density-bifurcation-structure}. 
As the dynamics conserve the total protein number (protein mass), the flow in phase space is constrained to subspaces (1-simplices) where $m+c=n$, henceforth called \emph{reactive subspaces} (Fig.~\ref{fig:total-density-bifurcation-structure}a).
The flow in phase space vanishes along the reactive nullcline (NC) $f = 0$.
For a given protein mass $n$, the fixed points $(m^*,c^*)$ are given by the intersections of the reactive nullcline $f(m^*,c^*) = 0$ with the reactive subspace $m^* \,{+}\, c^* = n$.
Because the fixed points are determined by a balance of reactive flows, we call them (\emph{reactive}) \emph{equilibria}.%
\footnote{The term equilibria in the sense of dynamical systems, as we use it here, is not to be confused with \emph{thermal equilibria}.} 
The dynamics (reactive flow) within each mass-conserving subspace is organised by the position, number, and stability of the reactive equilibria, as illustrated in Fig.~\ref{fig:total-density-bifurcation-structure}.

Using mass conservation, the reaction dynamics can be written solely in terms of $m(t)$:
\begin{equation*}
	\partial_t m(t) 
	= f \big( m(t), n - m(t) \big)
	\, .
\end{equation*}
This form makes explicit that in addition to the chemical rates also the total protein density $n$ is a control parameter.
In the vicinity of an equilibrium $m^*$ the linearised reactive flow reads
\begin{equation*}
	\partial_t m 
	\approx 
	(f_m - f_c) \cdot
	\big( m - m^*(n) \big)
	=:
	\sigma_\mathrm{loc}(n) 
	\cdot \big(m-m^*(n)\big) \,,
\end{equation*}
with the eigenvalue given by $\sigma_\mathrm{loc}(n) := f_m - f_c$ and the partial derivatives defined as $f_{m,c} := \partial_{m,c} f|_{(m^*, c^*)}$; in the above formulas we have also made explicit that both the position and the stability of the equilibria depend on the protein mass $n$.
The equilibria are stable if $\sigma_\mathrm{loc}(n) < 0$ and unstable if $\sigma_\mathrm{loc}(n) > 0$.
The stability condition can be given a geometric interpretation in terms of the slope $\snc(n)$ of the reactive nullcline that is given by
\begin{equation} \label{eq:snc-def}
	 \snc(n) 
	 =  \partial_m c^*(m) \big|_n 
	 = -\frac{f_m}{f_c}\Big|_n.
\end{equation}
For specificity consider the case of an attachment--detachment kinetics where $f_c = a(m) > 0$.
Then, the condition for linear stability ($\sigma_\mathrm{loc}(n) < 0$) can be written as
\begin{equation} 
	\label{eq:local-stability-slope-criterion}
	\snc(n) = -f_m/f_c > -1
	\, .
\end{equation}
Geometrically, this means that local equilibria are stable if the tangent to the reactive nullcline cuts the simplex of the reactive phase space from below, and unstable otherwise.
For the example shown in Figure~\ref{fig:total-density-bifurcation-structure}a, the dynamics is mainly monostable with only one stable fixed point ($\bullet$) except for a window of protein masses (near $n_2$) where the dynamics exhibits bistability with one unstable ($\circ$) and two stable fixed points ($\bullet$).

Given this geometric criterion it is straightforward to construct the bifurcation diagram of the (reactive) equilibrium $c^*(n)$ as a function of the total density $n$, cf.\ Fig.~\ref{fig:total-density-bifurcation-structure}b.
This reiterates a point we have made earlier, namely that the total protein density is a control parameter of the dynamics. 
Varying the total protein density, the position and number of equilibria as well as the stability of these equilibria change.
This fact will turn out to be a key element for  understanding the mass-conserving reaction--diffusion dynamics~\cite{Halatek:2018a,Brauns.etal2020c}, as we will discuss later in these lecture notes.

Above we have analysed the well-mixed system within the reactive subspace for a given fixed protein mass $n$.
For the discussion of the spatially extended systems it will turn out to be informative to perform this analysis in the two-dimensional phase plane $(m,c)$, as the total protein density might be spatially heterogeneous. 
Then, to study the linear stability one defines the displacement vector $\delta \vek{u} :=  (m\,{-}\,m^* ,c\,{-}\,c^*)^\mathrm{T}$ and considers the linearised system corresponding to Eq.~\eqref{eq:2compR},
\begin{equation*}
	\partial_t \, \delta \vek{u} 
	= \mathcal{J}  \delta \vek{u} \, ,
\end{equation*}
with the Jacobian given by
\begin{equation*}
	\mathcal{J} = \begin{pmatrix}
	  \hphantom{-} f_m & \hphantom{-} f_c \\
     -f_m & -f_c
	  \end{pmatrix} ,
\end{equation*}
with $f_{m,c}$ defined as above. 
Using the ansatz $\vek{u} (t) =  \mathrm{e}^{\sigma t} \, \bm{e}$ one finds the eigenvalues 
%
%\begin{subequations}
\begin{align*} 
	\sigma^{(1)} &= 0 \, ,\\
    \sigma^{(2)} &= f_m -f_c \, ,
\end{align*}
%\end{subequations}
%
and the corresponding (not normalized) eigenvectors, 
%
%\begin{subequations}
\begin{align*} 
	\vek{e}^{(1)}
	&=(f_c,-f_m)^\mathrm{T}
	 \, ,\\
    \vek{e}^{(2)}
    &=(1,-1)^\mathrm{T} \, .
\end{align*}
%\end{subequations}
%
The first eigenpair defines a \emph{center space} that is spanned by the eigenvector $\vek{e}^{(1)}$ which is tangent to the line of fixed points given by the nullcline $f(m,c) = 0$. 
This also explains why the associated eigenvalue $\sigma^{(1)}$ is zero. 
The second eigenpair defines the stability of the equilibrium (fixed point) against perturbations that preserve the total particle density $n$; 
note that the eigenvector $\bm{e}^{(2)}$ spans a simplex in phase space defined by the mass conservation constraint $m + c = n$. 
The eigenvalue agrees with the one obtained above, $\sigma^{(2)} = \sigma_\mathrm{loc}$.

\begin{exercise}
	Perform a bifurcation analysis of the mass-conserving two-component system with reaction kinetics given by Eq.~\eqref{eq:wave-pinning-nondim}. Make use of graphical constructions in the the $(m,c)$-phase portrait to find the qualitative bifurcation structure first (cf.\ Fig.~\ref{fig:total-density-bifurcation-structure}).
\end{exercise}

\begin{exercise}
Repeat the bifurcation analysis of the set of equations
\begin{align*}
	\partial_t m(t) &= \hphantom{-} \, \lambda \, m^2 \, c - \delta \, m 
	\, , \\
	\partial_t c(t) &= - \, \lambda \, m^2 \, c + \delta \, m 
	\, ,
\end{align*}
but now in $(m,c)$-phase space.
Calculate the eigenvalues as well as the eigenvectors using the methods explained above. Before performing the analysis write the set of equations in dimensionless form, such that the total protein number $n=m+c$ remains as the only control parameter.	
\end{exercise}

\subsection{General two-component systems}
\label{sec:2d_well-mixed}

In this section we consider the general properties of genuinely two-component nonlinear systems
\begin{subequations} \label{eq:2d_ode_general}
\begin{align}
	\partial_t u 
	&= f(u,v) \, , \\
	\partial_t v
	&= g(u,v) \, .
\end{align}
\end{subequations}
with $u$ and $v$ two independent dynamic variables.
The solution of this set of ordinary differential equations is uniquely determined by the initial conditions. Hence trajectories in phase space cannot cross each other, except at fixed points $(u^*, v^*)$ where $\partial_t u = \partial_t v = 0$, i.e.\ $f = g = 0$.

\paragraph{Stability analysis}

\begin{figure}[tb]
\centering
\includegraphics[scale=1.25]{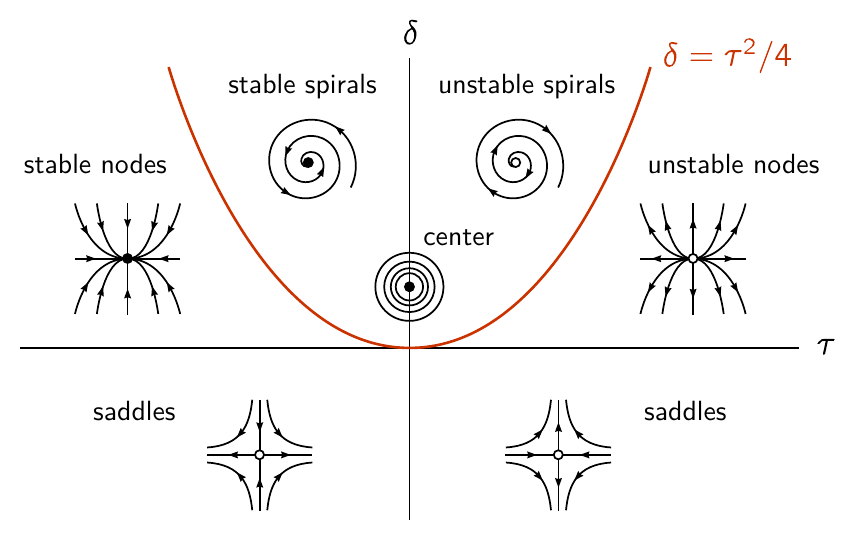}
\caption{Classification of fixed points for two-dimensional nonlinear systems. As a function of the Jacobian's trace $\tau$ and determinant $\delta$.}
\label{fig:2d_classification_linear}
\end{figure}

To study the flow field in the vicinity of a fixed point $\vek{u}^* = (u^*, v^*)$, one expands the dynamics Eq.~\eqref{eq:2d_ode_general} to linear order in the displacement $\delta \vek{u}(t) = \vek{u}(t) - \vek{u}^*$
\begin{equation*}
\partial_t \delta \vek{u} = \mathcal{J} \delta \vek{u}
\end{equation*}
where the Jacobian at the fixed point is given by
\begin{equation*}
\mathcal{J} = Df|_0 =
\left.
\begin{pmatrix} 
\partial_u f & \partial_v f\\
\partial_u g & \partial_v g
\end{pmatrix}
\right|_{\vek{u}^*} 
=: 
\begin{pmatrix} 
f_u & f_v\\
g_u & g_v
\end{pmatrix}
\, .
\end{equation*}
The solutions of this linear system can be fully classified. 
We are seeking solutions of the form $\delta \vek{u} (t) \propto  e^{\sigma t} \, \vek{e} $ with eigenvalues $\sigma$ and eigenvectors $\vek{e}$. 
The characteristic equation for the eigenvalues is given by $\det(\mathcal{J}-\sigma\mathbf{1})=0$ and results in a quadratic equation $\sigma^2 -\tau\sigma + \delta = 0$, where $\tau = \text{tr} \, J = f_u+g_v$ and $\delta = \det \mathcal{J} = f_u g_v -f_v g_u$. 
Hence the eigenvalues of $\mathcal{J}$ read
\begin{align*}
	\sigma_{1/2}
	=\frac{1}{2} \left(\tau\pm\sqrt{\tau^2-4\delta}\right)
	=\frac{1}{2}  \left(\tau\pm\sqrt{\Delta} \right)\, ,
\end{align*}
with $\Delta := \tau^2-4\delta$ the discriminant. There are three cases for the eigenvalues:
\smallskip\begin{center}
\begin{tabular}{cm{4cm}l}
	(i) & $\sigma_1,\:\sigma_2\in\mathbb{R} \quad (\sigma_1\neq\sigma_2)$ & if $\quad\Delta > 0$ \\
	(ii) & $\sigma_1 = \sigma_2\in\mathbb{R}$ & if $\quad\Delta = 0$ \\
	(iii) & $\sigma_{1,2}$ complex conjugate & if $\quad\Delta = 0$
\end{tabular}
\end{center}\smallskip
%\begin{align*}
%	&\;\text{(i)} && \;  \sigma_1,\:\sigma_2\in\mathbb{R}\quad (\sigma_1\neq\sigma_2) & \text{if} \quad \Delta >0 \\
%	&\,\text{(ii)}  &&\; \sigma=\sigma_1=\sigma_2\in\mathbb{R} & \text{if} \quad  \Delta = 0\\
%	&\text{(iii)} && \; \sigma_{1/2}\; \mbox{complex conjugate} & \text{if} \quad \Delta < 0
%\end{align*}
Moreover, the signs of $\Re \sigma_{1,2}$, which determine the stability of the fixed point, can be inferred from the signs of $\tau$ and $\delta$. 
The corresponding linear flows close to a fixed point are classified as follows (see Fig.~\ref{fig:2d_classification_linear}).
For $\delta < 0$, the eigenvalues $\sigma_{1/2}$ are real and have opposite signs; hence the fixed point is a \textit{saddle point}. For $\delta > 0$, the eigenvalues $\sigma_{1/2}$ are real with the same sign (\textit{nodes}), or complex conjugates (\textit{spirals and centers}). For $\Delta > 0$ the fixed point is a \textit{node}, for $\Delta < 0$ it is a \textit{spiral}.
The parabola $\Delta = 0$ marks the border between nodes and spirals, where so called \textit{star nodes} and \textit{degenerate nodes} are the respective border cases.
The stability of the nodes and spirals is determined by $\tau$. For $\tau <0$, both eigenvalues have negative real parts, so the fixed point is stable. Unstable spirals and nodes have $\tau > 0$.
On the half line $\tau = 0, \delta >= 0 $, the eigenvalues are purely imaginary such that the fixed point is neutrally stable (called a center).

\paragraph{Phase-portrait analysis: nullclines and invariant manifolds.}

While linear stability analysis facilitates a general classification as presented above, it only informs about the \emph{local} properties of the flow close to fixed points. 
How can one gain insight into the dynamics far away from fixed points, that is, the flow's \emph{global} structure?
It is instructive to consider the \emph{nullclines} $f(u,v) =0$ and $g(u,v) =0$ where the flow becomes fully vertical ($\partial_t u = 0$) or horizontal ($\partial_t v = 0$), respectively. 
Nullclines intersect at the system's fixed points. 
Moreover, they partition the phase space into regions where $\partial_t u$ and $\partial_t v$ have different signs.
This makes it possible to infer the qualitative structure of the phase space flow.

\begin{figure}[tb]
\centering
\includegraphics{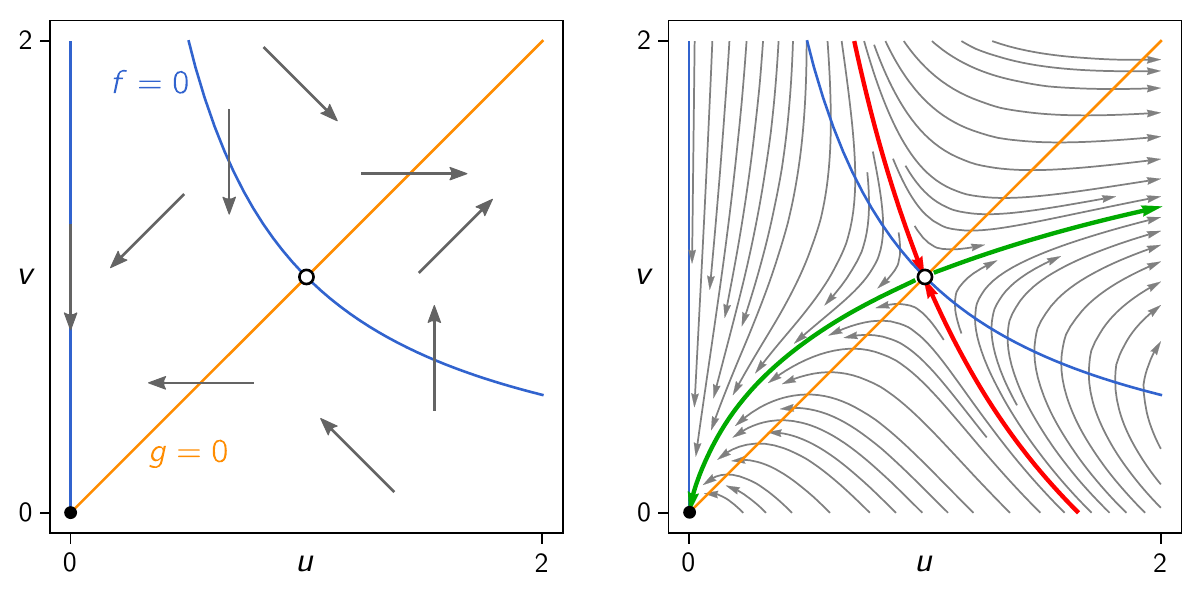}
\caption{
\textbf{Illustration of a phase portrait.} 
(\textit{Left}) Nullclines of the reaction kinetics Eq.~\eqref{eq:example_2d_dyn}, shown as blue ($f=0$) and orange ($g=0$) lines. Their intersections mark two fixed points, a saddle ($\circ$), and a stable node ($\bullet$). As $\partial_t u$ and $\partial_t v$ switch sign upon crossing the respective nullcline, the qualitative flow structure can be inferred from the nullclines as indicated by the gray arrows.
(\textit{Right}) Visualization of the phase space flow with nullclines and invariant stable (green) and unstable (red) manifolds corresponding to the saddle point.
}
\label{fig:example_2d_flow}
\end{figure}

In Fig.~\ref{fig:example_2d_flow}, we illustrate the basic ideas of such a \emph{phase-portrait analysis} for an elementary example
\begin{subequations} \label{eq:example_2d_dyn}
\begin{align}
	\partial_t u &= - \, u + u^2 v \, , \\
	\partial_t v &= \hphantom{-} \, u - v \, , 
\end{align}
\end{subequations}
The nullclines are given by $v=1/u$ and $v=u$; there is also a trivial branch of $f=0$ where $u=0$. 
Hence, the fixed points are given by $(u^*, v^*) = (0,0)$ and $(u^*, v^*) = (1,1)$.
The nullclines partition the phase space into four quadrants with the gray arrows indicating the direction of flow, i.e.\ the signs in the velocities $\partial_t u$ and $\partial_t v$.
These suggest that the fixed point at $(u^*, v^*) = (0,0)$ is stable, and the fixed point at $(u^*, v^*) = (1,1)$ is a saddle point.
Taken together a sketch of the flow as obtained from the nullclines (Fig.~\ref{fig:example_2d_flow}, \textit{left}) already gives a rather decent picture of the actual flow shown in Fig.~\ref{fig:example_2d_flow} on the right.

The phase portrait in Fig.~\ref{fig:example_2d_flow}b also shows another set of important geometric characteristics of the flow --- stable and unstable \emph{invariant manifolds}, shown as green and red lines respectively. 
The defining properties of these manifolds are that they are (i)  invariant under the flow and (ii) tangential to the stable / unstable eigenspaces at fixed points. These eigenspaces are spanned by the sets of eigenvectors associated to the sets of stable / unstable eigenvalues ($\Re \sigma < 0$ / $\Re \sigma > 0$) respectively.\footnote{When there are neutral eigenvalues ($\Re \sigma = 0$), there is also a center manifold, spanned by the eigenvectors associated to the neutral eigenvalues. The `reactive phase spaces' in Fig.~\ref{fig:total-density-bifurcation-structure} are a trivial example for center manifolds.} 
The flow in the vicinity of the unstable manifold is directed away from the manifold. 
It therefore plays the role of a \emph{separatrix} that separates the basins of attraction of different stable fixed points of the dynamics.
Saddle points lie at the intersection of stable and unstable manifolds.

Invariant manifolds play a paramount role in the mathematical analysis of dynamical systems  --- in particular in the classification and characterisation of their bifurcations --- in higher dimensions. Interested readers are referred to classical, advanced textbooks on dynamical systems theory~\cite{Wiggins:Book,Guckenheimer:Book}.

\paragraph{Nonlinear oscillators and limit cycles}

Nonlinear oscillators are genuinely different from harmonic oscillators we know from classical mechanics. 
To highlight the difference let's recall the basic results for a classical harmonic oscillator.

\paragraph{Harmonic oscillator}

Newton's equation of motion for a harmonic oscillator with mass $m$ and spring constant $k$ 
\begin{equation*}
	m \, \partial_t^2 x = -k x
\end{equation*}
can be rewritten as a set of two first order differential equations for the position $x$ and the velocity $v$:
%\begin{subequations}
\begin{align*}
	\partial_t x &= v \, , \\
	\partial_t v &= -\frac{k}{m} x \, .
\end{align*} 
%\end{subequations}
%
The steady state is $x^* = v^* = 0$, and its stability is given by the eigenvalues of the characteristic equation
\begin{equation*}
	\begin{vmatrix}
	-\sigma & 1 \\
	-\frac{k}{m} & -\sigma
	\end{vmatrix} = 0 \, 
	\quad \Rightarrow \sigma_\pm = \pm i\sqrt{\frac{k}{m}} \, .
\end{equation*}
In the terminology of the previous section this corresponds to a center. This is related to the fact that for a harmonic oscillator the total energy is strictly conserved,
\begin{equation*}
	\frac{1}{2}m v^2 + \frac{1}{2}k x^2 = E \, ,
\end{equation*}
and hence the orbits in the phase plane $(x,v)$ are cycles around the origin, each with a given energy. 
While the frequency $\omega = \sqrt{\nicefrac{k}{m}}$ is an intrinsic feature of the harmonic oscillator, the amplitude is \emph{not} since it depends on the initial conditions $x_0$ and $v_0$. Moreover, in real life there is nothing like a harmonic oscillator. There is always some kind of damping such that the sum of potential and kinetic energy is actually not conserved but transformed into heat.~\footnote{In the language of nonlinear dynamics, the equation of motion for the harmonic oscillator is structurally unstable. That means, upon adding a generic small term to the equation the dynamics change qualitatively.}
Therefore, in order to achieve sustained oscillations in a technical or a biological system the harmonic oscillator can not be used. 
In the following we will discuss how nonlinear systems give rise to robust self-sustained oscillations.

\paragraph{Hopf bifurcation}

Before discussing biological examples, we begin by analysing nonlinear oscillations in their simplest mathematical form, also known as the normal form \cite{Wiggins:Book} 
%
%\begin{subequations}
\begin{align*}
	\partial_t u 
	&= \rho u - \omega v + 
	   (\mu u - \lambda v) \,
	   (u^2 + v^2) \\
	\partial_t v  
	&= \rho v + \omega u + 
	   (\mu v + \lambda u) \, 
	   (u^2 + v^2) \, .
\end{align*}
%\end{subequations}
%
The origin $u=v=0$ is always a fixed point, with the Jacobian given by
\begin{equation*}
	\mathcal{J} = Df|_0 
	= 
	\begin{pmatrix}
	\rho & -\omega \\
	\omega & \rho
	\end{pmatrix} 
\end{equation*}
with eigenvalues $\sigma_\pm = \rho \pm i\omega$. Here, we can have a situation where the eigenvalue's real part passes through zero while the imaginary part $\omega \neq 0$.

The set of dynamic equations can be considerably simplified using polar coordinates $u = c \cos\theta$, and  $v = c \sin\theta$:
\begin{subequations}
\begin{align}
	\partial_t c 
	&= \rho \, c + \mu \, c^3 \, ,
	\label{eq:r_dynamics_hopf} \\
	\partial_t \theta 
	&= \omega + \lambda \, c^2 \, .
	\label{eq:theta_dynamics_hopf}
\end{align}
\end{subequations}
The time evolution of $c$ and $\theta$ depends only on $c$ but not on $\theta$. 
Hence we have reduced the dynamics of $c$ to a one-component problem as discussed in Section~\ref{sec:1d_well-mixed}.
Introducing a potential $V (c) = - \frac12 \rho \, c^2 - \frac14 \mu \, c^4$ corresponding to a Ginzburg-Landau free energy function, it can also be written as Model A dynamics (for a spatially uniform system) \cite{Hohenberg:1977a}
\begin{equation}
	\partial_t c 
	= - \, \partial_c V(c) 
	\, .
\end{equation} 
For $\rho >0$ and $\mu < 0$, it corresponds to the non-conserved gradient dynamics of an Ising system below the critical temperature exhibiting a second order phase transition at $\rho = 0$ with $\rho > 0$ corresponding to the low temperature phase (see Fig.~\ref{fig:find-fp}).
Changing the sign of $\mu$ to positive values leads to unstable potentials making it necessary to complement the potential by a $c^6$ term with a positive coefficient.
The ensuing phase transition is then a first order phase transition. 
As discussed next, these features have their analogues as sub- and supercritical pitchfork bifurcations.

\begin{figure}[tb]
\centering
\includegraphics{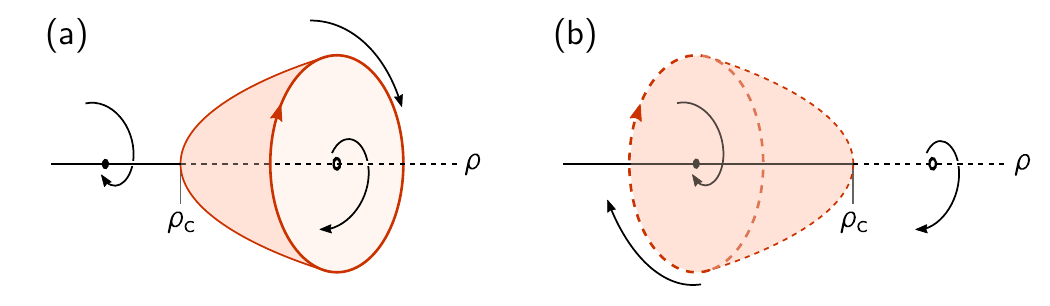}
\caption{\textbf{Hopf bifurcation}. Supercritical (a) and subcritical (b) Hopf bifurcation. Stable and unstable fixed points and limit cycles (red) are indicated by solid and dashed lines, respectively.
}
\label{fig:hopf-bifurcation}
\end{figure}

\textit{Supercritical Hopf bifurcation.\ ---} We start our discussion with the case $\mu<0$, where the potential $V(c)$ has a double well form for $\rho > 0$ and a single minimum for $\rho < 0$.
Then, there is always a fixed point $c^*_0 = 0$ whose stability changes from stable at $\rho < 0$ to unstable at $\rho > 0$; compare Fig.~\ref{fig:hopf-bifurcation}a for an illustration.
While for $\rho < 0$ this is the only fixed point of the dynamics, two new fixed points $c_\pm^* = \pm \sqrt{\nicefrac{\rho}{(-\mu)}}$ emerge for $\rho >0$, leading to a qualitative change in the flow. 
The point $\rho =0$ is called a \emph{pitchfork bifurcation}. 
In general, both branches may have significance. 
For the present case, however, only $c_+^*$ is relevant as $c$ denotes a radial variable.
Its linear stability can be determined by linearising Eq.~\eqref{eq:r_dynamics_hopf} with respect to the fixed point $c_+^*$. 
To leading order one finds for $\delta c = c-c_+^*$: 
\begin{equation*}
	\partial_t \delta c 
	= -2\rho \, \delta c 
	\, .
\end{equation*}
Hence the fixed point $c_+^*$ is stable for $\rho >0$; this is also evident from the form of the potential $V(c)$ that exhibits two local minima at $c_\pm^*$ (Fig.~\ref{fig:find-fp}).  
The defining feature of such a \emph{supercritical} Hopf bifurcation (or, supercritical pitchfork bifurcation if both branches are considered) is that a stable fixed point solution $c_+^* = \sqrt{\nicefrac{\rho}{|\mu|}}$ continuously branches off from the solution $c^*_0 = 0$ at $\rho=0$.
This is the same phenomenology as found in continuous (second order) phase transitions.
In the present case, combined with the  solution of the radial equation, a closed orbit emerges for $\rho \geq 0$  that traces out a circle with radius $c_+^* = \sqrt{\nicefrac{\rho}{|\mu}|}$ at an angular velocity $\omega + \lambda \, (c_+^*)^2$.

\begin{figure}[tb]
\centering
\includegraphics{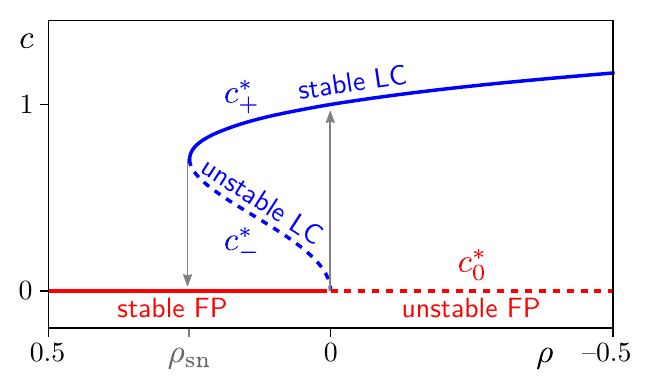}
\caption{\textbf{Subcritical Hopf bifurcation.}
Bifurcation diagram for the nonlinear system described by Eq.~\eqref{eq:subcritical_stabilized_c}. The fixed point $c_0^*=0$ (red line) corresponds to a fixed point of the full dynamical system, Eq.~\eqref{eq:subcritical_stabilized}. It is stable / unstable for $\rho < 0$ / $\rho > 0$. Including the angular variable $\theta$, the fixed points $c^*_\pm$ correspond to stable and unstable limit cycles, respectively. In the parameter window $\rho_\mathrm{sn} = -\frac14 < \rho < 0$ the dynamics is bistable: excitations of a finite magnitude are required to trigger limit cycle oscillations. Starting from $\rho < 0$ in the fixed point $c=0$ and slowly increasing $\rho$, a limit cycle of finite amplitude arises discontinuously upon passing through $\rho = 0$.}
\label{fig:subcritical_excitable}
\end{figure}

\textit{Subcritical Hopf bifurcation.\ ---} For the opposite case, $\mu > 0 $, the fixed point $c_0^*=0$ changes stability as before, but now the fixed points $c_\pm^* = \pm \sqrt{-\nicefrac{\rho}{\mu}}$ exist only for $\rho <0$. 
Moreover, as one can easily check, both branches $c_\pm^*$ are unstable. 
This is called a \emph{subcritical} Hopf (subcritical pitchfork) bifurcation, and is illustrated in Fig.~\ref{fig:hopf-bifurcation}b. 
The feature that the fixed point $c_+^*$ is unstable leads to a runaway flow for $c > c_+^*$.
In order to stabilize the dynamics one may generalise the radial equation by introducing a higher order term (for simplicity we set $\mu$ and the prefactor of the quintic term to unity)
\begin{subequations} \label{eq:subcritical_stabilized}
\begin{align}
	\partial_t c 
	&= \rho \, c + c^3 -c^5 \, , 
	\label{eq:subcritical_stabilized_c} \\
	\partial_t \theta 
	&= \omega + \lambda \, c^2 \, .
	\label{eq:subcritical_stabilized_theta}
\end{align}
\end{subequations}
This corresponds to a potential 
\begin{equation*}
	V(c) = - \frac12 \rho \, c^2 - \frac14 c^4 + \frac16 c^6 
	\, ,
\end{equation*}
i.e.\ a Landau free-energy potential describing a first order phase transition. 
The fixed points for non-negative $c$ are given by $c_0^*=0$ and $(c_+^*)^2 = \frac12  \pm \sqrt{\rho + \frac14}$. 
Performing a linear stability analysis (left as an exercise) yields Fig.~\ref{fig:subcritical_excitable}.
The additional quintic term ($c^5$) prevents the blowup of solutions and gives rise to a new stable branch $c_+^*$ shown as the solid blue line in Fig.~\ref{fig:subcritical_excitable}.
As a result, a limit cycle is created in a saddle node bifurcation at $\rho_\text{sn} = \nicefrac{-1}{4}$.
The fixed point at $c_0^*=0$ becomes unstable at $\rho_\mathrm{c} = 0$.
Taken together, this gives rise to two new phenomena:  
(i) \emph{Hysteresis}: Starting from the fixed point $c_0^*=0$, this fixed point becomes unstable at $\rho = 0$, and the amplitude of the limit cycle oscillations discontinuously jumps to a finite value $c_+^* = 1$; see gray arrow in Fig.~\ref{fig:subcritical_excitable}. Then, upon reducing the control parameter $\rho$ below $0$ these limit cycle oscillations persist until one reaches $\rho_\text{sn} = -1/4$, where it then jumps back to the stable state at $c_0^*=0$ (in a saddle-node backward bifurcation).
(ii) \emph{Bistability}: In the parameter window $ \rho_\text{sn}<\rho<\rho<0$ the dynamics is bistable --- a stable fixed point ($c_0^*=0$) and a stable limit cycle ($c_+^*$) coexist in phase space. The unstable limit cycle at $c_-^*$ acts as a separatrix between the basins of attraction of the fixed point and the stable limit cycle. If a system at $c_0^*$ is perturbed with a magnitude $\delta c > c_-^*$, it will leave the basin of attraction of $c_0^*$ and approach the limit cycle at $c_+^*$. In other words, a sufficiently large stimulus is required to trigger the limit cycle oscillations in this regime. 

\paragraph{Rho GTPase oscillators}

Limit cycle oscillations are important for a range of cellular processes, especially for controlling diverse cellular rhythms~\cite{Ferrell:2011a,Novak:2008a}. Here we discuss a basic model for limit cycle oscillations of Rho GTPases, an important class of proteins that play a central role in the regulation of cell polarity and signal transduction pathways~\cite{Etienne-Manneville_Hall:2002}. 
In Section~\ref{sec:intracellular_protein_patterns} we have already discussed the Cdc42 system of budding yeast which is the central protein responsible for the self-organised establishment of a defined location for daughter cell growth (bud formation).
Cdc42 belongs to the larger class of Rho GTPases which all share the GTPase cycle between a GDP-bound (`inactive') state and a GTP-bound (`active') state regulated by two main classes of proteins: Guanine nucleotide exchange factors (GEFs) promoting the exchange of GDP for GTP, and GTPase activating proteins (GAPs) that facilitate hydrolysis of GTP to GDP. 

\begin{figure}[tb]
\centering 
\includegraphics{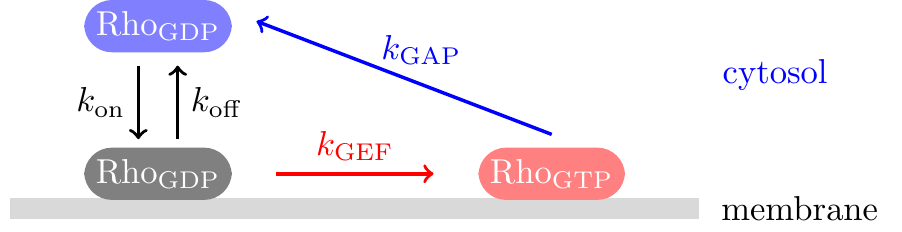}
\caption{
\textbf{Schematic of a Rho GTPase cycle.} Inactive Rho-GDP can attach to and detach from the membrane with rates $\kon$ and $\koff$, respectively. On the membrane, Rho gets activated by GEFs, effectively described by an autocatalytic process with rate  $\kGEF = \kbasal + \kauto \, \mTsq$. When Rho-GTP is hydrolysed with a rate $\kGAP$ it is assumed to detach from the membrane.
}
\label{fig:rho_schematic}
\end{figure}

Figure~\ref{fig:rho_schematic} shows a simplified reaction scheme for a Rho GTPase cycle where the action of the regulatory proteins is accounted for effectively~\cite{Wigbers.etal2020b}.
Cytosolic, inactive Rho can attach to and detach from the membrane with rates $k_\text{on}$ and $\koff$, respectively. 
On the membrane, the inactive Rho conformation can get activated, either with a basal rate  $\kbasal$ or mediated by GEFs. 
As these GEFs are typically recruited to the membrane by active Rho, this can be accounted for effectively by an autocatalytic process.
Here we choose $k_\text{GEF} (\mT)= \kbasal + \kauto \, \mTsq$, where $\mT$ denotes the density of membrane-bound Rho-GTP; other choices for the nonlinearity are possible as well. 
In the active state, Rho can undergo hydrolysis, mediated by a GAP, with rate $k_\text{GAP}$. 
We assume that Rho-GTP immediately detaches from the membrane after it is hydrolysed.
The reaction rates in the model are effective rates and depend on the concentrations of the regulatory proteins. 
The nucleotide exchange rates will depend on the GEF concentration, and the hydrolysis rate on the GAP concentration. 
The above reaction scheme is a simplification of actual biological systems for several reasons: 
First, the regulatory proteins are only accounted for effectively. 
Second, Rho as well as the regulatory proteins in general form interaction networks with a multitude of other proteins.
Here, we disregard all these effects, as our main interest is to present a biologically plausible but still pedagogical example for a limit cycle oscillator.

Mathematically, the reaction network (in a well-mixed system) can be written as 
\begin{subequations}
\begin{align}
	\partial_t \mT 
	& = - \kGAP \, \mT 
	    + \kGEF (\mT ) \, \mD  
	\, , \\
	\partial_t \mD 
	& = \hphantom{-} \kon \, \cD 
		- \koff \, \mD
	    - \kGEF (\mT) \, \mD 
	\, ,  \\
	\partial_t \cD 
	& = - \kon \, \cD +  \koff \, \mD + \kGAP \, \mT 
	\, , 
\end{align}
\end{subequations}
where $\cD$ denotes the cytosolic density of inactive Rho, an $\mD$ and $\mT$ the membrane density of inactive and active Rho, respectively. 
\begin{figure}[tb]
\centering
\includegraphics{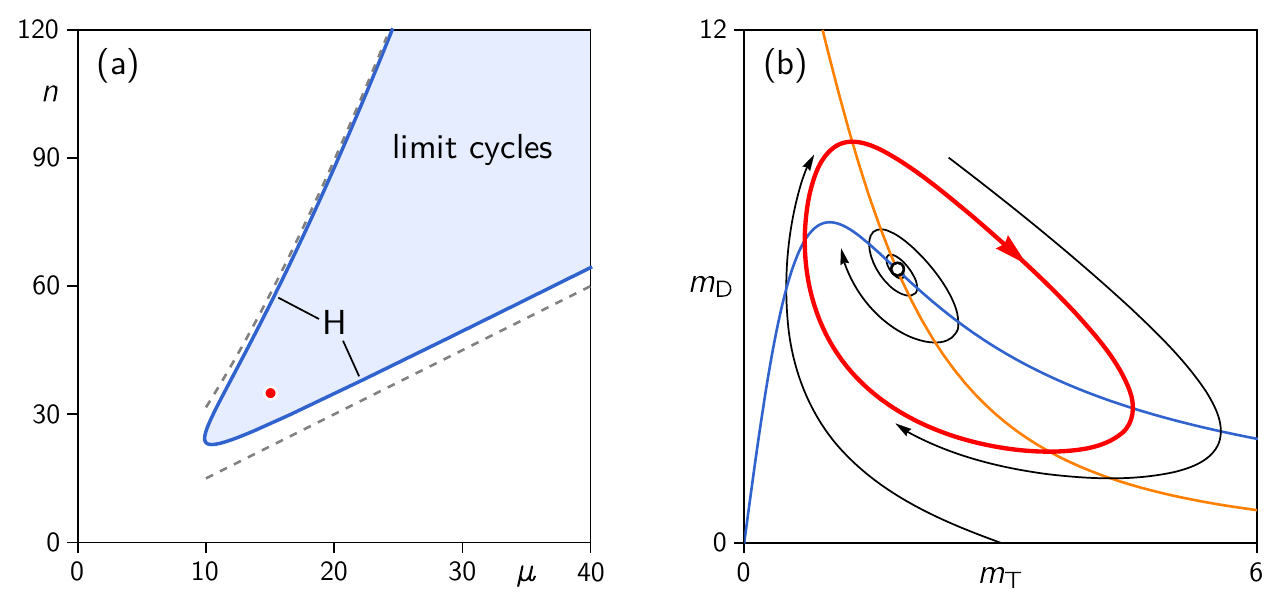}
\caption{\textbf{Limit cycle oscillations for Rho GTPases}.
(a) Bifurcation diagram as a function of the hydrolysis rate $\mu$ and the total protein density $n$. In the regime shaded in blue, enclosed by a Hopf bifurcation (H), one observes limit cycle oscillations; the gray, dashed lines indicate upper and lower bounds to the stability boundaries as discussed in the main text.
(b) Phase space portrait for $n=35$ and $\mu =15$ (red dot in (a)). The blue and orange lines indicate the nullclines $\partial_t \mT = 0$ and $\partial_t \mD = 0$, respectively. The thick, red line shows the limit cycle.
}
\label{fig:illustration_limit_cycle}
\end{figure}
Mass conservation requires that the total protein number $n$ remains fixed:
\begin{align}
	n = \cD + \mD + \mT \, .
\end{align}
In the following we analyse a simplified version with $\koff = 0$. 
We choose the time scale such that $\kon = 1$, and rescale $n \rightarrow n / \sqrt{\kauto}$. 
Then, using mass-conservation, the dimensionless form of the reaction kinetics reads
%\begin{subequations}
\begin{align*}
	\partial_t \mT 
	& = - \mu \, \mT 
	    + \big( \lambda + \mTsq \big) \, m_D 
	\, , \\
	\partial_t \mD 
	& =  n - \mD - \mT
	    - \big( \lambda + \mTsq \big) \, \mD
	\, ,
\end{align*}
%\end{subequations}
where $\mu = \kGAP/\kon$ and $\lambda = \kbasal/\kon$. For simplicity, we specify $\lambda =1$. Then the nullclines are given by
\begin{equation}
	\mD = \mu \frac{\mT}{1 + \mTsq} \quad \text{and} \quad
	\mD = \frac{n-\mT}{2 + \mTsq} \, ,
\end{equation}
respectively (see blue and orange lines in Fig.~\ref{fig:illustration_limit_cycle}b).
The intersection of the nullclines defines the fixed point (reactive equilibrium). 
Its position and stability can be changed by varying the overall protein number $n$ (protein mass) and the hydrolysis rate $\mu$.
As discussed above, the stability of the fixed point is determined by the sign of the trace of the Jacobian: 
\begin{equation*}
	\tau = - \mu + 2 m_\mathrm{T}^*  m_\mathrm{D}^* - 2 - (m_\mathrm{T}^*)^2 \, .
\end{equation*}
Solving the fixed point equations $\partial_t \mT = \partial_t \mD = 0$ together with the Hopf bifurcation criterion $\tau = 0$ one obtains the locus of the Hopf bifurcation in parametric form 
\begin{equation*}
	\mu_\mathrm{Hopf}^{} =
	\frac{2 + 3 \mTsq + m_\mathrm{T}^4}{\mTsq -1} \,, \quad
	n_\mathrm{Hopf}^{} = \frac{\mT \, (3 + 5 \mTsq + m_\mathrm{T}^4)}{\mTsq -1} \, .
\end{equation*}
The Hopf bifurcation encloses a parameter regime with $\tau >0$ where the reaction kinetics exhibits limit cycle oscillations (Fig.~\ref{fig:illustration_limit_cycle}); note that within this domain, the determinant of the Jacobian $\delta = (3 + 4 \mTsq + 4 m_\mathrm{T}^4 + m_\mathrm{T}^6)/(\mTsq-1) > 0$ since $m_T > 1$. 
The upper and lower branch of the stability boundary are approximated by $n = \mu^{3/2}$ and $n = \frac32 \mu$, respectively. 
In the limit of large hydrolysis rate ($\mu \to \infty$), one obtains $(\mT, \mD) \to (1, n/3)$ (lower branch) and  $(\mT, \mD) \to \nicefrac{n}{\mu} \, (1,1)$ (upper branch).
Hence, limit cycles are obtained in a parameter regime where the hydrolysis rates are large and proteins are mainly localised in the cytosol. 
Oscillations can be turned on and off by either varying the total protein mass or the hydrolysis rate via the number of GAP proteins.

\begin{exercise}[tb]
Perform a bifurcation analysis for the `Brusselator' \cite{Prigogine:1968a}
\begin{align*}
	\partial_t u(t) 
	&= \hphantom{-} u^2 \, v - (\lambda + 1) \, u + \mu  \, , \\
	\partial_t v(t) 
	&= - u^2 \, v + \lambda \, u  \, ,
\end{align*}
and the `Schnakenberg model' \cite{Schnakenberg:1979a}
\begin{align*}
	\partial_t u(t) 
	&= \hphantom{-} u^2 \, v - u  + \mu
	\, , \\
	\partial_t v(t) 
	&=  - u^2 \, v + \lambda
	\, ,
\end{align*}
where $\mu$ and $\lambda$ are positive parameters.
In both cases, determine the parameter domain for limit cycle oscillations. 
\end{exercise}

%% ================================
%% TWO COMPONENT REACTION DIFFUSION
%% ================================

\section{Spatially extended two-component systems}
\label{sec:two-component_MCRD}

This section discusses the dynamics of reaction--diffusion systems with two components in one spatial dimension, 
\begin{subequations}
\label{eq:turing_model}
\begin{align}
	\partial_t a 
	&= D_a \partial_x^2 a + f(a,b) \, ,\\
	\partial_t b 
	&= D_b \partial_x^2 b \kern0.2em + g(a,b) \, ,
\end{align}
\end{subequations}
where $a(x ,t)$ and $b(x ,t)$ denote the concentrations of components A and B, respectively. 
The analysis of such systems goes back to the pioneering work by Alan Turing \cite{Turing:1952a}.
The functions $f(a,b)$ and $g(a,b)$ are  nonlinear functions describing the reaction kinetics of the underlying biochemical system. 
In the original paper by Turing~\cite{Turing:1952a} and subsequent analysis~\cite{Gierer:1972a,Segel:1972a}, the functions $f$ and $g$ were assumed to be independent.
In these lecture notes, we are interested in mass-conserving systems, where the spatial average of the total protein density $n(x,t) := a(x,t) + b(x,t)$ is a conserved quantity:
\begin{equation*}
	\nbar = \frac{1}{L} \int_0^L \! \dd x \, n(x,t) = \text{const.}
\end{equation*} 
Then, the nonlinear functions can no longer be independent but must be related: $g(a,b)=-f(a,b)$. 
Such two-component \emph{mass-conserving reaction--diffusion} (MCRD) systems have been considered in the literature as `null-models' of protein pattern forming systems~\cite{Ishihara:2007a,Otsuji:2007a,Goryachev:2008a,Altschuler:2008a,Mori:2008a,Jilkine:2011a,Jilkine:2011b,Edelstein-Keshet:2013a,Trong:2014a,Chiou:2018a,Hubatsch:2019a}.
They have also been studied in the context of slime mold aggregation~\cite{Keller:1970a}, cancer cell migration (glioma invasion)~\cite{Pham:2012a}, precipitation patterns~\cite{Goh:2011a}, and simple contact processes~\cite{Wijland:1998a,Kessler:1998a}. Furthermore, non-isothermal solidification models~\cite{Caginalp:1986a} can also be rewritten in the form of MCRD equations; see e.g.\ Refs.~\cite{Morita:2010a,Pogan:2012a}.
As will be discussed in the following, these MCRD systems are qualitatively different from models where the functions $f$ and $g$ are genuinely independent.
More broadly, we will argue that reaction--diffusion systems with a mass-conserving `core'
\begin{subequations} 
\label{eq:two-comp-dyn_non-cons}
\begin{align}
	\partial_t a(x,t) 
	&= D_a \partial_x^2 a 
	 + f(a, b) + \varepsilon \, s_1 (a, b)
	 \, , 
	\label{eq:m-dyn_non-cons} \\
	\partial_t b(x,t) 
	&= D_b \partial_x^2 b \kern0.18em
	 - f(a, b) + \varepsilon \, s_2 (a, b)
	 \, ,
	\label{eq:c-dyn_non-cons}
\end{align}	
\end{subequations}
are a general class of models, whose dynamics is generically driven by the interplay between the spatiotemporal redistribution of total protein density $n(x,t)$ and moving local equilibria. 
Here, the parameter $\varepsilon$ describes the degree to which mass-conservation is broken by processes like production and degradation of proteins encoded in the source terms $s_{1,2}$. 
For example, the reaction kinetics of the Brusselator may be written as
\begin{subequations}
\label{eq:Brusselator-rewritten}
\begin{align}
	 \partial_t a
	 &= - \lambda  \, a + a^2 \, b 
	    + \varepsilon \, (\mu -  a)
	\, , \\
	\partial_t b
	 &= \hphantom{-} \lambda \, a  - a^2 \, b  \, ,
\end{align}
\end{subequations}
where the additional terms in the dynamics of species A can be interpreted as supply of particles A from an abundant resource ($\varepsilon \mu$) and degradation ($-\varepsilon a$).

As a preparation, we will recap Turing's classical linear stability analysis for general two-component reaction diffusion systems \eqref{eq:turing_model}. This analysis shows that diffusion can induce an instability in a system with locally stable reaction kinetics.
We then turn to mass-conserving two-component systems, starting with a brief description of their generic phenomenology, as obtained by numerical simulations. In particular, we illustrate how the dynamics of the spatially extended system can be visualized in the $(m,c)$-phase portrait introduced in Fig.~\ref{fig:total-density-bifurcation-structure} and point out several key observations that motivate the subsequent analysis in terms of phase-space geometry.
As a first step to gain a deeper understanding of the dynamics of MCRD systems, we then repeat the linear stability in this specific case, highlighting the effect of mass-conservation on the linear stability properties (dispersion relation and associated eigenvectors). 

Both the phenomenology observed in numerical simulations and the linear stability analysis suggest that lateral redistribution of the total density $n(x,t)$ plays a crucial role in the dynamics.
Based on this insight, we discuss a thought experiment that elucidates the key concept that arises from mass conservation and underlies our framework: local (reactive) equilibria.
 Revisiting the linear stability analysis once again, we use the local equilibria to explain the physical mechanism that underlies the Turing instability in MCRD systems.
 Finally, we show how an analysis based on phase-space geometry in $(m,c)$-phase space allows us to go beyond linear stability analysis and characterise the dynamics and final steady state (stationary pattern) in the highly nonlinear regime. 
The key geometric object will be a linear subspace, called flux-balance subspace, that represents the steady-state balance of diffusive fluxes in $(m,c)$-phase space.
Intersection points between the reactive nullcline and the flux-balance subspace will act as landmark points for the construction of stationary patterns.
Moreover, we will show that the balance of reactive turnovers that arises as a stationarity condition is (approximately) represented by a balance of areas in the $(m,c)$-phase portrait, akin to a Maxwell construction.

\subsection{Classical Turing system}
\label{sec:classical_turing}

\begin{figure}[t]
\centering 
\includegraphics{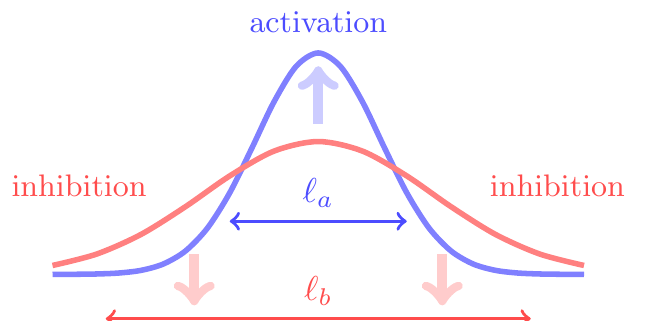}
\caption{Density profiles of species A and B after some initial local perturbation in the absence of reaction kinetics. Since the diffusion constants are different, these perturbations spread out to different length scales $\ell_b > \ell_a$, leading to `de-mixing' of the species. In an `activator-inhibitor' scenario, the density profile of the activator gets self-enhanced on short scales and inhibited on large scales, indicated by the arrows.} 
\label{fig:demixing_turing} 
\end{figure}

While an analysis of the classical Turing case can be found in many textbooks, we will still give a concise discussion here, mainly to contrast it with the mass-conserving systems.
Let us start with a heuristic argument. 
Assume that the spatially homogeneous state with uniform densities $a^*$ and $b^*$ is stable with respect to small spatially uniform perturbations, i.e.\ $(a^*,b^*)$ is a stable fixed point of the reaction kinetics. Henceforth, we will refer to this as \emph{local stability} as it is a property of the local reaction kinetics.
For the sake of the argument, we also assume that the reaction kinetics is characterised by a single typical time scale $\tau$.
Then, during that reaction time, components A and B will diffuse over diffusion length scales
\begin{equation*}
	\ell_{a/b} :=\sqrt{D_{a/b} \, \tau} \, .
\end{equation*}
If these length scales are approximately the same, $\ell_a\approx \ell_b$, components A and B diffuse at the same `speed' such that one expects that the system remains well-mixed and it will return back to a spatially homogeneous steady state after small spatially heterogeneous perturbations. 
If one species, however, spreads much faster (say $D_b\gg D_a$) then, after some reaction time $\tau$, a local spatial perturbation will lead to largely different spatial distributions for components A and B with widths $\ell_b \gg \ell_a$. In other words, diffusion leads to a `de-mixing' of the species as illustrated in Fig.~\ref{fig:demixing_turing}.

What emerges out of this `de-mixing' depends on the reaction kinetics. Consider the following \emph{`activator--inhibitor'} scenario \cite{Gierer:1972a}: 
(i) The `activator' A is auto-catalytic, and also enhances the inhibitor B. (ii) The inhibitor B slows down the activation process of A, i.e.\ there is a negative feedback. 
(iii) Finally, A and B also decay at some rate.
The consequences of this reaction scheme on the concentration profile of the activator are the following: 
In the core region of the spatial perturbation, $\ell \lesssim \ell_a$, there is a surplus of activator A since the inhibitor B has spread out faster than the activator. As a consequence, the activator concentration grows (and is finally levelled off by nonlinear terms in the reaction kinetics).
On the other hand, in the outer region $\ell \sim \ell_b$ there is a surplus of inhibitor leading to a depletion of the activator.
Taken together this leads to a density profile for the activator illustrated in Fig.~\ref{fig:demixing_turing}. 
Since the region of enhanced activator concentration is surrounded by a cloud of inhibitors, the ensuing spatial profile is argued to be rather robust to external perturbations. 
For such type of reaction kinetics, the Turing mechanism is sometimes also summarised as `short-range activation with long-range inhibition leads to pattern formation'~\cite{Gierer:1972a,Segel:1972a}.

This heuristic line of argument, however, does not specify the exact conditions on the diffusion constants and the reaction rates that lead to pattern formation. 
These arguments merely state that one needs some kind of `demixing' driven by unequal diffusion constants onto which a nonlinear feedback mechanisms can act to produce a stable pattern. 

To quantify the above ideas mathematically one has to use a formal \emph{linear stability analysis}, asking whether the spatially uniform steady state $(a^*,b^*)$ is stable or unstable with respect to \emph{spatially heterogeneous perturbations}, termed \emph{lateral (in)stability}. To this end, we expand the reaction--diffusion equations, Eq.~\eqref{eq:turing_model}, to linear order in the deviation from the steady state, $\delta \vek{u} := (a-a^*,b-b^*)^\mathrm{T}$:
\begin{equation}
	\partial_t \delta \vek{u}
	= \mathcal{J} \delta \vek{u} + 
	  \mathcal{D} \, 
	  \partial_x^2 \delta \vek{u}\, ,
\label{eq:turing_linear}
\end{equation}
where
\begin{equation}
	\mathcal{J}  = \left.
	\begin{pmatrix}
	\partial_a f & \partial_b f \\
	\partial_a g & \partial_b g
	\end{pmatrix} \right|_{(a_0,b_0)} 
	=: 
	\begin{pmatrix}
	f_a & f_b \\ g_a & g_b 
	\end{pmatrix} 
\end{equation}
denotes the Jacobian of the reaction kinetics at the uniform steady state $(a^*,b^*)$, and $\mathcal{D}= \diag \, (D_a,D_b)$ is the diffusion matrix.
Stability of $(a^*,b^*)$ to uniform perturbations (local stability) requires that both eigenvalues $\sigma_1$ and $\sigma_2$ of the Jacobian $\mathcal{J}$ are negative, i.e.\ (compare Section \ref{sec:2d_well-mixed})
%
%\begin{subequations}
\begin{align*}
	\tau_0 
	&= \Tr \, \mathcal{J}
	= \sigma_1+\sigma_2
	= f_a+g_b <0 
	\, , \\
	\delta_0 
	&= \Det \, \mathcal{J}
	= \sigma_1 \cdot \sigma_2
	= f_a g_b - g_a f_b >0 \, .
\end{align*}
%\end{subequations}

To study the stability against spatially inhomogeneous perturbations (lateral stability), the linear reaction--diffusion equations, Eq.~\eqref{eq:turing_linear}, are solved by expanding $\delta \vek{u} (x,t)$ in terms of the eigenmodes $\phi_q (x)$ of the Laplacian, fulfilling the eigenmode equation $\partial_x^2 \phi_q (x) = - q^2 \phi_q (x)$,\footnote{Note that the eigenmodes of the Laplacian depend on the domain geometry and the boundary conditions. This will play a critical role in Sec.~\ref{sec:LSA-box-geometry} where linear stability analysis is performed for bulk-surface coupled systems.} 
\begin{equation} \label{eq:mode-decomposition}
	\delta \vek{u} (x,t)
	= \sum_i \sum_q 
	   A_q^{(i)} \vek{e}_q^{(i)} 
	   \phi_q (x) 
	   \exp \bigl( \sigma_q^{(i)} t \bigr) 
	   \, ,
\end{equation}
resulting in the eigenvalue problem
\begin{equation} \label{eq:line-EV-problem}
	\sigma_q 
	\vek{e}_q
	= (\mathcal{J} - \mathcal{D} \, q^2) \vek{e}_q =: \mathcal{J}_q \vek{e}_q
	\, .
\end{equation}
The eigenvalues $\sigma_q^{(i)}$ are obtained from the characteristic equation
\begin{equation} \label{eq:line-characteristic-eq}
	\det
	\bigl(\mathcal{J}_q - \sigma_q \mathbb{I} \bigr)
	= 0 \, .
\end{equation}
For each Fourier mode $q$ one finds two eigenvalues
\begin{equation}
	\sigma_q^{(\pm)}
	= \frac{1}{2}
	  \left( \tau_q \pm   
	         \sqrt{\tau_q^2-4\delta_q}
	  \right) \, ,
\label{eq:eigenvalues_turing}
\end{equation}
where $\tau_q = \Tr\mathcal{J}_q$ and $\delta_q = \det \, \mathcal{J}_q$ are the trace and determinant of the matrix $\mathcal{J}_q = \mathcal{J} - \mathcal{D} q^2$, respectively. 
The eigenvalues $\sigma_q^{(i)}$ as a function of the wavenumber $q$ constitute the so called \emph{dispersion relation}. Often one is primarily interested in the eigenvalue with largest real part for each $q$ since this determines the stability of the $\cos(q x)$ mode.

\begin{figure}[tb]
\centering
\includegraphics{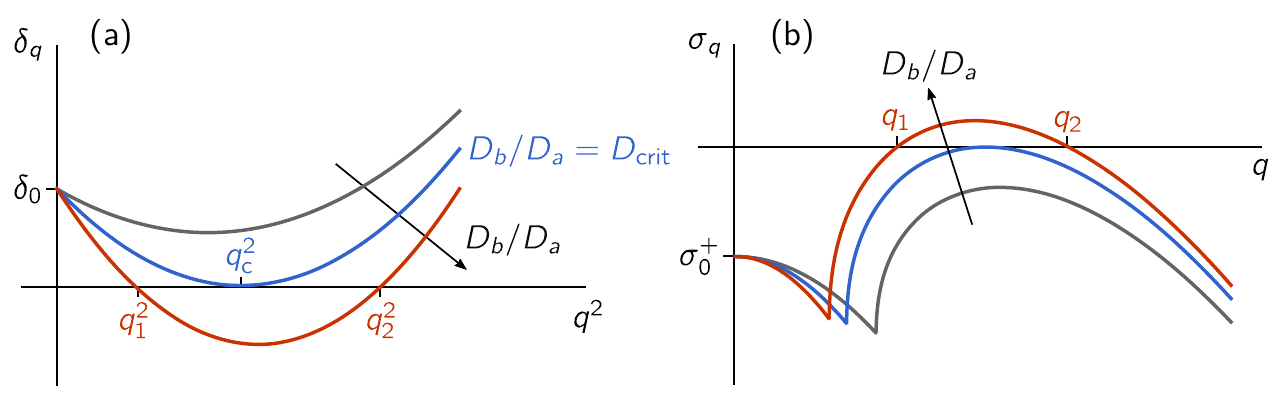}
\caption{
(a) Determinant $\delta_q \,{=}\, \det \big(\mathcal{J}-\mathcal{D}q^2 \big)$ as a function of $q^2$. 
(b) Dispersion relation $\sigma_q^+$. There is a critical value for the ratio of the diffusion constants, $D_\mathrm{crit}$, where the mode corresponding to the wavenumber $q_\mathrm{c}$ becomes marginally stable. For $D \,{>}\, D_\mathrm{crit}$, there is a band of wavenumbers $[q_1^2,q_2^2]$ where $\delta_q \,{<}\, 0$, and, therefore, one eigenvalue becomes positive and thereby the dynamics becomes unstable.
(The plots were generated for the Jacobian $\mathcal{J} = ((1,-1),(3,2))^\mathrm{T}$, $D_a = 1$, and $D_b = 5, 7.45, 10$ for the gray, blue, and red line respectively.)
}
\label{fig:determinant_turing}
\end{figure}

To find the parameter regime where the homogeneous steady state is unstable against spatial perturbations, we seek a condition that guarantees that one eigenvalue $\sigma_q^{\pm}$ is non-negative. 
Since we demand that the fixed point $(a^*,b^*)$ of the spatially uniform system is stable against uniform perturbations, we have $\Tr\mathcal{J} <0$, and thus
\begin{equation*}
	\tau_q :=
	\Tr \mathcal{J}_q 
	= \Tr \mathcal{J} - q^2 \, \Tr\mathcal{D} < 0 \, .
\end{equation*}
Hence the eigenvalue with the larger real part, $\sigma_q^+$, becomes positive only if 
\begin{equation*}
	\delta_q 
	= \det \mathcal{J}_q 
	= \big( f_a-D_a q^2 \big)
	  \big( g_b-D_b q^2 \big) - g_a f_b < 0 \, .
\end{equation*}
As shown in Fig.~\ref{fig:determinant_turing}a, the $\delta_q$ is a parabola in $q^2$ that opens upwards. At $q \,{=}\, 0$ it takes the value $\delta_0 \,{>}\, 0$ which is always positive since the spatially uniform state is stable. 

In order to determine when $\delta_q$ changes sign, we ask when the minimum value of this parabola first becomes negative. Setting the derivative of $\delta_q$ with respect to $q^2$ to zero, we learn that the minimum occurs at the wave number
\begin{equation*}
	q_*^2
	=\frac{f_a D_b+g_b D_a}{2 D_a D_b} \, .
\end{equation*}
The corresponding value of $\delta_q$ at this minimum is 
\begin{equation*}
	\delta_{q_*}
	= \delta_0 - 
	  \frac{\left(f_aD_b+g_bD_a\right)^2}
	       {4D_aD_b} \, .
\end{equation*}
This expression is negative if the inequality
\begin{equation}
	f_a D_b+g_b D_a > \sqrt{4 D_a D_b \delta_0} 
\label{eq:turing_condition}
\end{equation}
is satisfied (\emph{Segel--Jackson condition}~\cite{Segel:1972a}). 
This condition is necessary and sufficient for the emergence of a Turing instability.
Upon defining $D \,{:=}\, \nicefrac{D_b}{D_a}$ we may rewrite the \emph{Turing condition}, Eq.~\eqref{eq:turing_condition}, as $
	f_a \, D + g_b 
	> 2 \sqrt{D \, \delta_0}$, 
i.e.\ only the relative strength $D$ of the diffusion constants matters. 
There is a critical value for $D$ (at given kinetic parameters) where the Turing instability occurs
\begin{equation*}
	f_a \, D_\mathrm{crit}+g_b
	= 2\sqrt{D_\mathrm{crit} \, \delta_0} \, .
\end{equation*}
At this threshold value $D_\mathrm{crit}$, the critical wave number is 
\begin{equation*}
	q_\mathrm{c}^2
	=\frac{f_a D_\mathrm{crit}+g_b}{2D_\mathrm{crit}}
	=\sqrt{\frac{\delta_0}{D_\mathrm{crit}}} \, .
\end{equation*}
Above this threshold value, $D \,{>}\, D_\mathrm{crit}$, the dispersion relation $\sigma_q$ exhibits a band $[q_1,q_2]$ of unstable modes (see Fig.~\ref{fig:determinant_turing}b). 
While the above conditions mathematically specify the parameter space where pattern formation is possible, they do not provide conceptual insights into the underlying mechanism responsible for the Turing instability.

\begin{exercise}
Perform a linear stability analysis for the Schnakenberg model~\cite{Schnakenberg:1979a} 
\begin{align*}
	\partial_t a(x,t) 
	&= D_a \partial_x^2 a 
	+ \mu + a^2 b  - a \, , \\
	\partial_t b(x,t) 
	&= D_b \partial_x^2 b \kern0.2em
	+ \lambda - a^2 b \, ,
\end{align*}
where $\mu$ and $\lambda$ are positive constants. Discuss the regimes of local and lateral stability.	You may also want to write a computer program that simulates the dynamics of the model beyond the regime of linear instability. What does one find?
\end{exercise}

\goodbreak

\subsection{Mass-conserving two-component reaction--diffusion systems}
\label{sec:mcrd_definition}

Now we turn to the analysis of two-component mass-conserving reaction--diffusion (MCRD) systems.
As in Section~\ref{sec:1d_well-mixed}, we consider two chemical species, M and C, and refer to them as membrane-bound proteins and cytosolic proteins.
Compared to an actual cellular system or reconstituted experimental model system, we again use a simplified one-dimensional geometry where proteins are confined to a domain $[0,L]$.
The fact that membrane-bound and cytosolic proteins experience different fluid environments is accounted for by two different diffusion constants $D_m$ and $D_c$ respectively. 
With $m(x,t)$ and $c(x,t)$ denoting the local densities of components M and C, respectively, the reaction--diffusion equations read
\begin{subequations}
\label{eq:2compRD}
\begin{align} 
	\partial_t m (x,t)  
	&= D_m \partial_x^2 m + f(m,c) \, ,\\
	\partial_t c (x,t)  
	&=  \kern0.3em D_c \partial_x^2 c \kern0.5em - f(m,c) \, ,
\end{align}	
\end{subequations}
where the nonlinear function $f(m,c)$ accounts for all chemical reactions.
Without loss of generality we assume that the relative diffusion constant $D:=D_m/D_c \geq 1$.
In the following we will restrict ourselves to no-flux boundary conditions\footnote{Periodic boundaries can be treated analogously.}
%\begin{subequations}
\begin{align*}
	\partial_x m|_{0}
	&= \partial_x m|_{L}=0 \, , \\
	\partial_x c|_{0}
	&= \kern0.225em \partial_x c|_{L} \kern0.225em =0 \, .
\end{align*}
%\end{subequations}
Recall from our analysis of well-mixed two-component systems with mass conservation in Section~\ref{sec:1d_well-mixed} that the equilibria are given by the intersection of the nullcline ($f=0$) with the reactive phase space ($n=m+c$) for a given protein mass $n$.

\paragraph{Phenomenology of the spatiotemporal dynamics}

\begin{figure}
\centering 
\includegraphics{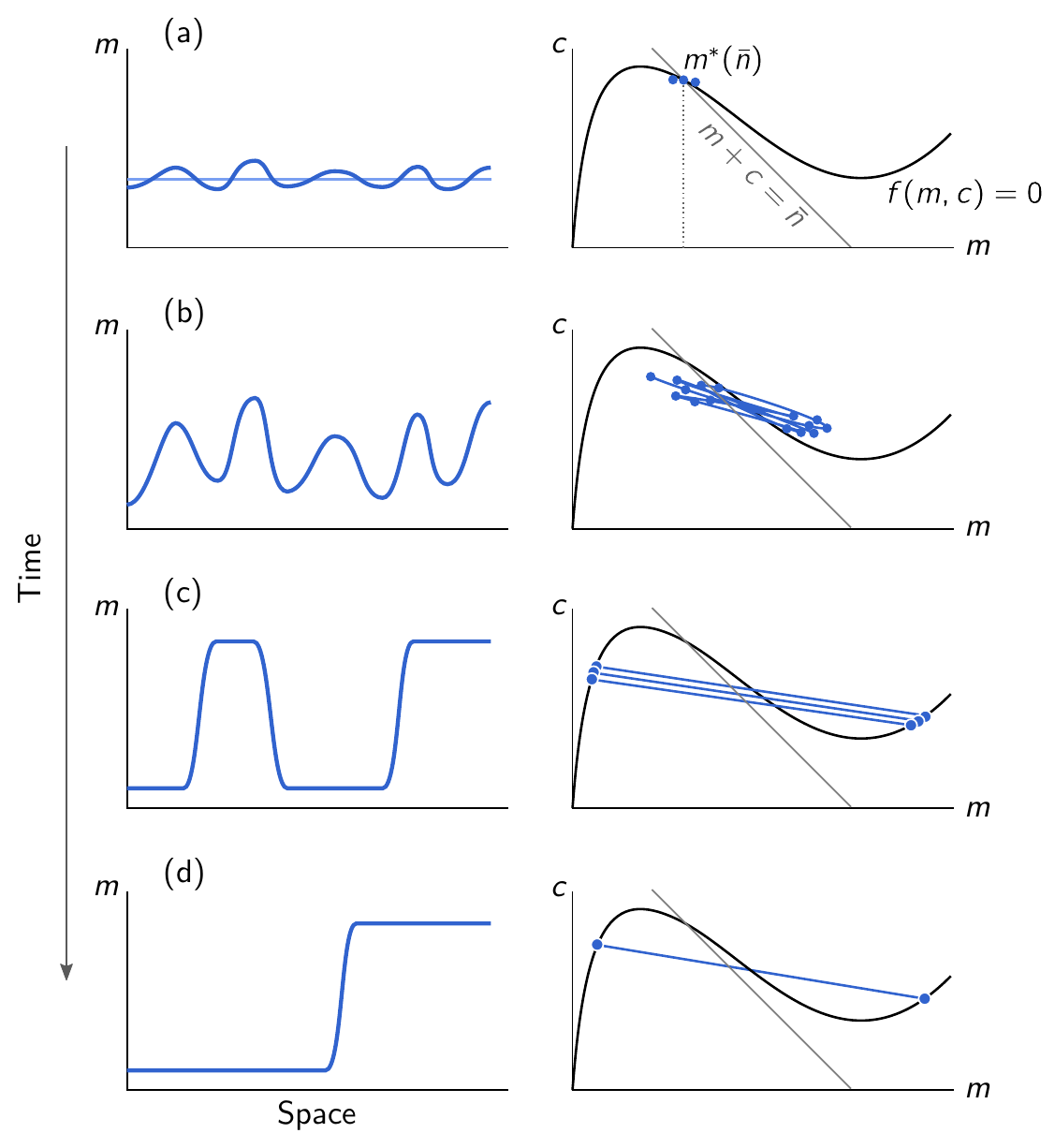}
\caption{\textbf{Mapping between pattern in real space and control space}.
Each pair of panels shows the concentration profile in real space next to the respective ``distribution'' in phase space; the reactive nullclines are shown in black. 
Mass conservation enforces that the center of mass of the ``phase-space distribution'' must always lie on the line $m+c=\bar{n}$ (light gray). 
The pattern growing initially exhibits a characteristic length scale, determined by the linearised dynamics close to the homogeneous steady state. 
Asymptotically --- via a coarsening process --- the steady state is a polar pattern with two plateaus separated by an interface.
}
\label{fig:real_space-phase_space}
\end{figure}

Before we embark on a mathematical analysis, let us look at the results of a numerical simulation of Eq.~\eqref{eq:2compRD} for a toy model with
\begin{equation} 
\label{eq:toy}
	f(m,c) 
	= 
	\left(
	k + \frac{m^2}{1+m^2}
	\right) c - m
	\, . 
\end{equation}
Our intention is to show the phenomenology of pattern formation in real space and how these patterns are represented in phase space. 
To this end, similar as above the one-dimensional domain is dissected into a set of compartments, and the time evolution of the membrane and cytosolic density in each of these compartments is monitored in $(m,c)$ phase space leading to a cloud of phase space points. 
Figure~\ref{fig:real_space-phase_space} shows a series of snapshots that sets side by side the spatial membrane density and the corresponding distribution of phase space points.

The initial concentration for the simulation displayed in Fig.~\ref{fig:real_space-phase_space}a is a small (random) perturbation with respect to a spatially homogeneous stationary state $(m^*,c^*)$, indicated by the dashed line in the real space plot on the right. This steady state corresponds to the fixed point determined by the intersection of the nullcline with the reactive phase space $m + c = \nbar$ (gray line) in the phase portrait on the right.
Initially, the dynamics exhibits a Turing instability which (exponentially) amplifies the small perturbations (Fig.~\ref{fig:real_space-phase_space}b).
The ensuing initial pattern is a periodic pattern with a characteristic wavelength that will later turn out to be the fastest growing mode as obtained from a linear stability analysis.
The growing pattern entails redistribution of protein mass, i.e.\ a spatially heterogeneous total density $n(x,t) := m(x,t) + c(x,t)$.
As a consequence, the corresponding cloud of points in the phase portrait starts deviating from the subspace $m + c = \nbar$ (gray line).

After this initial phase, one observes that the points in phase space become constrained to an affine subspace (i.e.\ a straight line) in phase space (Fig.~\ref{fig:real_space-phase_space}c), which will later turn out to be the so-called flux-balance subspace. Within that subspace, the pattern's phase-space distribution extends from one branch of the nullcline to the other.
On even longer timescales the pattern profile undergoes slow coarsening dynamics until it reaches a polarised steady state with a high and a low density separated by an interface, as shown in Fig.~\ref{fig:real_space-phase_space}d. During the coarsening process, the distribution in phase space remains very close to the flux-balance subspace.

These phenomenological observations raise a set of fundamental questions: 
What is the mechanism underlying the pattern-forming instability from a spatially homogeneous state?
What is driving the coarsening process from the initial periodic pattern to the final polar pattern?
Is there a way to generalise the geometric phase space ideas for well-mixed systems to spatially extended systems?
Is there a relation between the real space patterns and the attractors observed in phase space?
What are the physical principles that determine the attractors in phase space?
In the following we will address these questions. 

\begin{exercise}
	Perform numerical simulations with the reaction kinetics $$f = m^2 c - m$$ from Exercise~\ref{ex:conservative-brusselator} and visualize the dynamics in real space and in phase space as shown in Fig.~\ref{fig:real_space-phase_space}. What are the qualitative differences between the cases $D_m \ll D_c$ and $D_m \lesssim D_c$? How do your findings compare to a system with reaction kinetics given by Eq.~\eqref{eq:toy}?
\end{exercise}

\goodbreak 
 
\subsection{Linear stability analysis of mass-conserving systems} 
\label{sec:lateral-instability}

As we have learned in the previous section, linear stability analysis of a reaction--diffusion system is performed by linearising the equations with respect to the homogeneous steady state $(m^*,c^*)$ and expanding a spatial perturbation in the eigenbasis of the diffusion operator (Laplacian) in the geometry of the system.
For a one-dimensional system with reflective boundary conditions at $x=0$ and $x=L$, the eigenfunctions of the Laplacian are the discrete Fourier modes $\cos( q_k x )$ where $q_k = k \pi/L$ are a discrete set of wave numbers with $k \in \mathbb{N}$. For simplicity of notation, we suppress the subindex $k$ in the following.
The amplitudes of the Fourier modes, $\delta m_q(t)$, then obey the linearised dynamics
\begin{equation*} 
	\partial_t \begin{pmatrix} \delta m_q(t) \\ \delta c_q(t) \end{pmatrix} = \mathcal{J}_q \begin{pmatrix} \delta m_q(t) \\ \delta c_q(t) \end{pmatrix} 
\end{equation*}
with the Jacobian
\begin{equation*}
	\mathcal{J}_q = 
	\begin{pmatrix}
		-D_m q^2 + f_m & f_c \\
		- f_m & -D_c q^2 - f_c
	\end{pmatrix},		
\end{equation*}
where $f_{m,c} = \partial_{m,c} f |_{(m^*,c^*)}$.
The eigenvalues of the Jacobian yield the growth rates $\sigma^{(i)}_q$ of the respective eigenmodes such that the time evolution of a perturbation in the spatial eigenfunction $\cos(q x)$ is given by (cf.\ Eq.~\eqref{eq:mode-decomposition})
\begin{equation*}
	\begin{pmatrix} \delta m_q(t) \\ \delta c_q(t) \end{pmatrix}  = \sum_{i=1,2} A^{(i)}_q \, \mathbf{e}^{(i)}_q \exp(\sigma^{(i)}_q t) \cos(q x),
\end{equation*}
with the eigenvectors $\mathbf{e}^{(i)}_q$ associated to the eigenvalues $\sigma^{(i)}_q$. For a given initial condition (perturbation), the coefficients $A^{(i)}_q$ are determined by projecting the initial condition onto the eigenbasis $\mathbf{e}^{(i)}_q \cos(q x)$.
\begin{figure}
\centering
\includegraphics[scale=1.25]{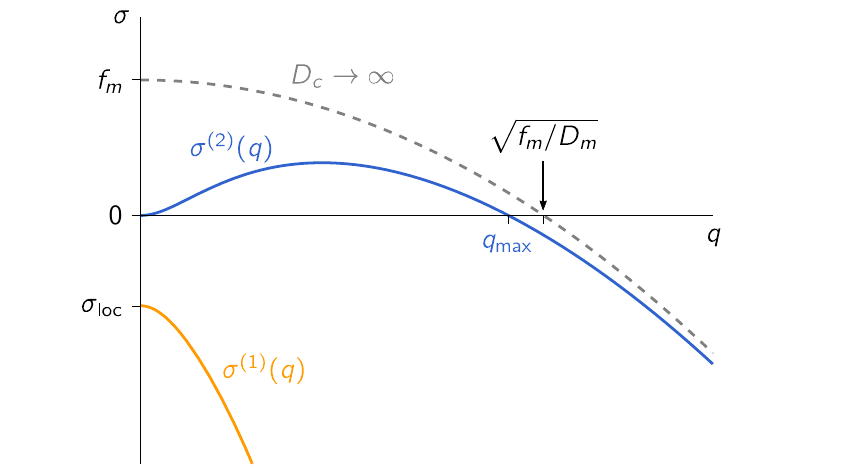}
\caption{
\textbf{Turing instability in a MCRD system.}
Generic dispersion relation of the two-component MCRD system in a laterally unstable regime, showing the two branches $\sigma^{(1,2)}_q$.
In the limit $q \rightarrow 0$, the spectra $\sigma^{(1,2)}_q$ reduce to the eigenvalues $\sigma_\text{loc} = f_m - f_c$ and $0$ of the local stability problem, as indicated in the graph.
For fast cytosolic diffusion $D_c \rightarrow \infty$ the second branch approaches $\sigma^{(2)}_q \rightarrow f_m - D_m q^2$ for $q > 0$ (dashed gray line), and the upper bound $q_\text{max}$ of the band of unstable modes is given by $\sqrt{f_m/D_m}$. Parameters: $f_m = 0.7$, $f_c = 1$, $D_m = 1, D_c = 10$. Adapted from \protect\cite{Brauns.etal2020c}.
	}
	\label{app-fig:disp-rel}
\end{figure}
Analogous to the analysis in Sec.~\ref{sec:classical_turing}, the eigenvalues of the Jacobian $\mathcal{J}_q$ can be expressed in terms of its trace $\tau_q$ and determinant $\delta_q$:
\begin{equation*}
	\sigma^{(1,2)}_q
	= \frac12 
	  \left( 
	  \tau_q \pm \sqrt{\tau_q^2 - 4 \delta_q} \,
	  \right) \, ,
\end{equation*}
where
%\begin{subequations} 
\begin{align*}
	\tau_q 
	&= f_m - f_c - (D_m + D_c) q^2 
	 = \sigma_\text{loc} - (D_m + D_c) q^2 \, , \\
	\delta_q 
	&= q^2 D_m D_c 
	   \left(
	   q^2 + \frac{f_c}{D_c} - \frac{f_m}{D_m}
	   \right)
	 = q^2 D_m D_c 
	   \left(q^2 - q_\text{max}^2\right) \, ,
\end{align*}
%\end{subequations}
and we have defined $q_\text{max}^2 
	= (f_m/D_m) - (f_c)(D_c)$.

We are looking for a Turing instability, i.e.\ for a parameter regime where one of the eigenvalues $\sigma^{(1,2)}_q$ becomes positive at a finite value of the wave number $q$, given that the homogeneous steady state is locally stable ($\sigma_\text{loc} < 0$).
Since $\tau_q = \sigma^{(1)}_q + \sigma^{(2)}_q < 0$ for all $q$, this is possible only if $\delta_q = \sigma^{(1)}_q \cdot \sigma^{(2)}_q < 0$. 
Hence, the discriminant $\Delta_q = \tau_q^2 - 4 \delta_q > 0$ such that all eigenvalues must be real.
This implies that an oscillatory instability for $q > 0$ is excluded for a locally stable homogeneous steady state. 
Moreover, the instability condition $\delta_q < 0$ immediately yields the band of unstable modes $[0,q_\text{max}]$.
Figure~\ref{app-fig:disp-rel} shows the two branches of eigenvalues $\sigma^{(1,2)}_q$ for such a laterally unstable case. 
In the limit $q \rightarrow 0$, the first branch connects to the eigenvalue of a well-mixed system $\sigma^{(1)}_0 = \sigma_\text{loc} < 0$ (cf.\ Sec.~\ref{sec:1d_well-mixed}). The corresponding eigenvector lies in the reactive phase space for $q = 0$, and hence respects mass conservation. 
The second branch $\sigma^{(2)}_q$ remains positive as it approaches zero in the limit $q \, {\rightarrow} \, 0$.\footnote{This case is called `type~II' instability in the Cross--Hohenberg classification scheme \cite{Cross:1993a}. While type~II instability is generic for two-component MCRD systems, this is not true for systems with more components and/or multiple conserved species, where the band of unstable modes can be bound away from zero (`type~I' in Cross--Hohenberg scheme); see e.g.\ \cite{Halatek:2018a}.}

What is the meaning of the neutral eigenvalue $\sigma^{(2)}_0 = 0$ at $q = 0$?
The associated eigenvector $\mathbf{e}^{(2)}_0$ points along the reactive nullcline, and thus represents homogeneous perturbations that change the average total density in the system. 
The perturbed system cannot relax to the former steady state, as this would break the conservation law. Instead, the steady state shifts along the reactive nullcline to a new steady state with mass $\bar{n} + \delta{n}$.
This behavior corresponds to the neutral eigenvalue. 
Importantly however, this eigenvalue is not relevant for the dynamics of a \emph{closed} system since perturbations in this case must conserve the average total mass. The homogeneous stability of closed systems is solely determined by the eigenvalue $\sigma^{(1)}_0$ whose associated eigenvector points along the reactive phase space where total density is conserved. 

Spatially inhomogeneous perturbations ($q \neq 0$), on the other hand, redistribute mass laterally, such that the masses change locally. As we shall see in the next section, introducing the notion of \emph{local reactive equilibria} based on these local masses enables us to understand the physical mechanism underlying the Turing instability in mass-conserving systems.

\begin{exercise}
Perform a linear stability analysis for the following MCRD system 
\begin{align*}
	\partial_t m(x,t) 
	&=  D_m \partial_x^2 m 
	+ \lambda \, m^a c - \mu \, m\, , \\
	\partial_t c(x,t) 
	&= \kern0.3em D_c \partial_x^2 c \kern0.5em 
	- \lambda \, m^a c + \mu \, m \, , 
\end{align*}
for exponents $a=1$ and $a=2$; the parameters $\mu$ and $\lambda$ are positive. 
Discuss the regimes of local and lateral stability.	You may also want to write a computer program that simulates the dynamics of the model beyond the regime of linear instability. What does one find? What happens if one weakly breaks mass conservation in the linear term of the reaction kinetics or by adding an external source?	
\end{exercise}

\subsection{Mass-redistribution instability}
\label{sec:mass_redistribution_instability}

From the above analysis we have learned how to formally calculate the conditions for patterns to form spontaneously using linear stability analysis. 
But what is the physics underlying the lateral instability driving pattern formation?

\begin{figure}
\centering
\includegraphics{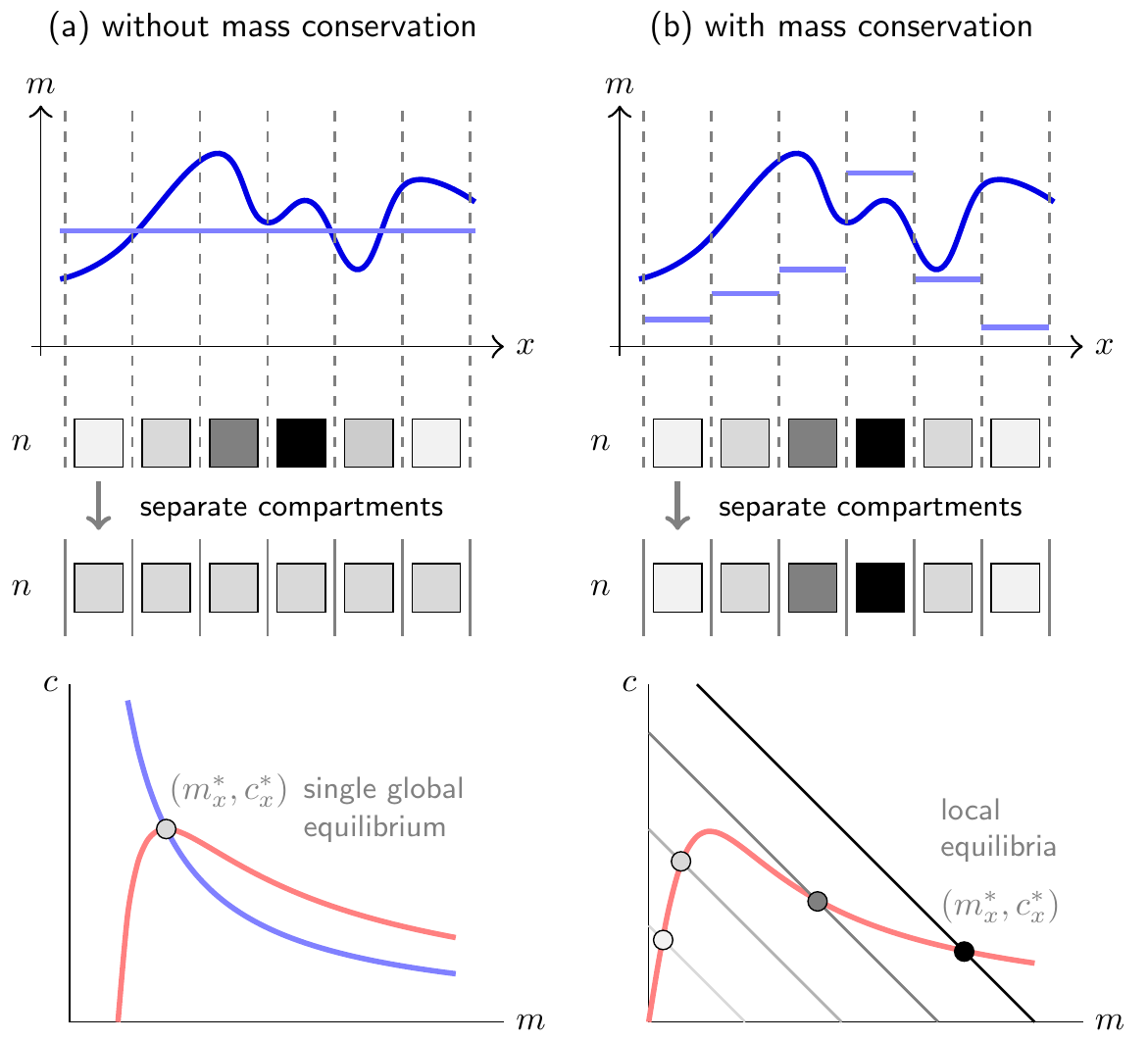}
\caption{
\textbf{Illustration of the concept of local equilibria.}  Spatially extended reaction--diffusion system dissected into small compartments (boxes) with gray scale indicating the local protein mass $n(x,t) = m(x,t) + c(x,t)$. Now suppose the compartments are isolated from one another by `shutting off' diffusion (indicated by solid lines between compartments). 
(a) In a (monostable) system without mass conservation, the concentrations in each compartment return to the same reactive equilibrium, corresponding to the homogeneous steady state of the system. 
(b) In contrast, in a mass-conserving system the equilibrium concentrations depend on the masses (total densities) in the compartments which will in general not be distributed homogeneously. 
This notion of \emph{local equilibria} is a key concept in understanding the dynamics of mass-conserving reaction--diffusion systems (local equilibria theory).
}
\label{fig:local_equilibria_sketch}
\end{figure}

\paragraph{Local reactive equilibria} 

From the mathematical linear stability analysis in the previous section, we learned that the branch of unstable modes emerging from $q = 0$ is associated with spatial redistribution of mass, as revealed by the direction of the corresponding eigenvector.
To gain some intuition into the consequences of lateral mass redistribution,
imagine the system to be dissected into a set of small, well-mixed compartments that are coupled by diffusion; see Fig.~\ref{fig:local_equilibria_sketch} for an illustration. 
To highlight the role of mass conservation, we contrast between systems that conserve mass, described by Eq.~\eqref{eq:2compRD}, and those that do not conserve mass, e.g.\ the Schnakenberg model or the Brusselator. 
At a given instant in time, individual protein densities as well as the total protein density will be spatially heterogeneous. 
In Fig.~\ref{fig:local_equilibria_sketch} this is illustrated by a snapshot of the density profiles for $m(x,t_0)$ and the grey scales of the compartments for the total density $n(x,t_0) = m(x,t_0) + c(x,t_0)$.\footnote{Note that we here use $m$ and $c$ as dynamic variables for systems with as well as without mass conservation.}

What happens if mass transport through diffusive coupling between these compartments is switched off?
For a system without mass conservation, each compartment relaxes back to one and the same equilibrium $(m^*,c^*)$ given by the fixed point of the reaction kinetics (assuming, for simplicity, locally monostable dynamics) with a spatially uniform mass given by $n^* = m^* + c^*$.
In contrast, for mass-conserving reaction--diffusion systems, each compartment will remain in the \emph{local phase space} determined by the local mass $n(x,t_0)$ at the time diffusive coupling was turned off. 
As the masses in the different reaction compartments are different, the dynamics will approach the reactive equilibria $(m^*_x, c^*_x)$ corresponding to the local masses $n(x,t_0)$ in each of these compartments; see Fig.~\ref{fig:local_equilibria_sketch}.
We term these equilibria \emph{local (reactive) equilibria}. 
They are obtained from the conditions of local stationarity $f (m^*_x, c^*_x) = 0$, and local mass conservation $m^*_x + c^*_x = n(x,t_0)$.
As in the analysis of well-mixed systems (Secs.~\ref{sec:1d_well-mixed} and~\ref{sec:2d_well-mixed}), local reactive equilibria and their linear stability act as proxies for the local reactive flow within each compartment.
Diffusive coupling redistributes mass laterally between the compartments, thus shifting their position in phase space and potentially changing their stability.
Understanding reaction--diffusion dynamics in terms of this interplay between lateral mass-redistribution and moving local equilibria is the defining idea behind local equilibria theory.
 
\begin{figure}[t]
\centering 
\includegraphics{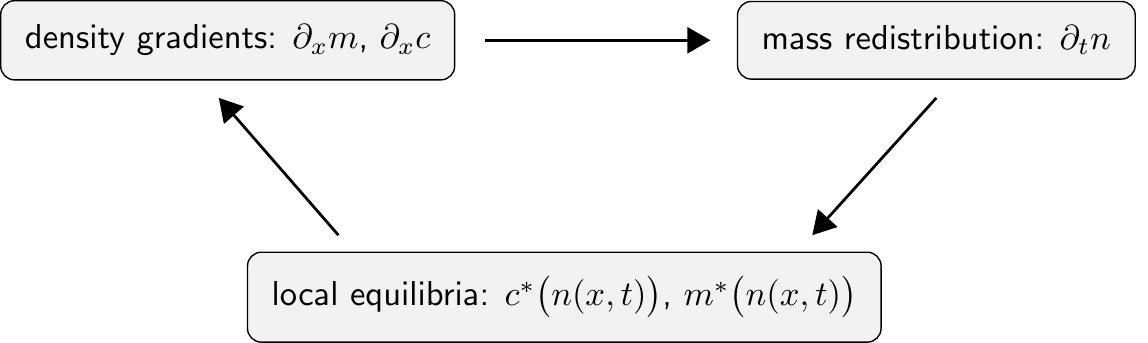}
\caption{\textbf{Schematic of the interplay between mass redistribution and moving local equilibria.} Gradients in the protein densities on the membrane and in the cytosol drive the redistribution of mass which leads to a change in the local reactive equilibria. Shifted local equilibria change the local reactive flows that drive the changes of the density gradients.
}
\label{fig:mass-redistribution_scheme}
\end{figure}

\paragraph{Mass-redistribution instability}
Equipped with the concept of local reactive equilibria, let us now return to the question of the physics underlying lateral instability in MCRD systems.
Consider the dynamics of the total (mass) density $n(x,t) = c(x,t) + m(x,t)$, simply obtained by adding the equations of the membrane and cytosolic density in Eq.~\eqref{eq:2compRD}.
As the reaction kinetics conserve the total density, the time evolution of the total density is driven only by diffusion due to spatial gradients in the concentrations $c(x,t)$ and $m(x,t)$:
\begin{equation} 
\label{eq:n-dynamics}
	\partial_t n(x,t) 
	=  D_c \, \partial_x^2 c(x,t) + D_m \, \partial_x^2 m(x,t) 
	\, . 
\end{equation}
The spatial redistribution of protein mass also leads to a change in the concentrations $c^*(n(x,t))$ and $m^*(n(x,t))$ of the local reactive equilibria.
These shifting local equilibria represent the change of the local reactive flow, which --- assuming stable equilibria --- is directed towards them.
The local reactive dynamics alter the spatial gradients in $c(x,t)$ and $m(x,t)$, which --- in turn --- drive the dynamics of the total protein density, closing the loop, as illustrated in Fig.~\ref{fig:mass-redistribution_scheme}.
This intricate coupling between redistribution of total mass, reactive flows, and diffusive flows drives pattern formation. 

The equation for the total mass density, Eq.~\eqref{eq:n-dynamics}, is not closed. 
In order to determine the time evolution of the total density, one needs to know the dynamics of both the membrane and the cytosolic densities.  
Let's assume that the relaxation of the local concentrations towards the local reactive equilibria is fast compared to the lateral redistribution of protein mass. 
In this limit, one can make the \emph{local quasi-steady state approximation} (LQSSA)
\begin{equation} \label{eq:LQSSA}
	c(x,t) \rightarrow c^*\big(n(x,t)\big)
	\, , \qquad
	m(x,t) \rightarrow m^*\big(n(x,t)\big)
	\, .
\end{equation}
which yields a closed equation for the mass density,
\begin{equation}	\label{eq:n-dynamics-slaved}
	\partial_t n(x,t) \approx  D_c \, \partial_x^2 c^*(n) + D_m \, \partial_x^2 m^*(n)
	\, . 
\end{equation}
While this local-quasi-steady state approximation is in general not strictly valid, it actually becomes exact close to the onset of lateral instability since we have a long wavelength instability where the mass redistribution dynamics becomes slow; cf.\ Section~\ref{sec:lateral-instability}.
Moreover and more importantly, it captures the essence of the dynamics even if there is no strict separation of timescales~\cite{Brauns.etal2020c}.

Let's start our analysis with the limiting case that membrane diffusion is much slower than cytosolic diffusion $D_m \ll D_c$ (which is typical for protein patterns).
Then, linearisation of Eq.~\eqref{eq:n-dynamics} for $n(x,t) = \nbar + \delta n(x,t)$ and applying the chain rule yields
\begin{equation*} 
	\partial_t \delta n(x,t) 
	\approx 
	\left( \partial_n c^*|_{\nbar} \right) 
	D_c \, \partial_x^2 \delta n
	\, .
\end{equation*}
This is a diffusion equation with the effective diffusion coefficient 
$D_c \, \partial_n c^*|_{\nbar}$
that depends on the slope of the nullcline $c^* (n)$ at $\nbar$, i.e.\ on how the equilibrium concentration $c^*$ depends on the mass $n$.
If the equilibrium cytosolic concentration increases with increasing mass, $\partial_n c^*|_{\nbar} > 0$, the effective diffusion constant is positive and hence the dynamics of the total mass is stable.
In contrast, if $\partial_n c^*|_{\nbar} < 0$, we have a negative effective diffusion constant which implies that one has a mass-redistribution instability.
This shows that the interplay between transport of mass and how the local masses changes the local equilibria is the core mechanism driving the dynamics and thereby pattern formation. 
This mass-redistribution instability is the physical mechanism underlying the Turing instability found in Sec.~\ref{sec:lateral-instability} by a formal linear stability analysis.

\begin{figure}[t]
\centering
\includegraphics[width=\textwidth]{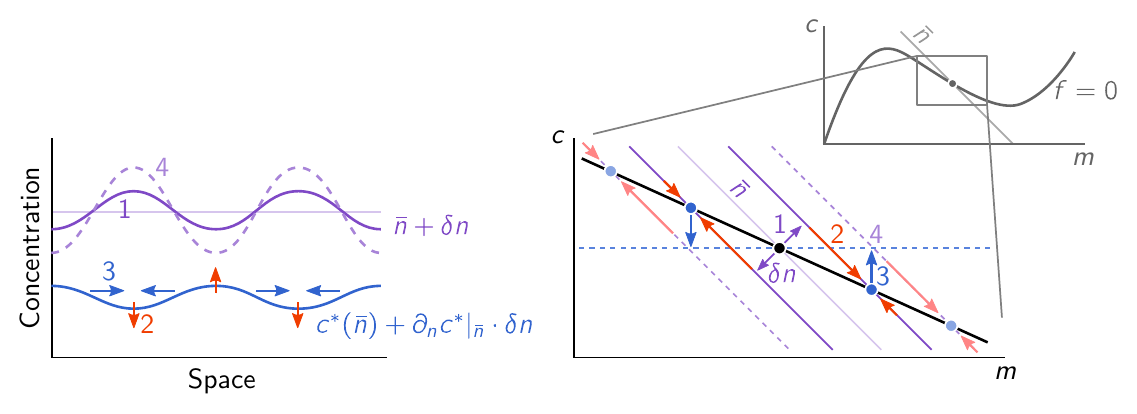}
\caption{
\textbf{Illustration of mass-redistribution instability.}
Consider a small amplitude modulation of the total density (purple line in the real space plot) on a large length scale~(1). 
As diffusion is slow on large scales, the system will locally relax to its reactive equilibrium~(2). The resulting cytosolic density profile is shown by the blue line in the real space plot. 
If the cytosolic equilibrium density decreases due to an increase of total density, the ensuing diffusive fluxes in the cytosol (3) will amplify the modulation of the total density profile (4), thus driving an instability. The membrane gradient is opposite to the cytosolic one, such that membrane diffusion counteracts the instability. However, in the limit $D_m \ll D_c$ this is negligible.
In the $(m,c)$-phase portrait, the criterion for this instability is that the slope of the reactive nullcline must be negative.
}
\label{fig:mass-redistribution-instability}	
\end{figure}

Figure~\ref{fig:mass-redistribution-instability} illustrates this mechanism in both real space (left panel) and in phase space (right panel) in a way that makes it intuitive.  
Consider a small-amplitude, long-wavelength density modulation $\nbar + \delta n(x)$, shown as the solid purple line in the left panel of Fig.~\ref{fig:mass-redistribution-instability}. 
In phase space, we show the reactive phase spaces  corresponding to the maximum and minimum values of this density modulation; see the solid purple lines in the right panel of Fig.~\ref{fig:mass-redistribution-instability}.
As diffusive relaxation for long wavelengths is slow compared to the reaction kinetics (adiabatic limit), both the membrane and the cytosolic densities will first relax towards the local equilibrium values determined by the respective local total protein density, as indicated by the red arrows in Fig.~\ref{fig:mass-redistribution-instability}, giving rise to a modulation of the cytosolic protein density $c^*(\nbar + \delta n) \approx c^*(\nbar) + \partial_n c^* |_{\nbar} \, \delta n$.
For a negative nullcline slope $\partial_n c^* |_{\nbar} < 0$ the modulation of the cytosolic density will be in anti-phase with respect to the modulation of the total density.
Hence, diffusive relaxation, according to Fick's law, will lead to a decrease of cytosolic density in the peaks of the total density and an increase of cytosolic densities where the total density is low.
In addition, since $D_m \ll D_c$, the membrane density remains nearly unchanged during diffusive relaxation of the cytosolic density, such that the cytosolic diffusive fluxes translate directly into changes of the total density; see blue arrows in Fig.~\ref{fig:mass-redistribution-instability}.
As a consequence, we have a further increase in regions of high total density and a further decrease in regions of low total density. 
In other words, the density modulation is amplified, thus driving the mass-redistribution instability.
Conversely, for positive nullcline slope, $\partial_n c^* |_{\nbar} < 0$, the cytosolic density modulation will be in phase with respect to the total density modulation, leading to diffusive fluxes that counteract the initial total density modulation.
On an even more heuristic level, mass-redistribution instability can be understood as the self-organised formation of spatially separated attachment and detachment zones~\cite{Halatek:2018b}.

\paragraph{Nullcline slope criterion for mass-redistribution instability}
In the general case, including the effect of membrane diffusion, Eq.~\eqref{eq:n-dynamics} can be rewritten as 	
\begin{equation*} 
	\partial_t n(x,t) \approx 
	\partial_x \, 
	\big[
	\big(
	D_c \, \partial_n c^* + 
	D_m \, \partial_n m^*
	\big) \, 
	\partial_x n 
	\big]
\end{equation*}
which can be interpreted as a diffusion equation for the total density $n(x,t)$ with an effective diffusion constant that depends on $n(x,t)$.
For specificity, consider the case of an attachment--detachment kinetics where $f_c > 0$; for a consideration of the general case please refer to \cite{Brauns.etal2020c}.
Recall that then the condition for linear stability ($\sigma_\text{loc}(n) < 0$) can be written as $
	\snc(n) = -f_m/f_c > -1$. 
Assuming locally stable equilibria, the effective diffusion constant will become negative if
\begin{equation} 
\label{eq:slope-criterion}
	\frac{\partial_n c^*}{\partial_n m^*} 
	= \snc(n) 
	= -\frac{f_m}{f_c} 
	< - \frac{D_m}{D_c}, 
\end{equation}
where $\snc(n) = \partial_m c^*(m)|_n$ is the slope of the reactive nullcline $c^*(m)$ (cf.\ Eq.~\eqref{eq:snc-def}).
Hence, starting from a homogenous steady state ($m^*(\nbar), c^*(\nbar)$), \emph{lateral instability} due to effective anti-diffusion takes place if (and only if) $\snc(\nbar) < - D_m/D_c$.
This inequality translates into the geometric criterion that the slope of the nullcline must be steeper than $-D_m/D_c$.
Below, we will see that this negative ratio of the diffusion constants is the slope of the so called diffusive flux-balance subspace that geometrically encodes a balance of diffusive fluxes on the membrane and in the cytosol.

How do these results compare with the instability criterion obtained from linear stability analysis? 
There we found that the dispersion relation exhibits a band of unstable modes for $0 < q < q_\text{max}$, with 
\begin{align} 
	\label{eq:qmax}
	q_\text{max}^2 
	= \frac{f_m}{D_m}-\frac{f_c}{D_c} 
\end{align}
if and only if $f_m/D_m > f_c/D_c$, which is exactly the above slope criterion, Eq.~\eqref{eq:slope-criterion}.
Importantly, the slope criterion for the mass-redistribution instability, Eq.~\eqref{eq:slope-criterion}, shows that the diffusion constants $D_m$ and $D_c$  are not required to be vastly different when the nullcline slope is sufficiently steep. In fact, when $\snc$ is close to $-1$, a ratio $D_m/D_c$ slightly  different from unity is sufficient.
This means that the mass-redistribution instability is not hard to achieve. Nullclines with a section of negative slope, for instance, N-shaped nullclines, generically arise as a consequence of nonlinear reaction kinetics and are encountered in a broad range of classical nonlinear systems. 

\begin{exercise}
Graphically analyse the following MCRD model
\begin{align*}
	f(m,c) 
	= 
	\big(
	k + m^2
	\big) \, c - m 
\end{align*}
using the above geometric conditions for the existence of a mass\--re\-dis\-tri\-b\-u\-tion instability, i.e.\ draw the nullcline in $(m,c)$ phase space and use the slope criterion Eq.~\eqref{eq:slope-criterion} for a Turing instability and the criterion for local stability Eq.~\eqref{eq:local-stability-slope-criterion}.   
In addition, use the mathematical conditions for lateral stability to calculate the stability diagram as a function of the total protein mass $n$ and the ratio of the diffusion constants $D=D_m/D_c$ for two typical values of $k$ in the locally monostable regime.
\end{exercise}

\subsection{Stationary states, flux balance, and landmark points}
\label{sec:MCRD_stationary-states}

The linear stability analysis performed in the previous two sections~\ref{sec:lateral-instability} and~\ref{sec:mass_redistribution_instability} informs about parameter regimes where the spatially homogeneous steady state is unstable to small spatial perturbations. 
This analysis, however, does not inform us about the dynamics beyond the initial transient close to a homogeneous steady state.
Nonlinear interactions become important with growing amplitudes of the linearly unstable modes.
What are the conditions that determine spatially inhomogeneous stationary patterns $\widetilde{m}(x)$, $\widetilde{c}(x)$?
Our starting point is the set of steady-state equations,
\begin{subequations} 
	\label{eq:stat-pattern}
\begin{align}
	D_m \partial_x^2 \widetilde{m} + f(\widetilde{m}, \widetilde{c}) &= 0
	\, , 	
	\label{eq:m-stat} \\
	D_c \partial_x^2 \widetilde{c} \kern0.5em - f(\widetilde{m}, \widetilde{c}) &= 0
	\, , 	
	\label{eq:c-stat}
\end{align}	
\end{subequations}
which have to be solved given the no-flux boundary conditions and the constraint of mass conservation.

\paragraph{Flux-balance subspace}
Upon adding the stationarity conditions, Eq.~\eqref{eq:stat-pattern}, one obtains $D_m \partial_x^2 \widetilde{m}(x) = -D_c \partial_x^2 \widetilde{c}(x)$.
Using the no-flux boundary conditions this can be integrated over the domain $[0,x]$ to give a local \emph{flux-balance condition}
\begin{equation*} 
	D_m \partial_x \widetilde{m}(x) 
	= 
	-D_c \partial_x \widetilde{c}(x)
	\, ,
\end{equation*}
which is stating that locally the diffusive fluxes on the membrane and in the cytosol are equal and opposite.
Integrating this relation once more over the interval $[0,x]$ results in a linear relation between $\widetilde{m}(x)$ and $\widetilde{c}(x)$
\begin{equation} 
\label{eq:flux-balance-subspace}
	\frac{D_m}{D_c} \, 
	\widetilde{m}(x) 
	+ 
	\widetilde{c}(x) 
	= \eta_0
	\, . 
\end{equation}
The constant of integration, $\eta_0$, is implicitly determined by integrating one of the stationarity conditions, Eq.~\eqref{eq:m-stat}, over the whole domain $[0,L]$
\begin{equation} \label{eq:balance_reactive_turnover}
	\int_{0}^{L} \dd x \,	f
	\big(
	\widetilde{m}(x), 
	\eta_0 - D_m\widetilde{m}(x)/D_c
	\big)	= 0
	\, .	
\end{equation}
This implicit equation for $\eta_0$ states that, in the stationary state, all reactive processes (``total reactive turnover'') must balance. 

\begin{figure}[tb]
\centering
\includegraphics{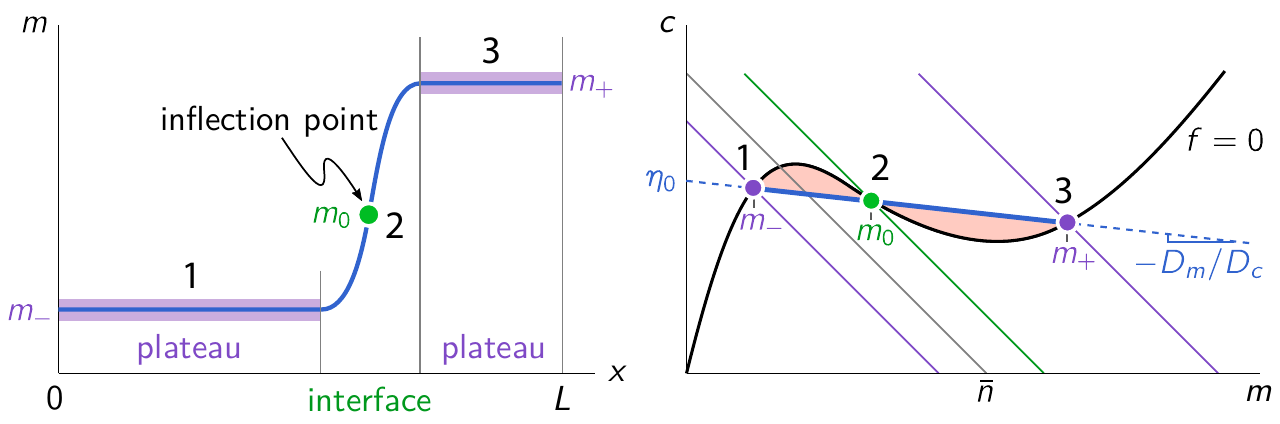}
\caption{
Illustration of the flux-balance construction for stationary patterns.
(\textit{Left}) Membrane density profile of a stationary ``mesa'' pattern composed of two plateaus (highlighted in purple) connected by an interface.
(\textit{Right}) In phase space, the stationary pattern (blue, solid line) is embedded a flux-balance subspace (blue, dashed line). 
The three intersection points between the reactive nullcline, $f = 0$, (black line) and the flux-balance subspace, correspond to the two plateaus (1,~3) and the inflection point (2) of the pattern profile. 
The offset of the flux-balance subspace, $\eta_0$, is determined by a balance of reactive turnovers on either side of the inflection point. This balance approximately corresponds to a balance of the areas enclosed by flux-balance subspace and the reactive nullcline (shaded in red).
}
\label{fig:2d_MCRD_phase_space}
\end{figure}

We conclude that in phase space, any stationary pattern must lie on a straight line with slope $-D_m/D_c$, given by the ratio of the diffusion constants on the membrane and in the cytosol; for an illustration see Fig.~\ref{fig:2d_MCRD_phase_space}. As it represents the balance of diffusive fluxes in steady state, we call this the \emph{flux-balance subspace} (FBS). Its intercept $\eta_0$ is determined by the balance of reactive turnovers.

An immediate consequence of the flux-balance condition is that, whenever the diffusion constants $D_m$ and $D_c$ are unequal, a stationary protein pattern entails spatially non-uniform distribution of the total density $\widetilde{n}(x) = \widetilde{m}(x) + \widetilde{c}(x)$, 
since the flux-balance subspace has a slope $-D_m/D_c$ different from $-1$, i.e.\ does not coincide with a reactive phase space.
Hence, the process leading from a spatially uniform initial state towards the final stationary state must redistribute mass.\footnote{The only exception to this is the case $D_m = D_c$, where $\eta = n$ such that the dynamics of the total density decouple and are purely diffusive, approaching $\widetilde{n}(x) = \nbar$.}
This is process is the mass-redistribution instability introduced in Sec.~\ref{sec:lateral-instability}.

\paragraph{Local equilibria and stationary patterns}
The intersection points of the FBS and the reactive nullcline in phase space correspond to the plateaus  and the inflection point of the density profile in real space (see labels 1--3 in Fig.~\ref{fig:2d_MCRD_phase_space}).
To understand this, consider the system dissected into small notional compartments.
The net diffusive flux $\partial_x (D_m \partial_x \widetilde{m})$ in and out of a compartment vanishes for compartments in the plateaus, where $\partial_x \widetilde{m} \approx 0$, and at the inflection point, where $\partial_x \widetilde{m}^2 = 0$. 
Therefore, in these compartments, stationarity requires that the reactive flow must vanish too, meaning that the concentrations lie on the reactive nullcline in phase space.
In other words, in the plateaus and at the inflection point, the real space pattern coincides with the local (reactive) equilibria $(m^*_x, c^*_x)$ corresponding to the local mass $\widetilde{n}(x)$. 
In regions where the concentration profile is curved (i.e.\ $\partial_x \widetilde{m} \neq 0$), the \emph{net} diffusive flux in and out of a compartment does not vanish and must be balanced by the reactive flow in that compartment (cf.\ Eq.~\eqref{eq:stat-pattern})
\begin{equation*}
	\partial_x \big( D_m \partial_x \widetilde{m} \big) 
	= - f (\widetilde{m}, \widetilde{c})
	\, .
\end{equation*} 
Hence, the reactive equilibria $(m^*_x, c^*_x)$ deviate from the stationary pattern $[\widetilde{m}(x), \widetilde{c}(x)]$ in these regions.
In phase space, this deviation is visible as the distance between the FBS and the reactive nullcline.

\paragraph{Peak patterns}
A pattern type qualitatively different from mesas forms if the third intersection point between the FBS and the nullcline is at much higher densities, or doesn't exist at all. Then maximum density of the pattern does not saturate in a plateau and forms a `peak' instead~\cite{Brauns.etal2020c}.
A typical case where peaks can form is when attachment to the membrane does not saturate for high membrane concentrations such that the reactive nullcline approaches $c = 0$ for large $m$ instead of curving up again.
In addition, the formation of peak patterns requires that the ratio of the diffusion constants $D_m/D_c$ is close to zero, such that the flux-balance subspace has a shallow slope.
Increasing the average mass in a system exhibiting a peak pattern will lead to an increase of the peak amplitude. 
Eventually, the maximum concentration will reach the third FBS-intersection point with the nullcline such that a plateau starts to form, giving rise to a mesa pattern.
Further increasing the average mass will shift the interface position of the mesa pattern.

\paragraph{Reactive turnover balance}
To close the graphical construction in phase space, we need a geometric interpretation of the FBS offset $\eta_0$, which implicitly ist determined by a balance of total reactive turnover, Eq.~\eqref{eq:balance_reactive_turnover}.
Can we understand this total-turnover balance in phase space? 
Above, we already argued that the deviation of the stationary pattern (embedded in the FBS) from the reactive equilibria (which lie on the nullcline in phase space), serves as a proxy for the local reactive flow. 
Intuitively, the total reactive turnover therefore (approximately) corresponds to the areas difference of the two areas enclosed by the FBS and the nullcline (shaded in red in Fig.~\ref{fig:2d_MCRD_phase_space}).
The stationarity condition of total reactive turnover balance therefore corresponds to an (approximative) balance of areas in phase space, akin to a Maxwell construction.

To obtain this approximate condition mathematically, we multiply Eq.~\eqref{eq:m-stat} with $\partial_x \widetilde{m}(x)$ before integrating, to obtain
\begin{equation}
	\label{eq:turnover-balance}
	\int_{\tilde{m}(0)}^{\tilde{m}(L)} \dd m \, \tilde{f}(m,\eta_0) = 0
	\, ,
\end{equation}
where $\tilde f (m,\eta) := f \big( m,\eta - (D_m/D_c) m \big)$.
This formulation of total turnover balance does not depend on the full density profile $\widetilde{m}(x)$, but only on the densities at the boundaries, $\widetilde{m}(0)$ and $\widetilde{m}(L)$. 
Geometrically, this total turnover balance can be interpreted as a kind of (approximate) Maxwell construction (balance of areas) in the $(m,c)$-phase plane, under the following approximations:
First, we linearise the reactive flow around the reactive nullcline (cf.\ Sec.~\ref{sec:2d_well-mixed}):
\begin{equation*} 
	\tilde{f}\big(m,\eta_0\big) \approx \sigma_\text{loc}\big(\widetilde{n}(m)\big) \cdot \big[m-m^*\big(\widetilde{n}(m)\big)\big],
\end{equation*}
where $\widetilde{n}(m) := \eta_0 + (1 - D_m/D_c)m$ because the pattern is embedded in the flux-balance subspace, cf.\ Eq.~\eqref{eq:flux-balance-subspace}.
The expression in the square brackets is simply the distance of the reactive nullcline from the flux-balance subspace measured along the respective local phase space. 
Further, suppose for the moment that the local eigenvalue $\sigma_\text{loc}(n)$ is approximately constant in the range of total densities attained by the pattern. 
Turnover balance, Eq.~\eqref{eq:turnover-balance}, is then represented by a balance of the areas between nullcline and flux-balance subspace on either side of the inflection point (see areas shaded in red (light gray) in Fig.~\ref{fig:2d_MCRD_phase_space}(b)):
\begin{equation*}
	\int_{\widetilde{m}(0)}^{\widetilde{m}(L)} \dd m 
	\big[m - m^*\big(\widetilde{n}(m)\big)\big] 
	= 0 \, . 
\end{equation*}
For a spatial domain of size $L$ much larger than the interface width, one can approximate the plateau concentrations by FBS-NC intersections: 
$\widetilde{m}(0) \approx m_-(\eta_0)$ and $\widetilde{m}(L) \approx m_+(\eta_0)$. 
In this case, Eq.~\eqref{eq:turnover-balance} is closed and can be solved for $\eta_0$, either numerically or geometrically using the approximate area-balance condition, akin to a Maxwell construction.

What is still missing from the above characterisation of stationary patterns is the width of the interface connecting the two plateaus. An argument based on the linearisation of the dynamics in the vicinity of the pattern's inflection point relates the interface to the right-hand edge of the dispersion relation. 
For details on this, we refer the interested reader to \cite{Brauns.etal2020c}.

\paragraph{Mass-redistribution potential} 
Motivated by the insight that the stationary state lies in a flux-balance subspace in phase space (cf.\ Eq.~\eqref{eq:flux-balance-subspace}), we define the field 
\begin{equation} \label{eq:eta-def}
	\eta (x,t) := \frac{D_m}{D_c} \, m (x,t) + c(x,t) \, ,
\end{equation}
describe the spatiotemporal dynamics in terms of $n(x,t)$ and $\eta(x,t)$ (cf.\ Refs.~\cite{Otsuji:2007a,Mori:2011a,Chiou:2018a}) instead of the membrane density $m(x,t)$ and the cytosolic density $c(x,t)$. 
The corresponding equations read
\begin{equation} 
	\label{eq:n-dyn-eta}
	\partial_t n(x,t) 
	= 
	D_c \partial_x^2 \eta(x,t)
	\, ,
\end{equation}
and
\begin{equation} \label{eq:eta-dyn-eta}
	\begin{split}
	\partial_t \eta(x,t) 
	= 
	(D_m + D_c) \, \partial_x^2 \eta(x,t) -
	D_m \partial_x^2 n(x,t) \\
	- \left( 1 - \frac{D_m}{D_c} \right) 
	f\big(m(n,\eta),c(n,\eta)\big).
	\end{split}
\end{equation}

As gradients in $\eta(x,t)$ drive the redistribution of mass  $n(x,t)$, we call it a \emph{mass-redistribution potential}. 
It plays a role analogous to the chemical potential in near equilibrium systems (e.g.\ Model~B dynamics \cite{Hohenberg:1977a}). 
However, it does not follow from a free energy density, but instead obeys a dynamic equation, Eq.~\eqref{eq:eta-dyn-eta}. 
The mass-redistribution potential allows us to write the condition for lateral instability Eq.~\eqref{eq:slope-criterion} in the compact form
\begin{equation}
	\partial_n \eta^*|_{\bar{n}} < 0.
\end{equation}
This follows immediately from applying the local quasi-steady state approximation $\eta(x,t) \approx \eta^*(n(x,t))$ to Eq.~\eqref{eq:n-dyn-eta}, and using the chain rule to obtain
\begin{equation}
	\partial_t n(x,t) = D_c \partial_x \big(\partial_n \eta^*(n)  \partial_x n\big).
\end{equation}
For $\partial_n \eta^* < 0$ the effective diffusion coefficient $D_c \partial_n \eta^*$ is negative, resulting in an instability.

An analogous reasoning can be employed to study the stability of periodic stationary patterns against competition for mass between neighboring high density domains (mesas or peaks) \cite{Brauns.etal2020}. Here, the local quasi-steady state approximation is replaced by a quasi-steady state approximation on the level of the individual \emph{elementary patterns} that constitute the periodic pattern (i.e.\ individual mesas or peaks, respectively). 
The stability of periodic stationary patterns is central to the question of wavelength selection far away from the onset of pattern formation.
It turns out that two-component MCRD systems generically exhibit uninterrupted coarsening, meaning that all periodic patterns with a wavelength shorter than the system size are unstable due to competition for mass between neighboring high density domains \cite{Brauns.etal2020}. As a result, the wavelength selected in the final steady state is equal to the system size.
Source terms that break mass conservation or coupling to additional components can counteract the competition for mass and thus interrupt and even reverse the coarsening process resulting in the selection of finite wavelength patterns.

\paragraph{Bifurcations of patterns}

\begin{figure}[tb]
\centering
\includegraphics[scale=1.25]{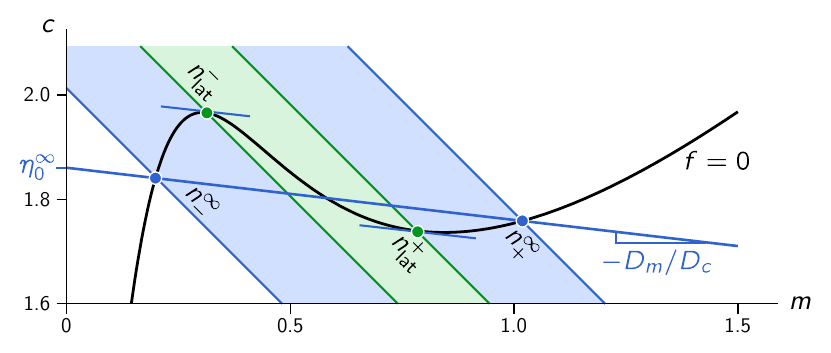}
\caption{
Bifurcations of mesa patterns (cf.\ Fig.~\ref{fig:2d_MCRD_phase_space}) in the large domain size limit ($L \rightarrow \infty$) can be constructed geometrically using the reactive nullcline.
Geometric construction of pattern bifurcation points for an example two-component system: Eq.~\eqref{eq:wave-pinning-nondim} with $k = 0.07$, $D_m=1$, and $D_c= 10$. 
The laterally unstable regime (shaded in green) is delimited by Turing bifurcations where the FBS is tangential to the NC (green dots, $n_\pm^\text{lat}$). 
FBS-NC intersection points (blue dots, $n_\pm^\infty$), delimit the range of pattern existence (shaded in blue), where the FBS-position $\eta_0^\infty$ is determined by global turnover balance, Eq.~\eqref{eq:turnover-balance}. 
Adapted from \protect\cite{Brauns.etal2020c}.	
}
\label{fig:n-bifurcation-mesa-patterns}
\end{figure}

Now that we have understood the relationship between stationary patterns and geometric objects in $(m,c)$-phase space, we turn to study bifurcations where the patterns change structurally or in stability.
We will perform the analysis using the phase space concepts introduced above, focusing on large systems, i.e.\ where the system size is much larger than the interface width and the distance of the interface from the domain boundaries.
In that case the position $\eta_0^\infty$ of the FBS is independent of $\nbar$ and $L$.

\textit{Bifurcations in average protein density.\ --- } 
We start our analysis by asking how patterns change as we change the average total density of proteins $\nbar$ in the system, through gene regulation, for instance. 
From the linear stability analysis as well as from the local quasi-steady state approximation we have learned that there is a Turing instability, i.e.\ a band of linearly unstable modes, $[0,q_\text{max}]$, if the NC-slope $\snc(\nbar)$ is negative and steeper than the FBS-slope, $-D_m/D_c$; cf.\ Eq.~\eqref{eq:slope-criterion}. 
Geometrically this corresponds to the mass range $(n_\text{lat}^-,n_\text{lat}^+)$, indicated by the green-shaded region in Fig.~\ref{fig:n-bifurcation-mesa-patterns}.

The range of average densities where stationary patterns exist is actually broader, as can be inferred from the phase space analysis performed in Sec.~\ref{sec:MCRD_stationary-states}.
There we have shown that stationary patterns are embedded in the FBS and the plateaus of a pattern are geometrically determined by the (outer) intersections of the FBS and the NC; see the blue points in Fig.~\ref{fig:n-bifurcation-mesa-patterns}(a). 
For an N-shaped nullcline, the protein masses $n_\pm^\infty$ corresponding to these intersection points define a range $n_-^\infty < \nbar < n_+^\infty$ that includes the Turing unstable range $(n_\text{lat}^-,n_\text{lat}^+)$. 
This implies that the onset of pattern formation is subcritical with regions of multi-stability in parameter space, where both stable stationary patterns exist and the homogeneous steady is stable; see regions shaded in blue in Fig.~\ref{fig:n-bifurcation-mesa-patterns}.

\begin{figure}[tb]
	\centering
	\includegraphics[scale=1.25]{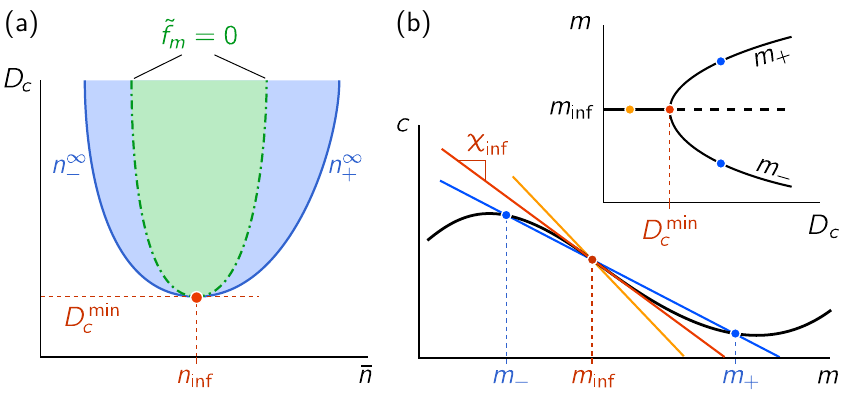}
	\caption{
	$(\nbar,D_c)$-bifurcation diagram of stationary patterns for a system with monostable kinetics (same color code as in Fig.~\ref{fig:n-bifurcation-mesa-patterns}).
	(a) The bifurcation diagram for a large system ($L \rightarrow \infty$) is obtained by tracking the geometrically constructed bifurcation points $n_\text{lat}^\pm$ and $n_\pm^\infty$ as $D_c$, and thus the slope and position of the FBS, are varied (cf.\ Fig.~\ref{fig:n-bifurcation-mesa-patterns}).  The onset of lateral instability (Turing bifurcation shown as green dash-dotted line) is generically subcritical since there exist stationary patterns outside the range of lateral instability $(n_\text{lat}^-,n_\text{lat}^+)$; in the blue regions, the system is multistable (both the stationary patterns and the homogenous steady state are stable). The regime where patterns exist, delimited by $n_\pm^\infty$ emerges from the critical point $(n_\text{inf},D_c^\text{min})$ (red point). 
	(b) This critical point corresponds to the inflection point of the nullcline at $n_\text{inf}$ where the nullcline slope $-f_m/f_c$ reaches its extremal value $\snc(n_\text{inf})$. This determines the minimal cytosolic diffusion $D^\text{min}_c = - D_m/\snc(n_\text{inf})$ (cf.\ Eq.~\eqref{eq:Dc_min}). At $D^\text{min}_c$, the plateaus' scaffolds $m_\pm$ bifurcate in a supercritical pitchfork bifurcation from the nullcline inflection point $m_\text{inf}$ (see inset).
	Adapted from \protect\cite{Brauns.etal2020c}.
	}
	\label{fig:n-Dc-diagram-monostable}
\end{figure}

\textit{Bifurcations in average mass and diffusion.\ --- }
We generalise now the above analysis by varying the protein mass and the diffusion constant in the cytosol $D_c$ while keeping $D_m$ fixed; changing $D_c$ rotates the FBS.
We construct the $(\nbar,D_c)$-bifurcation diagram by inferring $n_\text{lat}^\pm$ and $n_\pm^\infty$, as described above, as functions of $D_c$. 
Figure~\ref{fig:n-Dc-diagram-monostable}a shows the qualitative structure obtained by this graphical construction. 

\begin{exercise}
Determine the bifurcation diagram of a two-component MCRD system with reaction kinetics $f=\big(\nicefrac14+m^2\big)\,c \, {-} \, m$ as a function of the total protein mass $n$ and the ratio of diffusion constants $D_c/D_m$. Indicate the boundaries of the Turing unstable regime as well as the boundaries of the regime where stationary patterns exist. Find closed analytical expressions for both boundaries in the vicinity of the critical point where both boundaries touch, i.e.\ at the inflection point $m_\text{inf}$ of the nullcline $c^*(m) = m/(\nicefrac14+m^2)$. Before starting a formal mathematical analysis familiarize yourself with the system's behavior using a graphical analysis in $(m,c)$-phase space.
\end{exercise}

As $D_c$ is decreased, the flux balance subspace becomes steeper, and thus the bifurcation points $n_\text{lat}^\pm$ and $n_\pm^\infty$ start to converge (see Fig.~\ref{fig:n-Dc-diagram-monostable}; cf.~Fig.~\ref{fig:n-bifurcation-mesa-patterns}). 
They meet in the inflection point of the reactive nullcline, $n_\text{inf}$, where the nullcline slope $\snc(n_\text{inf})$ is extremal ($\partial_n \snc |_{n_\text{inf}} = 0$). 
The extremal nullcline slope at the nullcline inflection point determines the minimal cytosolic diffusion constant,
\begin{equation} 
\label{eq:Dc_min}
	D_c^\text{min} := \frac{D_m}{-\snc(n_\text{inf})} \, ,
\end{equation}
above which there are three FBS--NC intersection points. 
When the `critical' point $(n_\text{inf}, D_c^\text{min})$ is traversed in $D_{\kern-0.075em c}\kern0.075em$-direction, the FBS--NC intersections bifurcate in a (supercritical) pitchfork bifurcation; 
see Fig.~\ref{fig:n-Dc-diagram-monostable}(b).
Since the FBS--NC intersection points $m_\pm$ are the scaffolds for the plateaus (in short: \emph{plateau scaffolds}; 
cf.\ Fig.~\ref{fig:2d_MCRD_phase_space}), this bifurcation at the critical point $(n_\text{inf}, D_c^\text{min})$ is a bifurcation of the scaffold itself. Importantly, the actual pattern is bounded by the plateau scaffolds. 
Thus, if there are no plateau scaffolds (i.e.\ only one FBS-NC intersection point), there cannot be stationary patterns.
For $L \rightarrow \infty$, patterns emerge slaved to the plateau scaffold, such that the pattern bifurcation is supercritical at the nullcline inflection point ($\nbar = n_\text{inf}$). 
Away from the nullcline inflection point ($n \neq n_\text{inf}$), the lateral instability bifurcation is always subcritical for $L \rightarrow \infty$ because the range $(n_-^\infty,n_+^\infty)$ where patterns exist always exceeds the range $(n_\text{lat}^-,n_\text{lat}^+)$ of lateral instability, as we learned above  (cf.\ Fig.~\ref{fig:n-bifurcation-mesa-patterns}).

For clarity and simplicity we have confined the above analysis to the large system size limit. 
As shown in \cite{Brauns.etal2020c} using weakly nonlinear analysis, for finite $L$, the bifurcation is actually supercritical in the vicinity of the nullcline inflection point. 
The transition from super- to sub-criticality depends on a subtle interplay of diffusive and reactive flow.

Strikingly, our geometric reasoning shows that the physics implied by the bifurcation diagram is the same as in phase separation kinetics (binodal and spinodal regimes) \emph{for all N-shaped nullclines}, and far away from the critical point.
This includes also cases where a mathematical equivalence of the underlying dynamics (e.g.\ in terms of a mapping to gradient dynamics) has not yet been found. 
This suggests that there might be a deeper connection between two-component MCRD systems and phase separation kinetics near thermal equilibrium.

%% ================================
%% BULK BOUNDARY COUPLING
%% ================================

\section{The role of bulk-boundary coupling for membrane patterns} \label{sec:bulk-boundary-coupling}

The dynamics of most intracellular pattern forming systems involves a coupling between dynamics on the membrane surface and in the cytosolic volume (bulk). 
We exemplified this in Sec.~\ref{sec:intracellular_protein_patterns} for the Min-protein dynamics, where biochemical interactions (with the exception of nucleotide exchange) are confined to the immediate vicinity of the membrane. 
At first glance, one might therefore suppose that it is straightforward to eliminate the cytosolic bulk from the dynamics and consider it simply as a passive particle reservoir. 
Although there are certain limits where this is indeed possible, in general, the bulk dynamics plays a key role for the spatiotemporal dynamics and can not be disregarded. 
In the following we will illustrate the underlying physics for two simple analytically tractable cases, and discuss general implications for reaction--diffusion systems with bulk--boundary coupling.
Moreover, we will discuss the implications of this coupling for the linear stability analysis and explicitly show how such an analysis can be implemented for a box geometry.
Finally, in the last section we discuss how cytosolic gradients impact the formation of patterns and show how they lead to to \emph{geometry-induced pattern formation}.

\subsection{Elementary examples in column geometry}
\label{sec:column_examples}

\begin{figure}[t!]
\centering 
\includegraphics{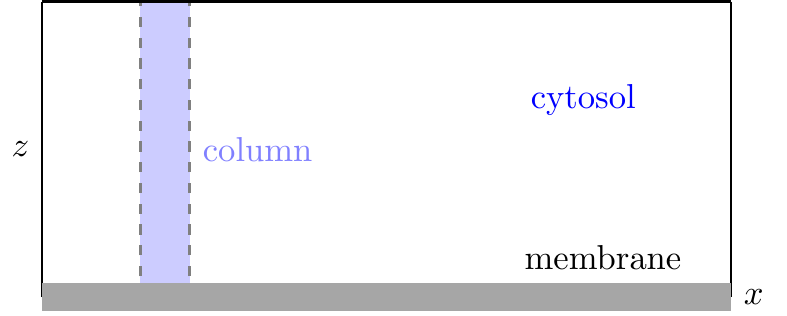}
\caption{
\textbf{Illustration of a box and column geometry.} For a conceptual analysis of mass-conserving reaction-diffusion systems it is instructive to consider a rectangular box geometry with the membrane considered as a one-dimensional line ($x$) and the cytosol as a two-dimensional rectangular box with vertical dimension $z \in [0,h]$. An even simpler geometry is a narrow column where the membrane is considered as a point and the cytosol as a line. 
}
\label{fig:box_geometry}
\end{figure}

This subsection illustrates the relevance of cytosolic density gradients for the dynamics of reaction-diffusion systems that couple reactions on a membrane and in the cytosol (bulk) through attachment and detachment of proteins.
To simplify the calculations and highlight the role of gradients normal to the membrane we consider a simplified one-dimensional geometry: a narrow `column' of height $h$, with a membrane `point' at the bottom ($z=0$) and a reflective boundary at $z=h$.
By laterally coupling such `columns' one obtains a `box geometry', with a one-dimensional membrane and a two-dimensional cytosolic volume as illustrated in Fig.~\ref{fig:box_geometry}. 
This is a simple, prototypical geometry to study reaction--diffusion systems with bulk-boundary coupling.

The exchange of proteins between membrane and cytosol (through various kinds of chemical reactions as discussed in Sec.~\ref{sec:biochemical_principles}) needs to be balanced by diffusive fluxes in the cytosol, otherwise local mass conservation would be violated. 
Hence, on these very general grounds, there must be spatial gradients in the cytosolic protein density. 
In this section we will answer the following questions: 
What determines the magnitude and spatial extent of these cytosolic gradients? 
Under what conditions   are these gradients essential for understanding the reaction-diffusion dynamics and when can they be neglected?
What, in particular, is the role of nucleotide exchange in the cytosol and the detachment and attachment kinetics at the membrane? 

\paragraph{Protein attachment causes cytosolic gradients} 

We start with analyzing the effect of protein exchange  between membrane and cytosol on the density profile $c(z,t)$ in the cytosol. 
To keep the analysis simple we consider a single protein species which diffuses in the cytosol
\begin{equation} 	\label{eq:attachPDE}
	\partial_t c (z,t)
	= D_c \, \partial_z^2 c (z,t) 
	\, , 
\end{equation}
and that can attach to the membrane with rate $k_\text{on}$
\begin{equation} \label{eq:attach_membrane}
	\partial_t m(t) 
	= k_\text{on} \, c(0,t) 
	\, , 
\end{equation}
where $m$ signifies the density (number) of the proteins on the membrane. 
To complete the dynamics, one needs to specify boundary conditions. 
The dynamics on the membrane and in the cytosol are coupled such that the overall protein density is conserved locally. 
This requires that at the membrane the diffusive flux of proteins in the cytosol, $j_\text{diff} = - D_c \, \partial_z c (z,t)|_{z=0}$, and the reactive flux of proteins from the cytosol to the membrane (due to protein attachment), $j_\text{on} = - k_\text{on} \, c(0,t)$, need to match:
\begin{equation} \label{eq:attachBC1}
	- D_c \partial_z c (z,t)|_{z=0}  
	= - k_\text{on} \, c(0,t) 
	\, . 
\end{equation}
This \emph{`reactive' boundary condition} is also called a \emph{`Robin'} boundary condition (see Sec.~\ref{sec:intracellular_protein_patterns}).
It implies that the protein flux from the cytosol to the membrane leads to a depletion zone of proteins in the cytosol close to the membrane, i.e.\ a gradient in the cytosolic density.
The length scale (diffusion length) on which the profile relaxes back to the bulk protein level can be read off (by simple dimensional analysis) from Eq.~\eqref{eq:attachBC1} as\footnote{Please note that the on-rate $k_\text{on}$ has dimension length/time since the cytosolic protein density is a line density and the membrane protein `density' is actually a protein number.} 
\begin{equation*}
	\ell_\text{on} := \sqrt{D_c / k_\text{on}} \, .
\end{equation*}
These considerations generalise to more complex reaction kinetics, for example by including a detachment flux $j_\text{detach} = k_\text{off} \, m(t)$, and clearly show that any reactive coupling between the cytosol and the membrane inevitably leads to density gradients in the cytosol.
The set of dynamic equations is completed by the non-flux (reflective) boundary condition at $z=h$:
\begin{equation} \label{eq:attachBC2}
	 D_c \, \partial_z c|_{z=h} 
	 = 0 \, . 
\end{equation}

Under what conditions can one disregard cytosolic gradients? 
Before starting with a formal analysis let's start with some heuristic argument:
the diffusive dynamics normal to the reactive membrane can be neglected if the time scale of cytosolic mixing, $\tau_\text{diff} \sim h^2/D_c$, is much smaller than the typical time a protein suspended in the bulk needs to attach to the membrane, $\tau_\text{on} \sim h/k_\text{on}$. 
This yields the condition
\begin{equation*}
 	h \ll \ell_\text{on} \, , 
\end{equation*}
Only if this condition is obeyed, cytosolic gradients can be disregarded and the cytosol be considered as a well-mixed particle reservoir. 

It is also instructive to show this with a formal analysis. 
We are looking for a dynamic equation for the spatial average of the cytosolic protein density
\begin{equation*}
	\bar{c} (t) 
	= \frac{1}{h} \int_0^h \! dz \, c(z,t) 
	\, .
\end{equation*}
One finds, combining the cytosolic diffusion equation, Eq.~\eqref{eq:attachPDE}, with both boundary conditions, Eq.~\eqref{eq:attachBC1} and \eqref{eq:attachBC2}: 
\begin{equation*}
	\partial_t \bar{c} (t) = - \frac{k_\text{on}}{h} \, c(0,t) \, .
\end{equation*}
This reduces to an equation for the   concentration at the membrane $c_0(t) := c(0,t)$ only if one can approximate the average density by the density at the membrane, $c_0(t) \approx \bar{c} (t)$. Then one gets a closed equation for the cytosolic density on the membrane (equivalently, the average cytosolic density)
\begin{equation} \label{eq:att_elim} 
	\partial_t c_0 (t) 
	\approx - \frac{k_\text{on}}{h} \, c_0(t) 
	\, .
\end{equation}
This equation is solved by $ c_0 (t)  = c_0 (0) \, e^{- t / \tau_\text{on}}$, an exponentially decaying reservoir of cytosolic particles, where the relaxation time is given by
\begin{equation*}
	\tau_\text{on}  
	:= h / k_\text{on} 
	\, .	
\end{equation*}
This approximation requires the absence of significant gradients in the cytosol and is valid only if the penetration depth $\ell_\text{on}$ is much larger than the height $h$ of the column, $h  \ll  \ell_\text{on}$. 
This can also be seen explicitly because the considered linear reaction-diffusion system is exactly solvable by the separation ansatz $c(z,t)  = T(t) Z(z)$.
With the separation constant $\sigma$, one finds a set of decoupled linear equations
%
%\begin{subequations}
\begin{align*}
	\partial_t T (t) 
	&= \sigma  \, T (t) 
	\, , \\
	D_c \partial_z^2 Z (z) 
	&= \sigma \, Z (z) 
	\, ,
\end{align*}
%\end{subequations}
%
which are solved by the ansatz
%\begin{subequations}
\begin{align*}
	T (t)  
	&= T (0) \, e^{\sigma t} 
	\, , \\
	Z (z)  
	&= Z (h) \, \cos \big( p (h-z) \big),
\end{align*}
%\end{subequations}
This shows that the separation constant $\sigma$ plays the role of a growth rate.
We have chosen the spatial part $Z (z)$ such that it already respects the reflective boundary condition at $z = h$, Eq.~\eqref{eq:attachBC2}. 
The wavenumber $p$ is connected to the growth rate $\sigma$ via the dispersion relation
\begin{equation*}
	\sigma_p = -D_c \, p^2 
	\, . 
\end{equation*}
The possible values of $p$ are determined by the reactive boundary condition Eq.~\eqref{eq:attachBC1}, which upon substitution of the above ansatz yields
\begin{equation} \label{eq:solvability-robin}
	h p \tan (h p) 
	= \frac{h}{\ell_\text{on}} 
	\, . 
\end{equation}
This type of transcendental equation makes the analysis of linear problems with reactive boundary conditions coupling two compartments more involved than boundary value problems with reflective or Dirichlet boundary conditions.

Closed form solutions to the characteristic equation~\eqref{eq:solvability-robin} exist in asymptotic limits only. 
For $h \ll  \ell_\text{on}$, one can  approximate $\tan (h p) \approx h p$ to find the smallest eigenvalue as $p_1  \approx  1/\sqrt{h \ell_\text{on}} $. 
Using the relation $\sigma_p = -D_c \, p^2$ from above, the corresponding relaxation rate is $\tau_1 = 1/|\sigma_1|$ turns out to be identical to $\tau_\text{on}$ defined above. 

After the decay of initial transients, corresponding to faster decaying smaller wavelengths, the dominant wavelength is 
\begin{equation*}
	\lambda_\text{max}  
	= \frac{2 \pi}{p_1}  
	=  2 \pi \sqrt{h \ell_\text{on}} 
	\, .
\end{equation*}
Since $h \ll l_\text{on}$, one has $\lambda_\text{max} \gg 2 \pi h$,  such that gradients on the scale of the column can be neglected on time scales comparable and larger than $\tau_\text{on}$.
On that time scale, the dynamics of the longest wavelength mode then corresponds to the dynamics of the average cytosolic concentration, Eq.~\eqref{eq:att_elim}, $c(z,t) \propto e^{- t / \tau_1} \cos ( p_1 (h-z) )$; note that  $ \cos ( p_1 (h-z) )$ hardly shows any dependence on the spatial coordinate $z$ as $p_1 \ll h$. 

In the opposite limit of a \emph{large} column height $h \gg \ell_\text{on}$, the eigenvalues are approximately given by $p_j  \approx  (j -1/2) \pi/h$ with $j \in \mathbb{N}$.
Then, the largest wavelength $\lambda_\text{max}  =  4h$ is comparable in size with column height $h$ and gradients are significant.
Hence, the cytosolic dynamics is an integral part of the system's dynamics and can not effectively be described as some passive reservoir. 
There is an intricate interplay between protein dynamics in the vertically extended cytosol and on the membrane.

A concrete biological example where this interplay plays an important role is the Min system.
The `bulk height' in the cell geometry, given by the cell radius of approximately $\SI{0.5}{\micro m}$, is magnitudes of order smaller than the typical bulk heights from tens to several hundreds of micrometers in most \textit{in vitro} setups~\cite{Loose:2008a,Ivanov:2010a,Zieske:2014a,Vecchiarelli:2014a,Caspi:2016a,Glock:2018b}.
This suggests that the bulk height is a key parameter that distinguishes the two scenarios and is responsible for the qualitative difference in phenomenology between \textit{in vivo} and \textit{in vitro}.
Indeed, qualitatively different mechanisms underlie pattern formation in these two regimes \cite{Brauns.etal2020a}.
Pole-to-pole oscillations in cells and the patterns found at low bulk height (in the range of 1--\SI{5}{\micro m}) in the reconstituted system are driven by the same lateral mass-redistribution instability.
In contrast, at large bulk heights the local equilibria can become \emph{locally} unstable. This oscillatory local instability is a consequence of the extended bulk volume, where the bulk far away from the membrane effectively acts as a reservoir that facilitates the local oscillations. In Sec.~\ref{sec:Min-control-space}, we discuss how the local instabilities, triggered by lateral mass redistribution, play a key role for the Min pattern formation at large bulk heights.

\paragraph{Nucleotide exchange causes cytosolic gradients}
Another important factor that limits the elimination of cytosolic protein dynamics is nucleotide exchange. 
As explained in Sec.~\ref{sec:biochemical_principles}, NTPases can switch between active (NTP-bound) and inactive (NDP-bound) states. 
The chemical reaction turning inactive into active states is called nucleotide exchange. 
For instance, in the Min system, after MinE-cyatalyzed MinD-hydrolysis, MinD is released into the cytosol in its ADP-bound form.
Reactivation by nucleotide exchange happens at a rate $\lambda$ that is estimated to be $\lambda \approx \SI{6}{s^{-1}}$; c.f.\ \cite{Meacci:2006a,Halatek:2012a}. 
Recently it has been shown that even MinE shows switch-like behavior that is relevant for the robustness of pattern formation in the Min system~\cite{Denk:2018a}.

Detachment of a protein from the membrane into the cytosol with a subsequent change in conformation can be viewed as a `source-degradation' process:
the membrane acts as a MinD-ADP source, and nucleotide exchange corresponds to the `degradation' process.
This implies that the MinD-ADP density decays exponentially into the cytosol with a \emph{penetration depth} given by the diffusion length
\begin{equation} \label{eq:penetration-depth}
	\ell := \sqrt{D_{c}/\lambda} \, .	
\end{equation}
Since typical values of the  cytosolic diffusion constant $D_{c}$ for proteins is \textit{in vitro} of the order $D_{c} \approx \SI{60}{\micro m^2.s^{-1}}$ \cite{Loose:2008a}, one estimates $\ell \approx \SI{3}{\micro m}$, which is much smaller than typical system heights $h$ used in \emph{in vitro} experiments \cite{Loose:2008a,Ivanov:2010a}.
As we will show next, these protein density gradients normal to the membrane critically influence the stationary density profiles at the membrane.

Consider again a one-dimensional column of height $h$ with a simplified form of the MinD dynamics to illustrate the effect of nucleotide exchange in the bulk.
The bulk dynamics of MinD-ATP and MinD-ADP are given by the reaction-diffusion equations
\begin{subequations}
\begin{align}
	\partial_t c_\text{DD}^{}  (z,t)
	& =  D_c \, \partial_z^2 c_\text{DD}^{} -\lambda \, c_\text{DD}^{}
	\, , \\
	\partial_t c_\text{DT}^{} (z,t)
	& =  D_c \, \partial_z^2 c_\text{DT}^{} +\lambda \, c_\text{DD}^{}
	\, ,
\end{align}
\end{subequations}
where $c_\text{DD}^{}  (z,t)$ and $c_\text{DT}^{}  (z,t)$ denote the cytosolic density of MinD-ADP and MinD-ATP, respectively. 
These are linear equations that can readily be solved by a separation ansatz as above. 
In general, active MinD-ATP can bind to the membrane through some nonlinear attachment process, specified by a function  $f_{\text{on}}(m, \cDT|_{z=0}^{})$, that depends both on the density of membrane-bound MinD-ATP, $m$, and the cytosolic density of MinD-ATP close to the membrane $\cDT|_{z=0}^{}$.
For simplicity, we do not model MinE explicitly but only effectively through a detachment rate from MinD-ATP to cytosolic MinD-ADP specified by the detachment rate $k_{\text{off}}$. 
Then, the dynamics on the well-mixed membrane `point' reads
\begin{equation}
	\partial_t m(t) 
	=  f_{\text{on}}\big(m, \cDT|_{z=0}^{}\big) - k_{\text{off}} \, m 
	\label{eq:ls4} \, .
\end{equation}
At the membrane ($z = 0$), we use boundary conditions that reflect local mass conservation: the diffusive fluxes for each protein state (active and inactive) have to match the corresponding reactive fluxes between membrane and cytosol:
\begin{subequations}
\label{eq:lsbc-reactive} 
\begin{align} 
	 - D_{c} \, \partial_z c_\text{DD}^{} \big|_{z=0} 
	 & = \phantom{+} k_{\text{off}} \, m 
	\label{eq:lsbc1} 
	\, , \\
	- D_{c} \, \partial_z c_\text{DT}^{} \big|_{z=0} 
	& = - f_{\text{on}}\big(m,  \cDT|_{z=0}^{}\big) 
	\, .
\label{eq:lsbc2} 
\end{align}
\end{subequations}
In addition, we assume reflective boundary conditions at $z = h$:
%
%\begin{subequations}
\begin{align*}
	D_{c} \, 
	\partial_z c_\text{DD}^{}  \big|_{z=h} 
	& =  0
	\, , \\
	D_{c} \, 
	\partial_z c_\text{DT}^{}  \big|_{z=h} 
	& =  0
	\, .
\end{align*}
%\end{subequations}
%

While the above analysis can be carried through for the full time-dependent problem, we now focus on the analysis of the steady state.
Then the set of stationary diffusion equations reads $\ell^2 \, \partial_z^2 c_\text{DD,DT}^{} = \mp c_\text{DD}^{}$ and, using the reflective boundary conditions at $z=h$, is solved by
\begin{subequations}
\begin{align}
	c_\text{DD}^*(z) 
	& = c_\text{DD,0}^* \; 
	      \frac{\cosh ( (h-z)/\ell )}
	             {\cosh (h/\ell)} 
	\, , \label{eq:cDD-bulk-profile} \\
	c_\text{DT}^*(z)
	& = c_\text{DT,0}^* + 
	       c_\text{DD,0}^*
	       \left(1 -  
	       \frac{\cosh \left( (h-z)/\ell \right)}
	           {\cosh \left( h/\ell \right)}
	       \right) \, ,
\end{align}
\end{subequations}
Note that the total cytosolic density of MinD is spatially uniform: $c_\text{D}^*(z) = c_\text{DD}^*(z)  + c_\text{DT}^*(z) = \mathrm{const}$. 
\begin{figure}[tb]
\centering
\includegraphics{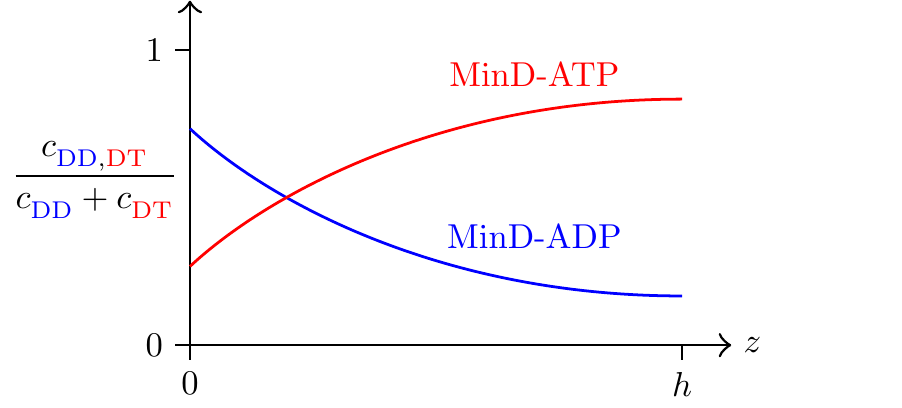}
\caption{
\textbf{Illustration of cytosolic density profiles of active and inactive MinD}. As one moves away from the membrane at $z = 0$, the density of inactive MinD is decreasing and the density of active MinD is increasing, mediated by nucleotide exchange, on a diffusive length scale $\ell = \sqrt{D_c/\lambda}$.
}
\label{fig:cytosol_profile_nucleotide_exchange}
\end{figure}
As shown in Fig.~\ref{fig:cytosol_profile_nucleotide_exchange}, the MinD-ADP concentration $c_\text{DD}^{} (z)$ has its largest value at the membrane and then decays exponentially into the bulk where it saturates at $c_\text{DD,0}^* / \cosh ( h / \ell )$.
In contrast, the MinD-ATP concentration increases from $c_\text{DT,0}^*$ to the saturation value $c_\text{DT,0}^* +  c_\text{DD,0}^* [1-1/\cosh ( h / \ell )]$.
Hence, active MinD --- the state of MinD that can attach to the membrane --- is depleted close to the membrane and reaches its saturation value only at a distance $\ell$ away from the membrane.
Taken together this shows that within a layer of size $\ell$,  active MinD is depleted while inactive MinD is enhanced with respect to their respective bulk values.
 
What determines the equilibrium protein concentrations ($c_\text{DD,0}^*$, $c_\text{DT,0}^*$, and $m^*$) at the membrane? 
It is the balance of the chemical reactions on the membrane, Eq.~\eqref{eq:ls4}, local mass conservation, i.e.\ the matching between the reactive dynamics and diffusive fluxes at the membrane, Eq.~\eqref{eq:lsbc-reactive}, and global mass conservation:
\begin{subequations}
\begin{align}
	f_{\text{on}} (m^*, c^*_\text{DT,0})
	&= k_{\text{off}} \, m^* 	
	\label{eq:stat_cond2} \, , 	\\
	\frac{\ell_\text{off}}{\ell} \, \tanh \left( h/\ell \right)  
	& =  \frac{m^*}{c_\text{DD,0}^* \, \ell_\text{off}}
	\label{eq:stat_cond1} \, , \\
	c_\text{DT,0}^*+c_\text{DD,0}^* + m^*/h 
	&= \nbar  \, ,  
\end{align}
\end{subequations}
where we have defined a second diffusive length scale $\ell_\text{off} := \sqrt{D_c / k_\text{off}}$, and $\nbar$ signifies the mean total protein density.

In the limit where the penetration depth $\ell$ is much smaller than the column height, $\ell \, {\ll}\,  h$, Eq.~\eqref{eq:stat_cond1} reduces to 
\begin{equation*}
	\frac{\ell_\text{off}}{\ell} 
    =  \frac{m^*}{c_\text{DD,0}^* \, \ell_\text{off}}
    \, .
\end{equation*}
Since the total mass density $\nbar$ is an upper bounded for the bulk densities, $\nbar \,{>}\, c_\mathrm{D,0}$, this implies that the stationary membrane density $m^*$ remains bound from above in the limit $h\rightarrow \infty$ 
\begin{equation*}
	\lim_{h\rightarrow \infty} m^* 
	< \nbar \, \frac{\ell_\text{off}^2 }{\ell} 
	=  \nbar \, \frac{\sqrt{D_c \lambda}}{k_\text{off}} 
	\, .
\end{equation*}
Hence, one concludes that the combined effect of detachment kinetics and nucleotide exchange leads to a saturation effect for the membrane density, even in the absence of an explicit saturation term in the reaction kinetics.\footnote{This emergent saturation effect explains the surprising observation that models using a simplified (unsaturated) recruitment term $\kdD \md \cDT$ reproduce both the Min dynamics \textit{in vivo} as well as \textit{in vitro} with essentially the same kinetic parameters --- despite the fact that membrane densities \textit{in vitro} are two orders of magnitude higher than \textit{in vivo}.} 
Notably, this \emph{emergent} saturation effect only depends on the linear reaction kinetics but is completely independent of the attachment dynamics $f_{\text{on}}$ that could be highly cooperative. 
Moreover, this also limits the stationary protein flux onto the membrane $f_{\text{on}}(m, c_\text{DT,0}^{}) = k_{\text{off}} \, m < \sqrt{D_c\lambda} \, \nbar$ \cite{Halatek:2018a}.

The saturation effect strictly requires the extended bulk dimension. 
To see this we consider the analogous set of reaction equations with reduced bulk dimension, obtained --- as in our first example above --- by assuming a spatially homogeneous cytosolic protein density. 
Then one finds for the spatially averaged cytosolic densities ${\bar{c}}_i (t)= \frac{1}{h} \int_0^h\! \dd z \, c_i (z,t)$ 
\begin{subequations}
\begin{align}
	\partial_t\bar{c}_\text{DD} (t)
	& = \hphantom{-}  k_{\text{off}} \,  m /h -
	       \lambda \, \bar{c}_\text{DD}
	\label{eq:redls1} \, , \\
	\partial_t\bar{c}_\text{DT} (t)
	& =  -f_{\text{on}}(m, \bar{c}_\text{DT})/h + 
	        \lambda \, \bar{c}_\text{DD}
	        \label{eq:redls2} \, \\
	\partial_tm  (t)
	& = \hphantom{-} f_{\text{on}}(m, \bar{c}_\text{DT}) - 
	       k_{\text{off}} \, m 
	       \label{eq:redls3} \, , 
\end{align}
\end{subequations}
For the stationary state, one obtains from Eq.~\eqref{eq:redls1} 
\begin{equation*}
	m^* 
	= h \, \frac{\lambda}{k_\text{off}} \, \bar{c}_\text{DD}^* \, .
\end{equation*}
which is obviously unbounded (for any finite $\bar{c}^*_\text{DD}$) as  $h\rightarrow \infty$.
Hence, for large bulk heights, mass-conserving diffusion-reaction systems with nucleotide exchange in the cytosol, prohibit a naive elimination of the cytosol. 
If, however, the bulk height is small compared to the reactivation length scale, $h  \ll  \ell$, one has $\tanh (h /\ell) \approx h /\ell$ such that using Eq.~\eqref{eq:stat_cond1} the reactive boundary conditions, Eqs.~\eqref{eq:lsbc1} and~\eqref{eq:lsbc2}, for the stationary bulk profiles reduce to 
%
%\begin{subequations}
\begin{align*}
	- D_{c}\partial_z c_\text{DD}^{} \big|_{z=0} 
	& = - \lambda \, c_\text{DD,0}^* \, h 
	  = - k_{\text{off}} \, m^*
	\, , \\
	- D_{c}\partial_z c_\text{DT}^{} \big|_{z=0} 
	& = \hphantom{-} \lambda \, c_\text{DD,0}^* \, h 
	  = \hphantom{-} f_{\text{on}}(m^*,c^*_\text{DT})
	\, .
\end{align*}
%\end{subequations}
%
This set of boundary conditions are identical to the stationarity condition in the reduced system obtained from Eqs.~\eqref{eq:redls1} and \eqref{eq:redls2} by setting the left hand sides to zero.
In summary, nucleotide exchange, a generic feature of all NTPase reaction cycles, prohibits an elimination of the cytosol unless $h  \ll  \ell$. 

For typical experimental setups studying the Min system \textit{in vitro} one has $h/ \ell = \mathcal{O}(10^3)$ \cite{Loose:2008a}.
Hence one is actually in the opposite limit $h  \gg  \ell$ where bulk dynamics is essential. 
Neglecting nucleotide exchange or reducing the bulk dimension if the condition $h  \ll  \ell$ is not met would be an invalid approximation as it misses the saturation of membrane attachment.

\subsection{Linear stability analysis in box geometry} \label{sec:LSA-box-geometry}

Let us not turn to laterally extended systems with bulk-boundary coupling and perform a linear stability analysis.
As we will see, there are two key differences to the linear stability analysis in one spatial dimension (see Sec.~\ref{sec:classical_turing}).
First, there may be vertical gradients in the bulk in steady state even if the membrane concentrations are laterally uniform. This has to be taken into account when calculating the laterally homogeneous steady state around which the dynamics is linearised for stability analysis.
Second, the algebraic problem that results by solving the linearised dynamics in terms of the spatial eigenmodes is no longer an eigenvalue problem. 
Instead, one obtains a solvability condition (`characteristic equation') which must be solved to find the growth rates corresponding to each eigenmode. 
Above, we already encountered an elementary form of such a solvability condition in the linear stability analysis for a single column (see Eq.~\eqref{eq:solvability-robin}).

As a simple geometry, consider the box-geometry that is formed by laterally coupling vertical bulk-columns, each with a membrane point at its bottom (Fig.~\ref{fig:box_geometry}).
This captures the physics of laterally extended systems with bulk-boundary coupling but avoids the mathematical complexity of the linear stability analysis in a more realistic geometry like an ellipsoid~\cite{Halatek:2012a} or a sphere~\cite{Levine:2005a,Klunder:2013a}.
In the box geometry, we simply have the diffusion operators $\nabla^2 = (\partial_x^2 + \partial_z^2)$ in the bulk and $\grad{m}^2 = \partial_x^2$ on the membrane, and the gradient operator along the membrane's inward normal $\nabla_{\!\perp} = \partial_z$.

\paragraph{Two-component system}
To illustrate the basic procedure of linear stability analysis in bulk-surface coupled systems, we first consider a system with only a single cytosolic component $c(x,z,t)$ and a single membrane component $m(x,t)$.
For simplicity, we consider purely diffusive bulk dynamics
\begin{equation} \label{eq:box-bulk-dyn-1c}
	\partial_t c(x,z,t) = D_c (\partial_x^2 + \partial_z^2) c,
\end{equation}
together with the reactive boundary condition (cf.\ Eq.~\eqref{eq:bulk-surface-general})
\begin{equation} \label{eq:box-bulk-surface-coupling-1c}
	-D_c \partial_z c |_{z=0} = -f(m, c|_{z=0}),
\end{equation}
and no-flux conditions $\partial_x c|_{x = 0,L} = \partial_z c|_{z = h} = 0$ at the remaining boundaries.
The membrane dynamics are 
\begin{equation} \label{eq:box-mem-dyn-1c}
	\partial_t m(x,t) = D_m \partial_x^2 m + f(m, c|_{z=0}),
\end{equation}
with reflective boundaries $\partial_x m_{x = 0,L} = 0$.
These dynamics conserve the average mass
\begin{equation} \label{eq:box-total-density-1c}
	\nbar = \frac{1}{L\cdot h} \int_0^L \dd x \, m(x,t) + \int_0^L \dd x \int_0^h \dd z  \, c(x,z,t) \, .
\end{equation}

\begin{exercise} \label{ex:mapping-to-1d}
	In the limit of small bulk height, map the dynamics in box geometry, Eqs.~\eqref{eq:box-bulk-dyn-1c}--\eqref{eq:box-mem-dyn-1c} to a 1d description (`line geometry') by approximating the cytosol concentration at the membrane by the vertical average $c(x,z=0,t) \approx \langle c(x,z,t) \rangle_z$. 
	Under which conditions is such an approximation valid?
\end{exercise}

The starting point of a linear stability analysis is the computation of a steady state around which the system can be linearised. 
For systems in 1d geometry we studied in Sec.~\ref{sec:lateral-instability}, we considered the stability of a homogeneous steady state.
The analog to this in a box geometry is a steady state that is spatially uniform along the membrane (i.e., in the lateral direction).
However, as we have learned from the analysis of the columnar geometry above, the corresponding cytosolic densities can have vertical gradients while they are laterally homogenous.

Because homogeneity along the membrane implies $\partial_x c_i = 0$ in the entire bulk, the vertical bulk profiles of a \emph{laterally homogeneous} steady state can be obtained by the same reasoning as in Sec.~\ref{sec:column_examples}.
In the case of purely diffusive bulk dynamics, Eq.~\eqref{eq:box-bulk-dyn-1c}, the steady state bulk profile is determined by $\partial_z^2 c^*(z) = 0$ and therefore homogeneous also in the vertical direction, $c^*(z) = c_0^*$; a constant gradient is prohibited by the no-flux boundary condition at the box's ceiling $z = h$).
The laterally homogeneous steady state of Eqs.~\eqref{eq:box-bulk-dyn-1c}--\eqref{eq:box-mem-dyn-1c} is therefore simply determined by 
\begin{equation}
	f(m^*,c_0^*) = 0 \quad \text{and} \quad c_0^* + m^*/h = \nbar.
\end{equation}

The linearised dynamics of a small perturbation $c(x,z,t) = c^*(z) + \delta c(x,z,t)$, $m(x,t) = m^* + \delta m(x,t)$ is given by
\begin{subequations} \label{eq:box-lin}
\begin{align}
	\partial_t \delta c(x,z,t) &= D_c (\partial_x^2 + \partial_z^2) \delta c \,,  \label{eq:box-lin-c} \\
	-D_c \partial_z \delta c |_{z=0} &= -f_m \delta m -f_c \delta c|_{z=0} \,, \\
	\partial_t \delta m(x,t) &= D_m \partial_x^2 \delta m + f_m \delta m + f_c \delta c|_{z=0} \,,
\end{align}
\end{subequations}
with the usual convention $f_{m,c} = \partial_{m,c} f|_{(m^*, c_0^*)}$.
To solve these equations, we look for solutions of the form
\begin{subequations} \label{eq:box-separation-ansatz}
\begin{align}
	\delta c(x,z,t) &= \delta \hat{m} \; e^{\sigma t} \cos(qx) \, Z(z; \sigma, q) \,, \\
	\delta m(x,t) &= \kern0.3em \delta \hat{c} \kern0.45em e^{\sigma t} \cos(qx) \, , 
\end{align}
\end{subequations}
where we used separation of variables for $x, z,$ and $t$, and anticipated that Fourier modes will diagonalize the Laplace operator in the lateral direction (along the membrane) as in 1d geometry; cf.\ Sec.~\ref{sec:lateral-instability}.
Our goal is to find pairs $(q,\sigma)$ for which this ansatz solves the linearised dynamics Eq.~\eqref{eq:box-lin}.
Rewritten in the form $\sigma(q)$ these pairs constitute the dispersion relation that informs about the growth rate(s) $\sigma$ of perturbations with lateral wavenumber $q$.

First, the vertical eigenmode profiles $Z(z)$ are obtained by substituting the separation ansatz into Eq.~\eqref{eq:box-lin-c}
\begin{equation} \label{eq:box-bulk-eigenmode-eq}
	\sigma Z(z; \sigma, q) = D_c \, q^2 \, Z(z; \sigma, q) + D_c Z''(z; \sigma, q) \,,
\end{equation}
which is solved by
\begin{equation} \label{eq:box-bulk-eigenmode-sol}
	Z(z;\sigma,q) = \cosh\left(\sqrt{q^2 + \sigma/D_c}\; (h-z)\right)\Big/\cosh\left(\sqrt{q^2 + \sigma / D_c} \; h\right) \,.
\end{equation}
With this, the linearised bulk-surface coupling and the membrane dynamics become a set of linear algebraic equations which are conveniently written in matrix form 
\begin{equation} \label{eq:box-linear-stability-problem-1c}
		\underbrace{\begin{pmatrix}
				D_c \, \Gamma(\sigma, q) - f_c & -f_m \\
				f_c & -\sigma -q^2 D_m + f_m 
		\end{pmatrix}}_{\displaystyle =: \mathcal{L}(\sigma, q)} \begin{pmatrix}
			\delta \hat{c} \\
			\delta \hat{m}
		\end{pmatrix} = 0 \,,
\end{equation}
with the bulk-surface coupling coefficient
\begin{equation*} 
	\Gamma(\sigma, q) := \partial_z Z(z = 0; \sigma, q)
	= -\sqrt{q^2 + \sigma/D_c} \, \tanh \left(\sqrt{q^2 + \sigma/D_c} \; h\right) \,.
\end{equation*}

Solutions to the Equation~\eqref{eq:box-linear-stability-problem-1c} exist only if the solvability condition (characteristic equation)
\begin{equation} \label{eq:box-characteristic-eq}
	\det \mathcal{L}\big(\sigma(q), q\big) = 0
\end{equation}
is fulfilled.
This is the analog to the characteristic equation Eq.~\eqref{eq:line-characteristic-eq} of the eigenvalue problem Eq.~\eqref{eq:line-EV-problem} in 1d geometry.
Because $\Gamma(\sigma, q)$ is not linear in $\sigma$ (here, it is actually non-algebraic), the matrix equation Eq.~\eqref{eq:box-linear-stability-problem-1c} does not constitute an eigenvalue problem, though. 
To determine the dispersion relation $\sigma(q)$, we solve the characteristic equation~\eqref{eq:box-characteristic-eq} numerically, using iterative methods like Newton's method. 
An implementation of the above analysis in Mathematica --- computing the laterally homogeneous steady state and its linear stability --- is provided in the repository \url{https://github.com/f-brauns/reaction-diffusion-stability} \cite{GitHubRepo}.

Note that Eq.~\eqref{eq:box-characteristic-eq}, has infinitely many solutions $\sigma_i(q)$ for each $q$, corresponding to the set of vertical bulk-eigenmodes which form basis functions for the bulk perturbations.
This was discussed in more detail for the elementary example of a simple attachment process from a cytosolic column to a membrane `point' above; see the discussion below Eq.~\eqref{eq:solvability-robin}. 
 In particular, it was shown that bulk perturbations with short wavelength modulations quickly decay owing to fast cytosolic diffusion. Thus the respective eigenvalues have a large and negative real part.
For linear stability analysis, we are typically only interested in the solution $\sigma_i(q)$ with the largest real part, which indicates the stability of the respective lateral mode $\cos(q x)$.

For growth rates close to zero, more precisely $|\sigma| \ll q^2 D_c$, one can expand the coupling coefficient $\Gamma$ to linear order in $\sigma$. Then Eq.~\eqref{eq:box-linear-stability-problem-1c} becomes an (approximate) eigenvalue problem the solutions of which provide good initial guesses for iterative (Newton) methods to solve the exact characteristic equation \eqref{eq:box-characteristic-eq}.

\begin{exercise}
	Show that in the limit $h \rightarrow 0$, Eq.~\eqref{eq:box-linear-stability-problem-1c} maps to the eigenvalue problem Eq.~\eqref{eq:line-EV-problem} of the two-component MCRD system in 1d geometry. The result should agree with the linear stability analysis of the approximate dynamics in 1d geometry obtained in Exercise~\ref{ex:mapping-to-1d}.
\end{exercise}

\paragraph{Min skeleton model}

\def\lD{\ell_\mathrm{D}}

As a practical and slightly more complex example, we repeat the calculation of laterally homogeneous steady states and their linear stability analysis for the Min skeleton model~\cite{Halatek:2012a,Halatek:2018a,Brauns.etal2020a}. 
The procedure is analogous to the two-component case above. 
Some slight modifications are required to account for the nucleotide exchange of MinD in the bulk.

To simplify the analysis, we rewrite the MinD bulk dynamics in terms of the total bulk density $\cD = \cDD + \cDT$ together with the density of inactive MinD, $\cDD$. Together, the cytosolic densities are $\mathbf{c} = (\cD,\cDD,\cE)$ while the membrane densities are $\mathbf{m} = (\md, \mde)$ as before.
With these new variables, the reactive bulk-boundary fluxes (cf.\ Eq.~\eqref{eq:Min-boundary-coupling}) read
\begin{equation*}
	\bm{f}(\vek{m}, \vek{c}) = \begin{pmatrix}
		-(\kD+\kdD \md)(\cD-\cDD)\\ \kde \mde\\ \kde \mde - \kdE \md \cE
	\end{pmatrix},
\end{equation*} 
and the membrane reactions (cf.\ Eq.~\eqref{eq:Min-membrane-dyn}) are given by
\begin{equation*}
	\vek{R}_\mathrm{mem}(\vek{m}, \vek{c}) = \begin{pmatrix}
		(\kD+\kdD \md)(\cD-\cDD) - \kdE \md \cE\\ \kdE \md \cE - \kde \mde
	\end{pmatrix}.
\end{equation*} 
The advantage of the new variables is that the bulk equations decouple
%\begin{subequations}
\begin{align*}
	\partial_t \cD(x,z,t) &= D_c \nabla^2 \cD \,, \\
	\partial_t \cDD(x,z,t) &= D_c \nabla^2 \cDD - \lambda \cDD \,, \\
	\partial_t \cE(x,z,t) &= D_c \nabla^2 \cE \, ,
\end{align*}
%\end{subequations}
which immediately yields that the bulk profiles of the laterally homogeneous steady state are uniform for $\cD$ and $\cE$: $c_\mathrm{D,E}^*(z) = c_\mathrm{D,E}^*(0)$. 
The bulk profile for $\cDD$ is given by
\begin{equation*}
	c_\mathrm{DD}^*(z) = c_\mathrm{DD}^*(0) \, \frac{\cosh \big((h-z)/\lD\big)}{\cosh(h/\lD)} \,,
\end{equation*}
where $\ell_\mathrm{D} = \sqrt{D_c/\lambda}$; cf.\ Eq.~\eqref{eq:cDD-bulk-profile}.
The steady state bulk concentrations at the membrane $c_\mathrm{D,DD,E}^*(0)$ and the membrane concentrations $m_\mathrm{d,de}^*$ are determined via the reactive boundary conditions; cf.\ Eq.~\eqref{eq:Min-boundary-coupling},
%\begin{subequations}
\begin{gather*}
	f_\mathrm{D} = 0 \,, \quad f_\mathrm{E} = 0 \,, \\
	f_\mathrm{DD} = -\frac{D_c}{\lD} \tanh(h/\lD) \,,
\end{gather*}
%\end{subequations}
and the condition that the membrane reactions are balanced 
\begin{equation*}
	\vek{R}_\mathrm{mem} = 0 \,.
\end{equation*}
Moreover, global mass conservation enforces
%
%\begin{subequations}
\begin{align*}
	\nbar_\mathrm{D}^{} 
	&= c_\mathrm{D}^*(0) + (m_\mathrm{d}^* + m_\mathrm{de}^*)/h \, , \\
	\nbar_\mathrm{E}^{} 
	&= c_\mathrm{E}^{*} + m_\mathrm{de}^*/h \, .
\end{align*}
%\end{subequations}

For linear stability analysis, we again consider perturbations $\vek{c} = \vek{c}^*(z) + \delta \vek{c}^*(x,z,t)$, $\vek{m} = \vek{m}^* + \delta \vek{m}^*(x,t)$, with the separation ansatz Eq.~\eqref{eq:box-separation-ansatz}.
The bulk eigenmode equations for $\delta \cD$ and $\delta \cE$ are identical to Eq.~\eqref{eq:box-bulk-eigenmode-eq}, while we have to replace $\sigma \rightarrow \sigma + \lambda$ to account for the linear degradation term in the linearised bulk dynamics of $\delta \cDD$.
The bulk eigenmodes are thus given by
\begin{equation*}
	Z_\mathrm{D,E}(\sigma,q) = Z(\sigma,q) \,, \quad Z_\mathrm{DD}(\sigma,q) = Z(\sigma + \lambda,q) \,,
\end{equation*}
with $Z(\sigma,q)$ defined by Eq.~\eqref{eq:box-bulk-eigenmode-sol}.
Substituting the separation ansatz with these bulk eigenmodes into the linearised bulk-boundary coupling and membrane dynamics (cf.\ Eqs.~\eqref{eq:Min-boundary-coupling} and~\eqref{eq:Min-membrane-dyn}), we obtain the matrix equation
\begin{equation*}
		\underbrace{\begin{pmatrix}
				D_c \, \bm{\Gamma}(\sigma, q) - \bm{f}_\vek{c} & -\bm{f}_\vek{m} \\
				\vek{R}^\mathrm{mem}_\vek{c}  & -\sigma -q^2 D_m + \vek{R}^\mathrm{mem}_\vek{m}  
		\end{pmatrix}}_{\displaystyle =: \mathcal{L}(\sigma, q)} \begin{pmatrix}
			\delta \hat{\vek{c}} \\
			\delta \hat{\vek{m}}
		\end{pmatrix} = 0 \,,
\end{equation*}
with the bulk-surface coupling coefficient matrix
\begin{equation*} 
	\bm{\Gamma}(\sigma, q) = \diag\big(\Gamma(\sigma,q),\Gamma(\sigma + \lambda,q),\Gamma(\sigma,q)\big) \,,
\end{equation*}
and the linearised reaction terms
\begin{align*}
	\bm{f}_{\vek{m},\vek{c}} &:= \partial_{\vek{m},\vek{c}} \bm{f} \big|_{(\vek{m}^*,\vek{c}^*|_\mathcal{S})} \,, \\
	\vek{R}^\mathrm{mem}_{\vek{m},\vek{c}} &:= \partial_{\vek{m},\vek{c}} \vek{R}_\mathrm{mem} \big|_{(\vek{m}^*,\vek{c}^*|_\mathcal{S})}\,.
\end{align*}

Analogously to the two-component system above, the dispersion relation is determined by the characteristic equation $\det \mathcal{L}\big(\sigma(q),q\big) = 0$, which can be solved numerically.
An implementation of this analysis in Mathematica is provided in the public repository \cite{GitHubRepo}.

\begin{exercise}
	As a further generalisation, repeat the linear stability analysis for the extended Min model that includes the MinE switch~\cite{Denk:2018a}.
	Explore how fast switching between two MinE states affects the regime of instability in the $(\nbar_\mathrm{D}, \nbar_\mathrm{E})$ parameter plane. 
	To that end, write a program to numerically determine the homogeneous steady states and solve the characteristic equation for the growth rates $\sigma(q)$. You can use the implementation for the Min skeleton model, provided in the online repository \cite{GitHubRepo}, as a starting point.  
\end{exercise}

\subsection{Pattern formation without a lateral instability}

\begin{figure}[tb]
\centering
\includegraphics[scale=1.25]{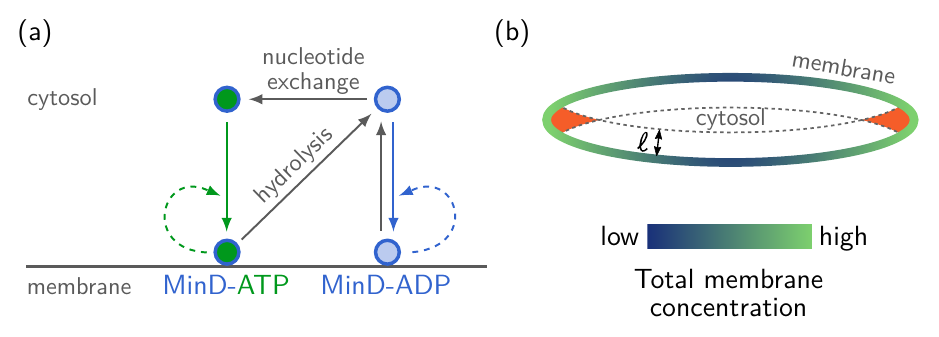}
\caption{ 
\textbf{Geometry-induced pattern formation by proteins cycling between membrane-bound and cytosolic states.} 
(a) Generic interaction network of an ATPase, like MinD, that can attach to the membrane, detaches from the membrane after hydrolysis, and undergoes nucleotide exchange in the cytoplasm. Self-recruitment (indicated by dashed arrows) enhances the geometry sensing but is not essential. 
(b) Nucleotide exchange in the cytosol with rate $\lambda$ results in a layer of depth $\ell = \sqrt{D_c/\lambda}$ in which MinD-ADP is enriched. 
In regions where the radius of membrane curvature is smaller than the penetration depth $\ell$ (illustrated by orange shading), the MinD-ADP enrichment is enhanced due to the lower ratio of bulk-volume to membrane-surface. 
This enhancement translates to increased (resp.\ decreased) membrane concentration in regions of high curvature if the membrane attachment of MinD is stronger (resp.\ weaker) in the ADP-bound state compared to the ATP-bound state.
}
\label{fig:geometry-sensing}
\end{figure}

This section serves to illustrate the importance of bulk-boundary coupling by showing that it can facilitate the formation of patterns even in the absence of any lateral instability~\cite{Thalmeier:2016a}.
To this end, we consider a basic reaction module that comprises a single type of NTPase which cycles between a NDP-bound and a NTP-bound state; see Fig.~\ref{fig:geometry-sensing}a.
It is assumed that both protein conformations can attach to the membrane either directly with rates $k_\text{D}^{+}$ and $k_\text{T}^{+}$, or cooperatively by forming homo-dimers with corresponding recruitment rates $k_\text{dD}$ and $k_\text{tT}$ for the NDP- and the NTP-bound state, respectively.
Membrane-bound, inactive proteins can detach into the cytosol with detachment rate $k_\text{d}^{-}$, while active proteins detach only after hydrolysis with rate $k_\text{t}^{-}$.
In the cytosol, inactive proteins are transformed into active ones by nucleotide exchange with rate $\lambda$.
As discussed in \cite{Thalmeier:2016a}, this reaction network serves as a model for the bipolar pattern of AtMinD in \textit{E.~coli} cells~\cite{Zhang:2009a}.

In elliptical geometry, as illustrated in Fig.~\ref{fig:geometry-sensing}b, this reaction module is described by the following set of (mass-conserving) reaction-diffusion equations
%\begin{subequations}
\begin{align*}
	\partial_t c_\text{T}^{}
	&= D_c \nabla^2 c_\text{T}^{}  
	 + \lambda \, c_\text{D}^{}  \, , \\
	\partial_t c_\text{D}^{}
	&= D_c \nabla^2 c_\text{D}^{}  
	 - \lambda \, c_\text{D}^{}  \, , \\
	\partial_t m_\text{t}^{}
	&= D_\text{m} \grad{\mathcal{S}}^2 m_\text{t}^{} \,  + \big( 
	   k_\text{T}^{+} +
       k_\text{tT}^{} \, 
	   m_\text{t}^{}  
	   \big)
	   \; c_\text{T}^{} |_\mathcal{S}^{} -   
       k_\text{t}^{-} \, 
       m_\text{t}^{}
   	\, , \\
	\partial_t m_\text{d}^{}
	&= D_\text{m} \grad{\mathcal{S}}^2 m_\text{d}^{}  + \big( k_\text{D}^{+} +
	   k_\text{dD}^{} \, 
	   m_\text{d}^{} 
	   \big) 
	   c_\text{D}^{} |_\mathcal{S}^{} - 
       k_\text{d}^{-} \, m_\text{d}^{}
    \, ,
\end{align*}
%\end{subequations}
where $c_\text{D}^{}$ and $c_\text{T}^{}$ denote the bulk and $m_\text{d}^{}$ and $m_\text{t}^{}$ the membrane concentrations of inactive and active proteins, respectively; cytosolic concentrations right at the membrane are indicated by $|_\mathcal{S}^{}$ as in Sec.~\ref{sec:protein_patterns}.
The exchange of particles between the cytosol and the membrane is accounted for by the reactive boundary conditions
%\begin{subequations}
\begin{align*}
	- D_c \nabla_{\!\perp} c_\text{T}^{}
	&= - \big(
	   k_\text{T}^{+} + 
	   \, k_\text{tT}^{} \; m_\text{t}^{}
	   \big) \,
	   c_\text{T}^{} |_\mathcal{S}^{}
	   \, , \\
	- D_c \nabla_{\!\perp} c_\text{D}^{}
	&= - \big(
	   k_\text{D}^{+} + 
	   k_\text{dD}^{} \, m_\text{d}^{}
	   \big) \,  
	   c_\text{D}^{} |_\mathcal{S}^{} + 
	   k_\text{d}^{-} \, m_\text{d}^{} +
	   k_\text{t}^{-} \, m_\text{t}^{}  \, .
\end{align*}
%\end{subequations}
%
Numerical studies of these equations have shown that the elliptical geometry of the cell induces pattern formation, even when there is no lateral instability~\cite{Thalmeier:2016a}. 

Figure \ref{fig:geometry-sensing}b illustrates how this finding can be explained heuristically. 
Consider a system where there is no recruitment ($k_\mathrm{dD} = k_\mathrm{tT} = 0$) and where the attachment and detachment rates for the active and inactive protein are the same ($k^+_\mathrm{d} = k^+_\mathrm{t}$).
In this case, the dynamics of the total density on the membrane $m = m_\mathrm{d}^{} + m_\mathrm{t}^{}$ and in the cytosol $c = c_\mathrm{D}^{} + c_\mathrm{T}^{}$ decouple. Because the dynamics of $c$ is purely diffusive, and the steady state condition $\partial_t m = 0$ implies that boundary fluxes $\nabla_{\!\perp} c|_\mathcal{S}$ vanish, the total protein density on the membrane must be spatially uniform in steady state.

What determines the cytosolic gradients of the active and inactive proteins? 
As the inactive form is released into the cytosol and reactivated there with rate $\lambda$, the dynamics corresponds to a classical source-degradation problem~\cite{Wartlick:2009a}: the membrane acts as a source of inactive proteins that are transformed into active proteins in the cytosol. 
This implies that there is a gradient of inactive proteins in the vicinity of the membrane with the thickness of this depletion layer given by the diffusion length  (reactivation length, cf.\ Eq.~\eqref{eq:penetration-depth} and Fig.~\ref{fig:cytosol_profile_nucleotide_exchange})
\begin{equation*}
	\ell	= \sqrt{D_c/\lambda}
	\, .
\end{equation*}
Since these gradients overlap at the cell poles, one expects an increased concentration of inactive proteins there; for an illustration see Fig.~\ref{fig:geometry-sensing}b.
The magnitude of this effect depends on the ratio of this diffusion length with respect to the cell length $L$, and should vanish as $\ell \gg L$.
These arguments explain why the local ratio of the reaction volume for nucleotide exchange (cytosolic volume) to the available membrane surface facilitates geometry-induced pattern formation.

\begin{figure}[t]
\centering
\includegraphics{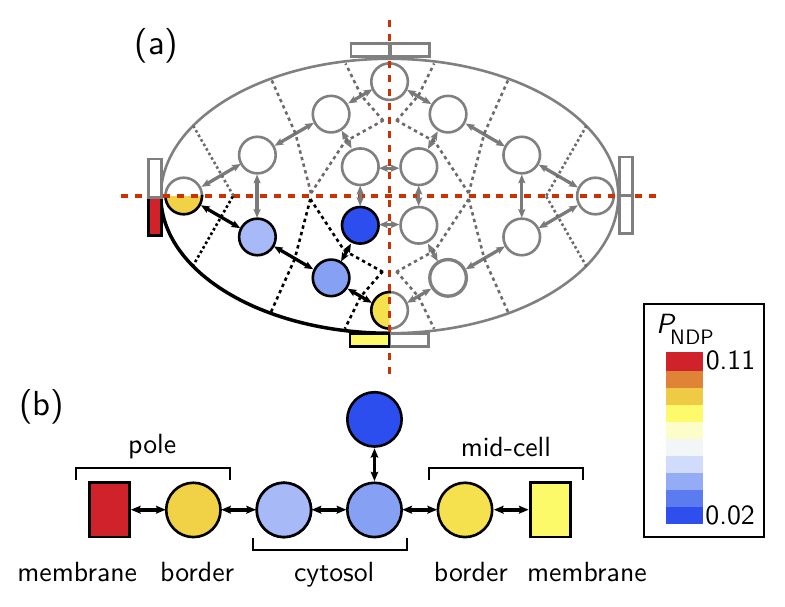}
\caption{\textbf{Spatial domain reduction to set of nodes.} 
(a) The dynamics of a spatially extended system, described in terms of partial differential equations, can be approximately reduced to a set of ordinary differential equations describing the dynamics on a \emph{network of coupled nodes}. The nodes are considered as well mixed and diffusion corresponds to exchange of proteins between these nodes. The set of nodes is chosen such that it approximately represent the bulk-to-surface ratio at the cell poles and at midcell. Here, we only consider symmetric stationary states, such that we only need to consider a single quadrant explicitly. 
(b) Illustration of the reduced model with two membrane nodes at the pole and at mid-cell, two border nodes connecting membrane and cytosol, as well as three cytosolic nodes.
Note that the top-right cytosolic node (dark blue) accounts for the increased ratio of membrane area to bulk volume at mid-cell. 
This network of nodes polarises with the color of the nodes indicating the density of inactive proteins $P_\text{NDP}$.
}
\label{fig:ellipse-to-nodes}
\end{figure}

To gain a deeper conceptual understanding of these heuristic results it is instructive to map the spatially extended model onto a system with a set of discretized nodes (Fig.~\ref{fig:ellipse-to-nodes}a). 
The idea of the mapping is that the placement and number of nodes is chosen such that it faithfully accounts for the different ratios of membrane area to bulk volume at the cell poles and at mid-cell.
Due to the elliptical symmetry, this system can be further reduced to one quadrant of the ellipse, leaving one with a network of one membrane node at a pole and one at mid-cell, two nodes representing the interface between membrane and cytosol, and three cytosolic nodes.
The corresponding set of ordinary differential equations can be found in the Supplementary Material of \cite{Thalmeier:2016a}.
Solving these equations leads to two main insights: 
First, the simple NTPase model indeed leads to polarisation between cell pole and mid-cell, corresponding to a bipolar pattern in the whole cell (Fig.~\ref{fig:ellipse-to-nodes}b). 
Second, one can perform a standard bifurcation analysis which shows that the steady states do not undergo any bifurcation.
One finds only one physically possible solution with positive protein density on the membrane and this density increases with recruitment rate.\footnote{ 
Only in the limiting (non-generic) case where the attachment rate of inactive protein vanishes, does one find a transcritical bifurcation. 
Hence one concludes, that a genetic NTPase reaction module shown in Fig.~\ref{fig:geometry-sensing}a generically has an imperfect transcritical bifurcation.}
This implies that the reaction module exhibits a robust mechanism for geometry-induced pattern formation amplified by nonlinear feedback but without an underlying Turing instability.

%% ================================
%% CONTROL SPACE
%% ================================

\clearpage

\section{Control space dynamics} \label{sec:control-space}

\subsection{The control space concept}

Consider a multi-component, mass-conserving reaction-diffusion (MCRD) system as discussed in section \ref{sec:MCRD_cell_geometry}. 
It is described in terms of the densities of all the components (states of protein species) in the biochemical reaction network $u_i (\vek{x},t)$ with $i \in \{ 1, \ldots, N \}$. 
In general, the number $N$ of these components is large compared to the number $S$ of protein species with mass densities $n_\alpha (\vek{x},t)$ where $\alpha \in \{ 1, \ldots, S \}$; see the reaction networks shown in Fig.~\ref{fig:biochemical-networks}.
How can one reduce the complexity of such large multi-component systems?

\begin{figure}[b!]
	\centering
	\includegraphics{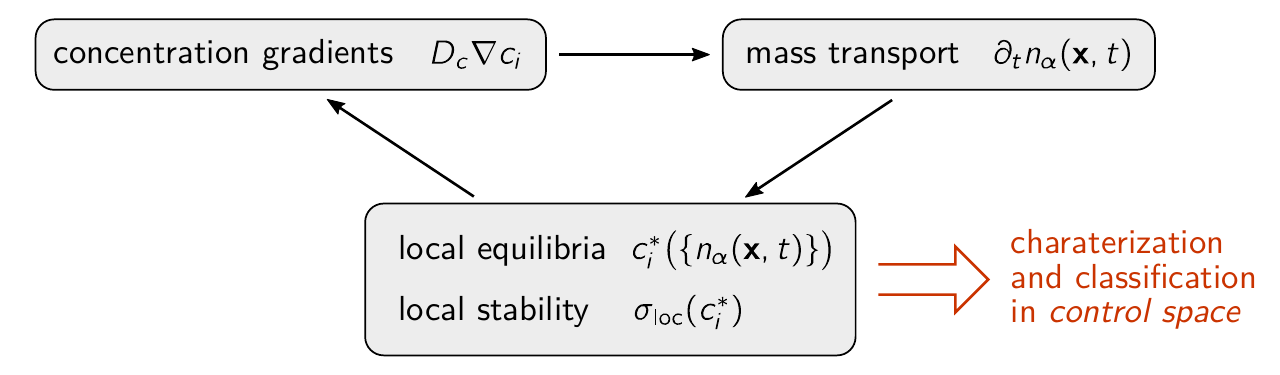}
	\caption{\textbf{Conceptualization of MCRD dynamics}. Deneralisation of the scheme shown in Fig.~\ref{fig:mass-redistribution_scheme}, to an arbitrary number of components $c_i$ and conserved masses $n_\alpha$. In addition to moving, i.e.\ changing position in phase space, the local equilibria $c_i^*$ can also change their stability as indicated by the leading local eigenvalue $\sigma_\mathrm{loc}$.
	The information about the position and stability of local equilibria as a function of the masses $n_\alpha$ is encoded in the bifurcation structure in control space; see Fig.~\ref{fig:control_space_flow} for an example. 
	}
	\label{fig:mass-redistribution-generalised}
\end{figure}

From the analysis of two-component MCRD systems in Section~\ref{sec:two-component_MCRD} we gained the  following key insights.
Pattern formation is mainly driven by the \emph{spatiotemporal dynamics of protein masses}.
While the average values for each protein mass $\nbar_\alpha$ are fixed by mass conservation, their local values $n_\alpha (\vek{x},t)$ are in general spatially inhomogeneous and can change over time.
Dynamics of protein masses inevitably leads to \emph{moving local equilibria}  since at a given spatial location $\vek{x}$ (and given time $t$) the value of the fixed point (equilibrium) $u_i^* (\vek{x},t)$ of the underlying biochemical network is determined by the corresponding set of local values for the protein masses: $u_i^* (\vek{x},t) = u_i^* \left( \{ n_\alpha (\vek{x},t) \} \right)$.
In other words, the local protein masses act as \emph{dynamic control variables}, in contrast to the reaction rates of the biochemical network that are global and fixed control parameters.  
These local equilibria drive local reactive fluxes that change the gradients in the densities of the various components of the protein network, which in turn drive diffusive fluxes that lead to a  redistribution of protein masses (Fig.~\ref{fig:mass-redistribution_scheme}).
This suggests that the essential dynamics of a multi-component system should be captured already in terms of the local densities of the conserved protein masses. 
We will refer to the space spanned by these control variables as \emph{control space}; for an illustration see Fig.~\ref{fig:mass-redistribution-generalised}.

\begin{figure}[tb]
	\centering
	\includegraphics{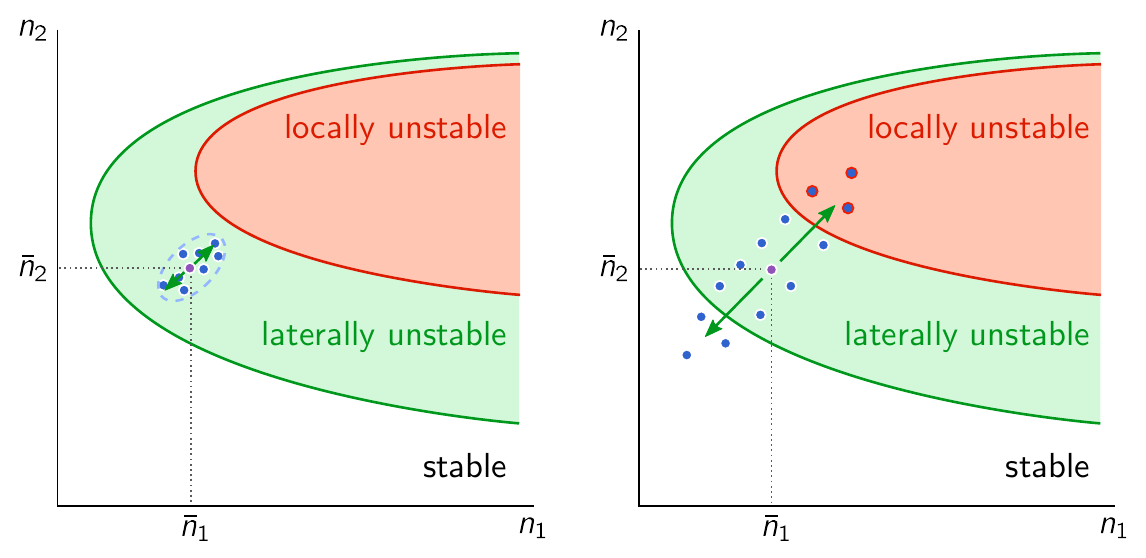}
	\caption{\textbf{Concept of control space.} Performing a linear stability analysis of a spatially extended system (using the proper eigenmodes for the geometry under consideration) one finds domains in control space $(n_1, n_2)$ which are locally unstable (red region) or laterally unstable (green region). Starting from a spatially homogeneous system in a domain of lateral instability but local stability (lilac point corresponding to average masses $(\nbar_1, \nbar_2)$) there are two distinct scenarios: Either the amplitude of the patterns stays bounded and the system does not enter the domain of local instability (\textit{left}, supercritical case), or the amplitude grows large and enters the domain of local instability (\textit{right}, subcritical case).
	}
	\label{fig:control_space_flow}
\end{figure}

In that control space it is straightforward to perform a linear stability analysis of both the local and lateral stability of the spatially homogeneous state. 
Figure~\ref{fig:control_space_flow} illustrates the result of such a bifurcation analysis for a system with two conserved protein species with mass densities $n_1$ and $n_2$. 
The regions shaded in red and green show the domains of local and lateral stability, respectively.
The lilac spot indicates the average masses $\nbar_1$ and $\nbar_2$ of the two protein species.
Imagine now that the spatially extended system is dissected into a set of small (well-mixed) compartments which are represented as points in control space.
While for a spatially uniform state all of these points are condensed into the point  corresponding to the average protein densities, they become spread out in phase space as soon as there is a spatially heterogeneous distribution of protein masses.

In Fig.~\ref{fig:control_space_flow}, the average mass densities $\nbar_1$ and $\nbar_2$ are chosen such that it lies in a domain of lateral instability but local stability; see lilac point in Fig.~\ref{fig:control_space_flow}. 
Hence, any small perturbation of a spatially uniform state will lead to a pattern that initially is dominated by the wavelength (in the band of unstable modes) with the largest (positive) eigenvalue.
In control space this implies that the set of phase space points representing the spatially extended system form a cloud with a center of mass given by the phase space point marking the spatially homogeneous state.
As the amplitude of the pattern grows, there are two distinct scenarios.
Either, the amplitude of the pattern (deviation of the local masses from the average masses) stays bounded and remains in the domain of local stability (Fig.~\ref{fig:control_space_flow}, \textit{left}), or the amplitude becomes so large that eventually the dynamics becomes locally unstable in some of the compartments (Fig.~\ref{fig:control_space_flow}, \textit{right}).
The first case is typical for a supercritical instability, where upon approaching the onset of instability the amplitude of the pattern continuously goes to zero. 
In contrast, for a subcritical system, the amplitude of the pattern grows to large amplitude even at onset.
Hence, it does not remain bounded in a small domain of control space but may extend into the regime of local instability.   
This can lead to a complex spatiotemporal dynamics, as we will illustrate in the next section for the specific example of the Min system.

\subsection{Control space attractors of the \emph{in vitro} Min system} \label{sec:Min-control-space}

We illustrate the control space concept for the Min system in box geometry, which is frequently used in experimental studies of the reconstituted Min system on supported lipid bilayers \cite{Loose:2008a,Ivanov:2010a,Zieske:2014a,Vecchiarelli:2014a,Caspi:2016a,Denk:2018a,Glock:2018b,Glock:2019a}.
The spiral patterns observed in experiments are reproduced by numerical simulations of the Min skeleton model as discussed in Section~\ref{sec:MCRD_cell_geometry}, see Fig.~\ref{fig:Min-simulation-examples}.
The membrane is located at the bottom of the box and the height $h$ of the bulk volume is used as a control parameter.

For simplicity we consider a two-dimensional slice geometry with the membrane given by a line of finite length $L$ and the bulk (cytosol) as a slice with the same length $L$ and height $h$; see Fig.~\ref{fig:Min-simulation-examples}.
For the Min system there are two globally conserved chemical species, MinD and MinE, with local densities $\nD (x,z,t)$ and $\nE (x,z,t)$, respectively.
We define the local total densities in the immediate vicinity of the membrane ($z = 0$), normalised with respect to their time-independent spatial averages as  
\begin{align*}
	D (x,t) 
	&= \frac{\nD (x,z=0,t)}{\nbar_\mathrm{D}}
	\, , \\
	E (x,t) 
	&= \frac{\nE (x, z=0, t)}{\nbar_\mathrm{E}}
	\, .
\end{align*}
The total densities right at the membrane, rather than, for instance, their vertical averages, are the proper control variables, because only these densities participate in the nonlinear dynamics at the membrane at each instant $t$. For details on this so called `adiabatic extrapolation' please refer to \cite{Halatek:2018a}.

As coordinates in the control space spanned by the total densities, we use the local MinD-MinE ratio
\begin{equation*}
	\Delta (x,t) 
	:=
	E(x,t)/D(x,t)
\end{equation*}
and the local root mean-square sum of protein densities $\Sigma(x,t)$, defined as
\begin{equation*}
	\Sigma (x,t) 
	:=
	\frac{1}{\sqrt{2}}
	\left( 
	E^2 (x,t) + D^2 (x,t)
	\right)^{1/2}
	\, .
\end{equation*}

\begin{figure}[t!]
\centering
\includegraphics{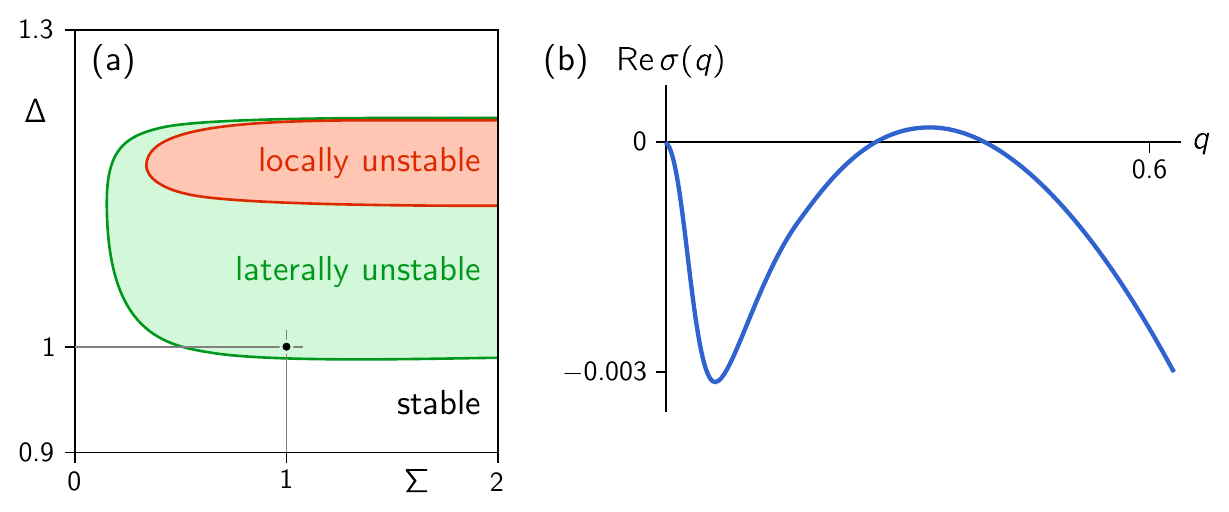}
\caption{\textbf{Linear stability analysis of the Min system.}
(a) Stability diagram in the parameter space of total densities of MinD and MinE, parametrized by $\Sigma$ and $\Delta$ (defined in text). Note that the ratio of MinE to MinD mass, $\Delta$, is the key parameter, while $\Sigma$ is almost irrelevant.
(b) Dispersion relation at the point $(\Sigma,\Delta) = (1,1)$, which lies very close to the onset of lateral instability.
 (Bulk height $h = \SI{20}{\micro m}$, remaining parameters as in Table~\ref{tab:Min-parameters}.)
}
\label{fig:MinDE_LSA}
\end{figure}

For specificity, we choose the Min skeleton model aas the reaction scheme (Fig.~\ref{fig:biochemical-networks}), resulting in the set of dynamic equations specified in Section~\ref{sec:MCRD_cell_geometry}; the parameters for the reaction rates, diffusion constants, and the average protein densities are given in Table~\ref{tab:Min-parameters}. 
Performing a linear stability analysis with respect to a spatially homogeneous state in box geometry~\cite{Halatek:2018a,Brauns.etal2020a}, one obtains a stability diagram in control space that contains domains of local and lateral instability.
Figure~\ref{fig:MinDE_LSA}a shows the stability diagram obtained for $h = \SI{20}{\micro m}$ which is chosen such that the spatially homogeneous state is slightly above the rim of lateral instability.  
The corresponding dispersion relation (for the real part of the largest eigenvalue), shown in Fig.~\ref{fig:MinDE_LSA}b, exhibits a narrow band of unstable modes, where $q$ denotes the wave vector in lateral direction $x$.
This dispersion relation implies that a spatially homogeneous state is (barely) unstable and initially a periodic pattern will form with a wavelength determined by the fastest growing mode. 
However, it does not inform us about the ensuing spatiotemporal dynamics far away from the homogeneous steady state.

\begin{landscape}
\begin{figure}
\includegraphics[width=1.3\textwidth]{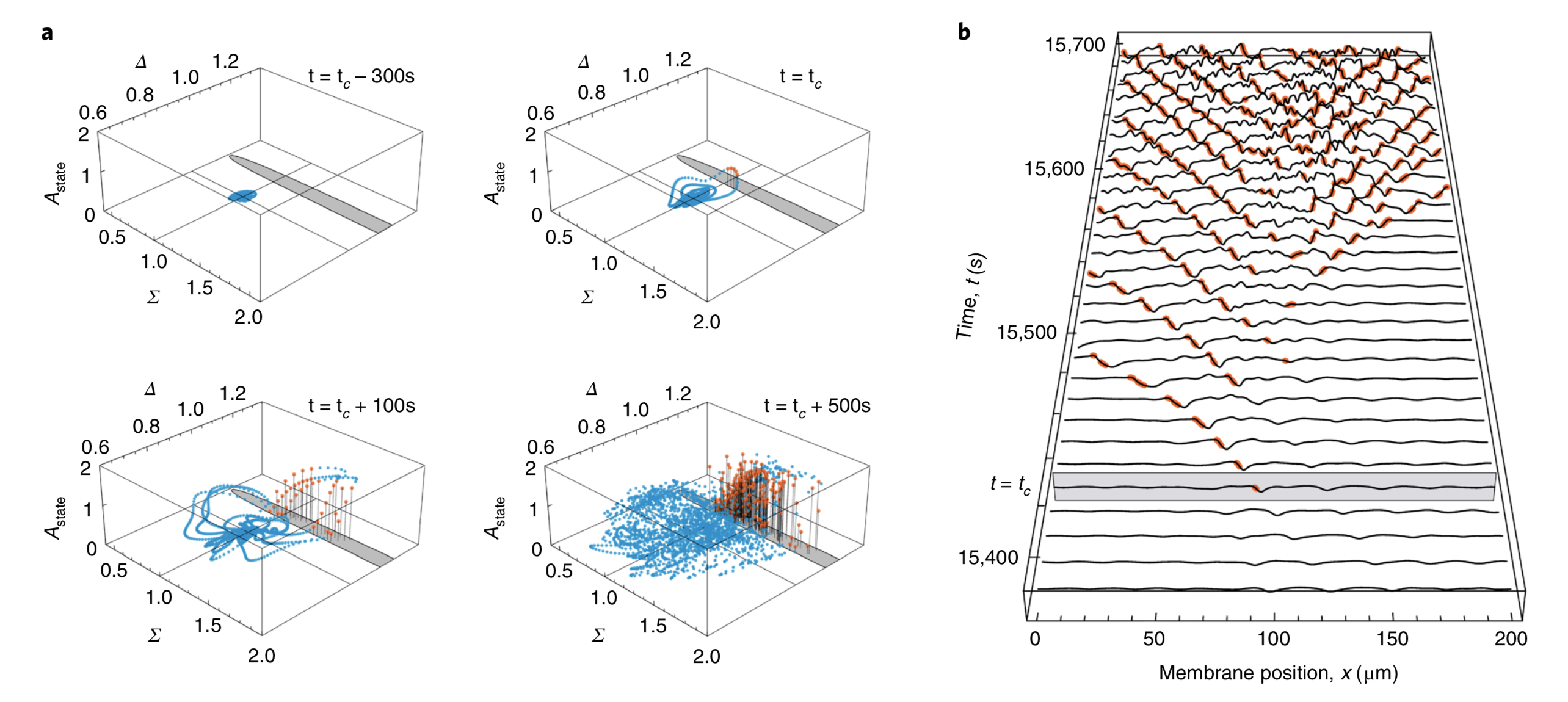}
\caption{
\textbf{Time evolution of a chaotic pattern in control space.}
(a) Snapshots of the control space distribution at four times. Starting from a nearly homogeneous initial state (i.e.\ a point in control space), the distribution expands driven by lateral instability alone until, at a time $t_\mathrm{c}$, it starts to enter the regime of local instability (top right panel). In the locally unstable regime (shaded in grey), the pattern is driven away from the local equilibria resulting in a strong increase in $A_\mathrm{state}$.
(b) Spatiotemporal waterfall plot of the MinD membrane density. Red points mark locations where the local equilibrium is unstable.   
(Parameters as in Fig.~\ref{fig:MinDE_LSA}; adapted from \protect\cite{Halatek:2018a}.) 
}
\label{fig:MinDE_time_evolution_example}
\end{figure}
\end{landscape}

To learn about the nature of these highly nonlinear dynamics, we follow the dynamics both in real space and in control space starting from a slightly perturbed, homogeneous steady state, similar as for the two-component system.
Snapshots of the control space distribution at four different time points together with a spatiotemporal `waterfall plot' of the MinD membrane density are shown in Fig.~\ref{fig:MinDE_time_evolution_example}.
The control space plots show the euclidean distance $A_\text{state}$ of the pattern from the respective local equilibrium as a function of the local dynamic control variables $\Delta(x,t)$ and $\Sigma(x,t)$; for a precise definition of $A_\text{state}$ see \cite{Halatek:2018a}. 
Initially, the amplitude of the pattern grows, which corresponds to an expanding cloud of points in control space; see upper left panel in Fig.~\ref{fig:MinDE_time_evolution_example}a.
As long as this cloud remains in the domain of local stability, the local deviation of the pattern from the local equilibria, $A_\text{state}$, remains small. 
Even though the system is very close to onset (cf.\ Fig.~\ref{fig:MinDE_LSA}), the pattern does not saturate at small amplitude but continues to grow, corresponding to a \emph{subcritical} bifurcation. 
Therefore, the extension of the cloud will eventually, at some time $t_c$, be so large that for some points in space the dynamics becomes locally unstable.
This is indicated by the red points in Fig.~\ref{fig:MinDE_time_evolution_example}a that have entered the area of local instability shaded in grey.
At these spatial positions, the distance between the pattern and the local equilibria strongly increases. Concomitantly, a propagating wavefront forms in the real space pattern; compare Fig.~\ref{fig:MinDE_time_evolution_example}a and Fig.~\ref{fig:MinDE_time_evolution_example}b.
As the cloud in control space further expands into the locally unstable domain in control space an increasing number of locally unstable spatial domains emerge (marked in red in Fig.~\ref{fig:MinDE_time_evolution_example}b).
Because the local instabilities are oscillatory, they act as local pacemakers sending out propagating wave fronts that start to interact with each other finally leading to a steady state that shows clear signatures of spatiotemporal chaos (see \cite{Halatek:2018a} for a detailed characterisation).
This is reflected in the geometry of the corresponding attractor in control space that appears rather disordered as well.

Taken together, this shows that the interplay between lateral mass redistribution --- driving the expansion of the cloud of representative points in control space --- and the changes in position and stability of local equilibria --- driving the reactive fluxes --- is the essential dynamics that determines the geometry of the attractor in control space, and thereby also the spatiotemporal dynamics of the patterns.

\paragraph{Control modes and pattern formation}

What is the relationship between patterns and the dispersion relation?
In the case of a supercritical bifurcation close to the transition, the pattern is well characterised by the fastest growing mode.
As we have learned above, this is not at all the case here. 
Strikingly, one finds spatiotemporal chaos close to the onset of lateral stability, where only very few modes are unstable.
Generally, patterns emerging from a subcritical bifurcation cannot be characterised in terms of the dispersion relation alone.
However, a systematic numerical study of the dynamics as one increases the bifurcation parameter $h$ shows that there is a remarkable relation between the dispersion relation and the observed pattern.

\begin{figure}[tb]
\centering
\includegraphics{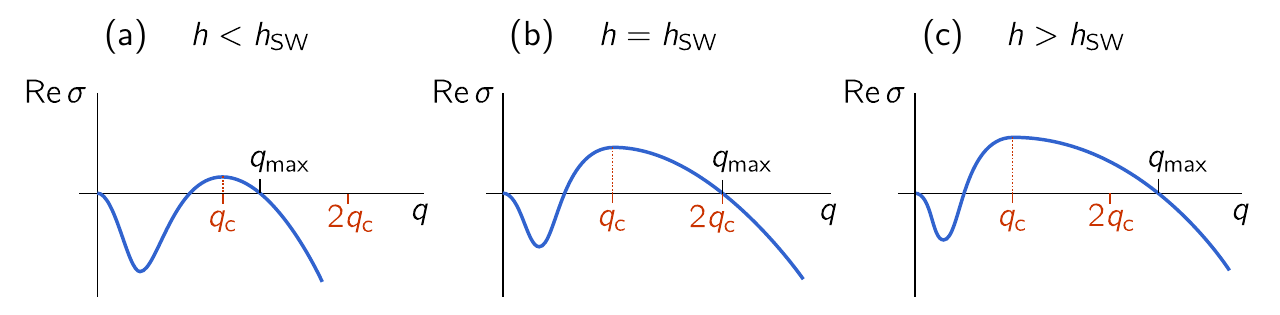}
\caption{
\textbf{Illustration of the dispersion relation as a function of bulk height}. At $h = h_\mathrm{SW}$, the right edge of the band of unstable modes (unstable mode with the shortest wavelength), $q_\mathrm{max}$, becomes commensurate with the fastest growing mode $q_\mathrm{c}$.
}
\label{fig:Min-commensurability}
\end{figure}

Upon increasing the height, one observes a transition from turbulence-like patterns to spatially coherent standing wave patterns, at threshold value $h_{\text{SW}} \,{\approx}\, \SI{23.5}{\micro m}$.
This transition coincides precisely with the point where the unstable mode $q_\text{max}$ with the shortest wavelength (right side of the band of unstable modes) becomes commensurable with the critical (fastest growing) mode $q_{\mathrm{c}}$:  $q_{\text{max}} \,{=}\, 2q_{\mathrm{c}}$; see Fig.~\ref{fig:Min-commensurability}.
Increasing the height even further, both the critical mode $q_{\mathrm{c}}$ and its commensurate mode $q_{\mathrm{r}} \,{=}\, 2 q_{\mathrm{c}} < q_{\text{max}}$ are linearly unstable and one observes travelling waves. 
Heuristically, this can be explained as the coordinated interplay of the fastest growing mode $q_{\mathrm{c}}$, which controls the distribution of wave sources and sinks by triggering local instabilities, and its commensurate mode $q_{\mathrm{r}}$ which drives the mass-redistribution between the wave nodes \cite{Halatek:2018a}.
However, a systematic understanding of this route from chaos to order and the associated commensurability criterion remains elusive.
The role of the local instabilities that act as pacemakers and are periodically triggered by lateral mass redistribution suggests that the local oscillatory instability is a key requirement.

\begin{landscape}
\begin{figure}
\includegraphics[scale=0.85]{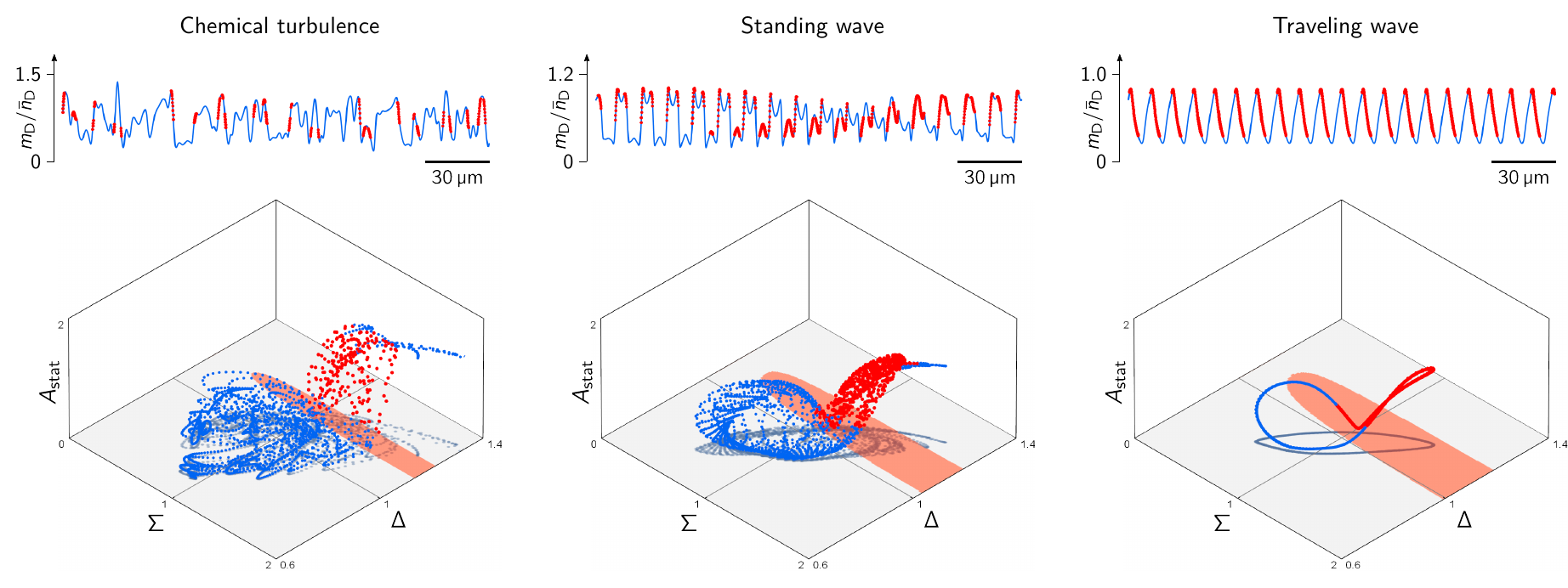}
\caption{
Snapshots of the density profiles and the corresponding control space distributions of the \emph{in vitro} Min system with bulk heights $h = 21, 30$, and \SI{35}{\micro m}, from left to right; remaining parameters as in Table~\ref{tab:Min-parameters}. The regime of local instability is shaded in red in the control space plots. Note how the increasing spatial coherence in the pattern profiles is reflected in an increase in order and decrease in dimensionality of the control space attractors.
}
\label{fig:in-vitro_control_space}
\end{figure}
\end{landscape}

In control space, the transition between patterns is reflected in a change in the type of attractor (Fig.~\ref{fig:in-vitro_control_space}).
The control-space attractor corresponding to the turbulence-like state for $h < h_{\text{SW}}$ has no discernible structure. 
Standing waves undergo amplitude modulations on large spatial scales. Therefore the distribution in control space appears as a broad band. In contrast to the control attractor corresponding to chemical turbulence, the standing wave attractor is highly ordered.
The spatial and temporal periodicity of traveling waves is reflected in a collapse of the control space distribution onto a single closed orbit, which corresponds to the cycle traversed by the concentrations at each point in space during one wave period.

The prediction of turbulence-like dynamics at the onset of lateral instability and coherent patterns deeper in the unstable regime has recently been confirmed in experiments with the \emph{in vitro} Min system~\cite{Denk:2018a}.
Moreover, local oscillations of the Min proteins between membrane and bulk have been observed in vesicles (radius $\sim \SI{10}{\micro m}$)~\cite{Litschel:2018a,Godino:2019a}.
Together, these experimental findings suggest that the mechanisms identified by the theoretical analysis indeed underlie the dynamics in the experimental system.

%% ================================
%% CONCLUSIONS AND OUTLOOK
%% ================================

\section{Conclusions and Outlook}

In this final section we will not only give an overview of the theoretical concepts that we introduced, but also provide an outlook on open questions.
The conclusions will summarise generic design aspects of protein-interaction networks for pattern formation, and emphasise the overarching ideas underlying the theoretical analysis of the mathematical models describing such systems: MCRD equations. 
In a nutshell this theory includes the local reactive concept, the mechanism of the mass-redistribution instability, a flux-balance construction in the phase plane of the reaction kinetics, and the control space concept.
In addition, we briefly discuss the consequences of cytosolic gradients normal to the membrane that arise generically from the bulk-boundary coupling between cytosol and membrane.

In the outlook we will discuss some recent ideas on how to apply and generalise the theory of MCRD systems. 
In the context of (conceptual) two-component systems, we discuss forthcoming works on the question of length scale selection (coarsening), the role of cytosolic flow, the role of heterogeneity in the spatial domain serving as a `template', and the role of a vertically extended cytosol. 
Moreover, we give a perspective on the generalisation of local equilibria theory beyond two-component models and strict mass conservation, focussing on the topics of model classification and model reduction.
Finally we propose to revisit classical pattern-forming systems like the Belousov--Zhabotinsky reaction, treating them as nearly mass conserving systems.

\subsection{Conclusions}

\paragraph{Generic design of pattern-forming protein networks}
The reaction-diffusion equations describing the spatiotemporal dynamics of proteins in cells exhibit several characteristic features:
\begin{itemize}
\item The underlying biochemical reaction networks have one or several NTP\-ase cycles as core modules. These cycles consume free energy by converting NTP to NDP thereby breaking detailed balance such that the dynamics is genuinely non\--equilibrium in nature. Additional key elements of the reaction networks include proteins that activate and deactivate the NTPases, termed NEFs and NAPs, as well as various types of feedback mechanisms. 
\item On the timescale of pattern formation and maintenance, protein production and degradation that break the conservation of mass for each protein species are typically negligible. Hence protein dynamics is described by mass-conserving reaction--diffusion equations.
 Production and degradation of proteins slowly change the average protein masses that play the role of control parameters for these dynamics. 
 \item The proteins diffuse on the membrane and in the cytosol, i.e.\ on a two-dimensional closed surface and in a three-dimensional confined bulk volume. Most chemical reactions are confined to the membrane, while the cytosol mainly serves as a (spatially non-uniform) particle reservoir. Chemical reactions in the cytosol are largely linear processes like nucleotide exchange such that the reaction--diffusion equations in the cytosol can be solved in closed form using eigenfunctions of the diffusion (Laplace) operator appropriate for the particular cell geometry. 
\item The diffusion constant for proteins in the membrane and the cytosol differ by orders of magnitude. While this is important for the quantitative analysis of a given system, the pattern formation mechanism that we discussed only require that these diffusion constants are different.  
\item Proteins are exchanged between membrane and cytosol respecting local conservation of protein masses. Reactive fluxes between membrane and cytosol necessarily lead to spatial protein gradients in the cytosol. Due to these gradients, the spatial geometry of the system (e.g.\ the cell shape) can play an essential role for the formation of patterns. This entails that the cytosol can, in general, not be considered as a well-mixed particle reservoir, but rather one has to account for its full spatiotemporal dynamics. 
 \end{itemize}

\paragraph{Phase-space geometry of nonlinear systems}
The overarching idea in the description of nonlinear systems is their characterisation in terms of `geometric structures' in phase space, a key mathematical insight going back to the work of Poincar\'e \cite{Poincare:Book}.
Instead of trying to solve the set of nonlinear equations in closed form --- which is not possible anyway except in few exceptional cases --- the mathematical analysis aims at identifying and charactertizing the key geometric structures in the phase space.
In these lecture notes, we have first illustrated the underlying basic ideas on nonlinear systems of well-mixed reaction networks with an emphasis on mass-conserving systems. 
A major focus there were one- and two-component systems, where the key geometric objects are nullclines, defined as curves in phase space along which one of the system's two variables is in equilibrium. 
Points at which nullclines intersect mark equilibria (fixed points) of the system. 
These geometric objects organise phase-space flow, and thereby allow us to infer the qualitative dynamics from the shapes and intersections of the nullclines.
Key concepts like linear stability, excitability, multi-stability, and limit cycles can be understood at this geometric level. 
Transitions (bifurcations) between qualitatively different regimes are revealed by structural changes of the flow in phase space as the control parameters are varied. 
One key advantage of such a geometric approach to nonlinear dynamical systems is that it yields systematic physical insights into the processes driving dynamics without requiring the explicit solution of the full set of equations. 

\paragraph{Local equilibria theory}
In these lecture notes we have discussed how these phase space ideas for well-mixed systems can be generalised to spatially extended reaction-diffusion systems.
Our discussion builds on recent work~\cite{Brauns.etal2020c} that introduces a \emph{`local equilibria theory'} for mass-conserving two-component systems.
Based on the insight that mass transport is the essential driver of pattern formation, this theory introduces the following central concepts:
\begin{itemize}
\item \textbf{Local (reactive) equilibria.} The total density (mass) at each position in space determines a \emph{reactive equilibrium} at this position. These local equilibria serve as proxies for the local reactive flow in the same way as fixed points and their stability are used in the analysis of ODE systems to infer qualitative features of the phase space flow.
\item \textbf{Mass-redistribution instability.} The general physical mechanism underlying the Turing instability in MCRD systems is a mass-redistribution instability. It is based on a feedback loop between lateral redistribution of mass and moving local equilibria. As local masses change, the local equilibria shift, thus driving the formation of gradients, which in turn lead to further redistribution of mass. In two-component MCRD systems, the condition for this instability can be phrased as a simple geometric criterion: negative slope of the reactive nullcline in the phase-portrait of the reaction kinetics. generalising the mechanism of a mass-redistribution instability beyond reaction--diffusion systems reveals an analogy between a Turing instability and Ostwald ripening~\cite{Wagner:1961a}.
\item \textbf{Diffusive flux balance.} In a stationary pattern, the diffusive fluxes on the membrane and in the cytosol have to balance at each position in space. This condition implies that a stationary pattern must lie in an affine subspace (i.e.\ a straight line) in phase space.
\item \textbf{Flux-balance construction.} Combining the concepts of local equilibria and diffusive flux balance enables one to construct stationary patterns (and their bifurcation diagrams) in the phase portrait of the reaction kinetics. This construction can be pictured as an analog to the Maxwell construction (also known as `common tangent construction'~\cite{Cates:2018a}) used to find the equilibrium phases in the phase separation of binary mixtures.
\item \textbf{Concept of regions.} In general, patterns can be understood as consisting of distinct spatial regions (e.g.\ plateaus with high/low density, interfaces, etc.), which correspond to different regions in phase space. This general idea can be employed to rigorously derive specific properties of the pattern, like the interface width and the threshold perturbation for stimulus-induced pattern formation, from the geometric phase-space analysis. For a more in-depth analysis of this idea we refer the interested reader to \cite{Brauns.etal2020c}.
\item \textbf{Control variables and control space.} Total densities play the role of control parameters for the local reactive dynamics while they are variables of the spatially extended system. We therefore refer to them as \emph{control variables} and introduce the \emph{control space}, which is the phase space spanned by the control variables. The control space and the bifurcation structure in it (obtained by linear stability analysis) can be used to characterise and classify the MCRD dynamics.
\end{itemize}

\paragraph{Bulk-boundary coupling} 
A general property of intracellular protein patterns is the coupling between protein dynamics on the membrane and protein dynamics in the cytosol. 
There is an exchange of proteins between both compartments due to linear and cooperative attachment and detachment processes.
An immediate consequence of this coupling is the existence of protein gradients in the cytosol, whose extent depends on the cytosolic diffusion constant as well as on some characteristic rates for the exchange of proteins between membrane and cytosol or the reactivation of proteins in the cytosol (Sec.~\ref{sec:bulk-boundary-coupling}).
For example, nucleotide exchange in the bulk leads to depletion zones of active proteins within a layer of size $\ell = \sqrt{D_\text{c}/\lambda}$.
These cytosolic gradients are responsible for the geometry dependence of the ensuing protein patterns~\cite{Thalmeier:2016a,Wu:2016a,Gessele:2020a}.
A reduced description in terms of surface variables (membrane concentrations and cytosolic concentrations at the membrane) by elimination of the cytosolic dynamics normal to the membrane is possible only if the diffusion length corresponding to the gradients is much larger than the size of a cell or the height of a microfluidic device \textit{in vitro}~\cite{Brauns.etal2020a}.
Finally, we introduced the basic elements of linear stability analysis for systems with bulk-boundary coupling which complicates the linear stability analysis of the \emph{laterally} homogeneous steady states (Sec.~\ref{sec:LSA-box-geometry}). The key step there is to find a common set of spatial eigenmodes that diagonalises the diffusion operator on the membrane, in the cytosol, and in the bulk-boundary coupling simultaneously. These eigenmodes then play an analogous role to the Fourier modes in the classical linear stability analysis of one-dimensional systems.

\paragraph{Dynamics in control space}

The control space concept can be used to characterise and to rationalise numerically obtained dynamics as we demonstrated in Sec.~\ref{sec:Min-control-space} for the \textit{in vitro} Min system.
There, we have summarised the findings of recent papers~\cite{Halatek:2018a,Denk:2018a,Brauns.etal2020a}.
The analysis in control space reveals that the lateral redistribution of mass locally triggers destabilisation of the local equilibria.
This local destabilisation is responsible for the development of spatiotemporal chaos close to the onset of lateral instability.
This prediction from numerical simulations has recently been confirmed in experiments~\cite{Denk:2018a}.
Further away from onset, standing and travelling waves emerge from an interplay between commensurable `control modes' that synchronise the points in time and space where local instabilities are triggered.
A systematic understanding of this route from spatiotemporal chaos to order remains elusive.
Nonetheless, the progress made so far demonstrates how the control space concept helps to bridge the gap between linear stability analysis and highly nonlinear dynamics far away from the homogeneous steady state. 

\paragraph{The Min system}
Throughout these lecture notes, the Min system of \emph{E.~coli} has served as a paradigmatic example for protein-based pattern formation.
This reflects the important role of the Min system in the field, both for experimental and theoretical research on intracellular pattern formation; see e.g.\ \cite{Halatek:2018b} and references cited therein.
A major experimental milestone was the reconstitution of the Min system \textit{in vitro} by \cite{Loose:2008a}. 
These experiments clearly demonstrated, that only two proteins, MinE and MinD, and chemical fuel in the form of ATP are required for pattern formation on a lipid bilayer that mimics the cell membrane.
This reconstitution provides a minimal system that enables precise control of reaction parameters and geometrical constrains \cite{Loose:2008a,Ivanov:2010a,Zieske:2014a,Vecchiarelli:2014a,Vecchiarelli:2016a,Caspi:2016a,Denk:2018a,Glock:2018a,Glock:2018b,Litschel:2018a,Kohyama:2019a,Glock:2019a}.
These studies show that, despite its molecular simplicity, the Min system exhibits a remarkable variety of phenomena.

The rich phenomenology of the Min system, and the relative simplicity of its protein-interaction network, also make it an ideal testing ground for theoretical ideas such as the control space analysis~\cite{Halatek:2018a,Denk:2018a} and the role of cell geometry~\cite{Halatek:2012a,Wu:2016a,Wu:2015a}.
In addition, upon engineering the interaction network by modifying the protein domains one can address the relevance of specific processes in the reaction network for the ability of the system to robustly form patterns~\cite{Denk:2018a} and design minimal pattern-forming systems~\cite{Glock:2019a}.

Despite the significant progress there still remain several important open questions.
First, the precise biomolecular mechanism underlying MinD's self-recruitment still remains elusive.
A possible mechanism based on indirect mechano-chemical interactions has recently been suggested by \cite{Goychuk:2019a} but has yet to be tested experimentally.
Second, there are striking differences between the dynamics in cells and in reconstituted systems, both qualitatively (pole-to-pole- and stripe-oscillations vs.\ spiral waves, uniform oscillations, quasi-stationary patterns, etc.) and quantitatively concerning the characteristic length- and timescales of the patterns.
In a forthcoming publication, we will address this long standing puzzle and show that the ratio of bulk-volume to surface-area (i.e.\ the bulk height) is the key parameter that distinguishes the \emph{in vivo} regime from classical \emph{in vitro} setups~\cite{Brauns.etal2020a}. 
A characterisation of the dynamics in control space shows that there is no local instability at low bulk heights, such that patterns are driven by lateral instabilities alone.
Local instabilities only come into play at sufficiently large  bulk heights, where they give rise to the phenomena characteristic for the \emph{in vitro} system, as discussed in Sec.~\ref{sec:Min-control-space}.
A third open question is the mechanism underlying the quasi-stationary patterns observed in recent experiments using reconstituted MinE with the purification tag removed~\cite{Glock:2018b}.
Further extensions of the skeleton model~\cite{Huang:2003a,Halatek:2012a} may be required to account for this phenomenon.

The future looks bright! With new advances in microfluidics and protein chemistry, we expect the Min system as well as other reconstituted protein systems to even gain importance as model systems for testing new theoretical ideas on self-organisation in biological systems. 
In particular, these systems will be instrumental in further extending and generalising the ideas and concepts presented in these lecture notes and our recent publications~\cite{Halatek:2018a,Brauns.etal2020c}.

\subsection{Outlook}

%\subsubsection{Conceptual two-component models}

Two-component MCRD systems were instrumental in developing general phase space concepts like the flux-balance subspace and the physics of the underlying pattern forming mechanism, namely the mass-redistribution instability. 
These systems have also been instrumental in identifying the total protein mass and the mass-redistribution potential as key quantities for a conceptual understanding of the mechanisms driving pattern formation and leading to the formation of stationary states. 
However, there still remain a series of open questions. 
These include the coarsening dynamics, the role of heterogeneities in the spatial domain that act as `templates' or `pre-patterns' and cytoplasmic advective flows, and generalisations that explicitly take into account the extended bulk and the accompanying reactive boundary conditions (bulk-boundary coupling).

\paragraph{Coarsening in two-component models}
While we now have a rather complete understanding of the instability mechanism and the final polar pattern of two\-/component MCRD systems, the mechanism responsible for the coarsening process from an initially periodic pattern towards the polar pattern remains elusive. 
The discussion in Sec.~\ref{sec:MCRD_stationary-states}, where we introduced the mass-redistribution potential hints towards a possible solution. 
Indeed, coarsening can be understood as a mass-redistribution instability driven by gradients in the mass-redistribution potential between spatial regions, each containing an elementary pattern (a single `peak' or `mesa')~\cite{Brauns.etal2020}.
Using elementary geometric arguments that build on the phase-portrait analysis presented here, we show that uninterrupted coarsening is indeed generic in two-component reaction--diffusion systems with mass conservation~\cite{Brauns.etal2020}.

The question of coarsening is a particular case of the more general problem of length-scale selection far away from the homogeneous steady state. In our view, this is one of the key open challenges in the field. 
Previous attempts to find a general theory remain restricted to one-component systems~\cite{Politi:2004a,Politi:2006a}.
Our analysis of two-component MCRD systems shows that the spatial redistribution of globally conserved quantities plays a key role in the coarsening dynamics. 
Combining mathematical tools like multi-scale analysis, renormalisation group theory, and singular perturbation theory with the physical concepts of local equilibria theory a new approach to coarsening dynamics has to be developed that can be generalised to multi-component systems.

\begin{figure}
	\centering
	\includegraphics[scale=1]{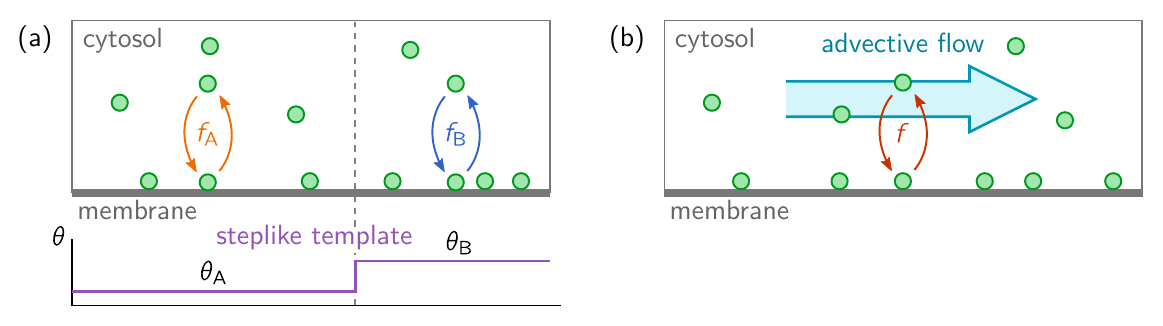}
	\caption{Illustration of two-component MCRD systems with cytoplasmic advective flow (a) and with a steplike spatial template (b).
	}
	\label{fig:template-and-flow}
\end{figure}

\paragraph{Cytoplasmic advective flows} 
Cytoplasmic advective flows play an important role in cellular systems~\cite{Mogilner:2018a,Mittasch:2018a,Gross:2018a}.
The influence of bulk flow on protein-based pattern formation has also been studied in the reconstituted Min system \cite{Ivanov:2010a,Vecchiarelli:2014a}.
Conceptually this can be studied by straightforward generalisation of the two-component system discussed in Sec.~\ref{sec:two-component_MCRD}, illustrated in Fig.~\ref{fig:template-and-flow}a, namely by adding an advection term to the cytosolic dynamics:
\begin{equation}
	\partial_t c(x,t) + v \, \partial_x c(x,t) = D_c \partial_x^2 c(x,t) + f(m,c) \, .
\end{equation}
How does this affect the formation of patterns? Does this induce moving patterns and how is the direction of pattern motion correlated with the direction of cytoplasmic flow?
In \cite{Wigbers.etal2020a}, these questions are addressed using the concepts of local equilibria theory~\cite{Halatek:2018a,Brauns.etal2020c} discussed in Sec.~\ref{sec:two-component_MCRD}.
In particular, it is shown that in two-component systems, cytoplasmic flow generically induces motion of high-density regions upstream (i.e.\ against the flow direction). In contrast, preliminary results indicate that systems with more components exhibit both upstream and downstream propagation in response to cytoplasmic flow. The direction of propagation induced by the flow is found to depend sensitively on the underlying protein-interaction network, suggesting that externally imposed flows can be used to probe the molecular details of pattern-forming systems in experiments.

\paragraph{Patterns forming patterns} 
In his seminal paper on reaction-diffusion systems Turing pointed out \cite{Turing:1952a} that ``\textit{most of an organism, most of the time, is developing from one pattern into another, rather than from homogeneity into a pattern.}''
While it is not fully clear what he was referring to with this statement, there are several biological scenarios for such a hierarchy or sequence of patterns.
In one scenario, previously formed protein patterns can acts as a \emph{spatial template} for the formation of subsequent protein patterns by affecting their local reaction kinetics.
A more intriguing scenario is a fully \emph{self-organised temporal sequence} of protein patterns with increasing complexity, as recently observed in starfish ooctyes~\cite{Wigbers.etal2020b}.
Another striking example of this is the morphogenesis of the drosophila embryo, where the positional information provided by the (maternal) bicoid gradient is precisely read out and translated into a complex gene expression pattern via a hierarchy of gene regulatory circuits \cite{Nusslein-Volhard:1980a}.
While, patterns in drosophila appear to be highly regulated by sophisticated gene networks, the pattern hierarchies in starfish oocytes are much simpler as they are based on protein--protein interactions only.
One may therefore speculate that these are evolutionary predecessors of the more complex morphological programs as observed in drosophila.

In a recent paper~\cite{Wigbers.etal2020} we have analysed a particularly simple case of pattern formation in a heterogeneous spatial domain: a two-component MCRD systems with a step-like `template' that defines two spatial subdomains with different reaction kinetics (Fig.~\ref{fig:template-and-flow}b).
generalising the phase space ideas discussed in Sec.~\ref{sec:two-component_MCRD} enabled us to give a geometric construction for the ensuing stationary patterns. 
Interestingly, we found that step-like templates can trigger a regional mass-redistribution instability near the template edge, leading to the accumulation of protein mass, which eventually results in a stationary peak at the template edge.
This edge-localised peak is stable when  the template edge is slowly moved, but lost for  rapidly moving templates.

There are still many open questions. For example, how do these results generalise beyond a single, stationary, step-like template in one spatial dimension?
Most challenging, how do sequences of patterns form in a fully self-organised manner?
This would require to account for the dynamics of the template itself, and include a feedback from the downstream pattern to the template.
Such feedback may give rise to complex spatiotemporal behaviour like oscillatory patterns and traveling waves that can then be characterised by building on the basic local equilibria theory presented in Sec.~\ref{sec:two-component_MCRD}.
Finally, the shape of the membrane of a cell or of channels in a microfluidic device containing a reconstituted pattern-forming system may affect pattern formation, as we will discuss next.

\paragraph{The role of geometry and an extended cytosol}

Cells have different geometries. They may be spherical, ellipsoidal, or be deformed into other shapes by internally or externally acting forces.
In synthetic cell applications, protein pattern formation can be reconstituted in flow chambers \cite{Ivanov:2010a,Vecchiarelli:2014a} and microfluidic devices of arbitrary shape \cite{Caspi:2016a,Brauns.etal2020a}. 
How does shape affect pattern formation?
There are conceptual answers to this question that have recently been discussed in specific biological applications~\cite{Thalmeier:2016a,Gessele:2020a}.
In a nutshell, the reactivation rate (rate of switching an inactive NDP-bound into an active NTP-bound state) defines a depletion zone of active NTPases close to the membrane; see Sec.~\ref{sec:column_examples}.
It will then depend on the local ratio of membrane surface to cytosolic volume how  likely proteins of the different conformations reencounter the membrane, which affects pattern formation.
How can one formalise this heuristic argument such that the local equilibria theory can be used to study pattern formation?
This requires to extend the techniques explained in Sec.~\ref{sec:two-component_MCRD} to explicitly account for an extended cytosolic volume and the reactive coupling between membrane and cytosol; cf.\ Sec.~\ref{sec:bulk-boundary-coupling}.
Is it possible to find a reduction scheme that eliminates the extended cytosol and reduces the reaction-diffusion dynamics to an effectively one-dimensional system?
What are the conditions for such a reduction scheme to work?
Preliminary studies indicate that it is actually possible to answer these questions informing about the role of geometry and bulk-boundary coupling for MCRD systems~\cite{Ziepke:2016a,Ziepke_Frey:2020}. 
In this context, it is the geometry induced spatial distribution of masses that drives or inhibits pattern formation. 
As another distinct mechanism, a mechano-chemical feedback between biochemical reaction networks and the shape of the surrounding geometry can lead to a variety of complex dynamics~\cite{Goychuk:2019a}.

%\subsubsection{Beyond two components and strict mass conservation}

\subparagraph{Model classification and reduction}

One of the key insights underlying local equilibria theory is that mass-redistribution is the key driver of the dynamics. In other words, total densities are the essential degrees of freedom.
This provides a new perspective for model reduction and the classification of pattern forming systems. 

For instance, it has been suggested that the generic bifurcation scenario for cell polarity is a cusp bifurcation~\cite{Trong:2014a}. In fact we were recently able to show that patterns in two-component MCRD systems --- which are the conceptual and elementary case for polar pattern formation --- generically unfolds from a cusp bifurcation~\cite{Brauns.etal2020c}. This bifurcation scenario generally arises for N-shaped reactive nullclines (or, more generally, nullclines with a section of negative slope), which are necessary and sufficient for pattern formation in theses systems.

More complex phenomenology can arise in multi-component models with more than one conserved quantity --- such that the control space becomes multi-dimensional --- and/or due to more complex bifurcation scenarios in control space. 
The Min system encompasses both these scenarios. The pole-to-pole oscillations \textit{in vivo} arise due to mass redistribution of MinE between the cell poles, periodically switching the direction of MinD polarisation. Importantly, the local equilibria remain stable at all times during this oscillation cycle. This allows one to map the \textit{in vivo} Min-protein dynamics to relaxation oscillations in the two-dimensional control space by generalising the principle of scaffolding by local equilibria~\cite{Brauns.etal2020b}. 
The much richer phenomenology of the reconstituted Min system (\textit{in vitro}) emerges as a consequence of a more complex bifurcation scenario in control space, involving the destabilisation of local equilibria in Hopf bifurcations that give rise to local limit cycle oscillations (cf.\ Sec.~\ref{sec:Min-control-space}).
Motivated by these results, we anticipate that new model reduction approaches and a systematic classification scheme for MCRD systems can be derived based on the dynamics and bifurcation scenarios in control space.

\paragraph{Beyond strict mass conservation: revisiting classical systems}

The class of nearly mass-conserving reaction--diffusion systems comprises many widely studied (classical) pattern-forming systems, including the Belusov--Zhabotinsky (BZ) reaction \cite{Belousov:1959a,Zhabotinsky:1964a}, oxidation of carbon monoxide on platinum surfaces (short PtCO system) \cite{Eiswirth:1986a,Rotermund:1990a}, and intracellular calcium (Ca\textsuperscript{2+}) oscillations \cite{Berridge:2000a,Falcke:2004a}.
It is likely that each of these systems exhibits a mass-conserving core, meaning that their capability to form patterns and key features of these patterns are captured by a mass-conserving subset of their reaction kinetics. 
This makes these systems potentially amenable to local equilibria theory and the tools and concepts presented in Sec.~\ref{sec:two-component_MCRD} and Sec.~\ref{sec:control-space}.
The reactions breaking mass-conservation may then be treated as perturbations to the core system that modify the patterns.

As an example, take the Brusselator model, which can be rewritten as a mass-conserving two-component system (its `core'), perturbed by a constant production and a linear degradation term (cf.\ Eq.~\eqref{eq:Brusselator-rewritten}).
Pattern formation is driven by a mass-redistribution instability of the core system, and the stationary patterns of the core system can be constructed by the phase-portrait analysis presented in Sec.~\ref{sec:two-component_MCRD}. The production and degradation terms perturb these core patterns on large length- and timescales. In particular, they interrupt the coarsening process at a finite length scale \cite{Kolokolnikov:2006a,Kolokolnikov:2007a}. 
Applying this approach to more classical pattern-forming systems is an exiting perspective for future research and might provide a deeper understanding of their rich phenomenology.

%% ================================
%% ACKNOWLEDGEMENTS
%% ================================

\paragraph{Acknowledgements}

We would like to thank all of our collaborators that over the years have shaped our understanding of pattern formation in biological systems, especially Silke Bergeler, Jonas Denk, Raphaela Ge\ss ele, Andriy Goychuk, Jacob Halatek, Tobias Hermann, Ben Kl\"under, Ching Yee Leung, Tobias Reichenbach, Steffen Rulands, Dominik Thalmeier, Henrik Weyer, Manon Wigbers, Laeschkir W\"urthner, and Alexander Ziepke. 
Without them it would have been impossible to get to the state of writing lecture notes on this topic.  
Our special thanks also goes to our experimental collaboration partners from the laboratories of Cees Dekker, Nikta Fakhri, Liedewij Laan, and Petra Schwille. They are a continuous source of stimulation and inspiration for our theoretical work.
Finally, we would like to gratefully acknowledge financial support from the Deutsche Forschungsgemeinschaft (DFG, German Research Foundation).  

%% ================================
%% BIBLIOGRAPHY
%% ================================

%\bibliography{library}
%\input{./master.bbl}

\end{document}